\documentclass[10pt, a4paper, bibtotoc, english, idxtotoc, liststotoc]
{scrbook}

\usepackage{amsfonts}

\usepackage{amssymb}

\usepackage[english]{babel}

\usepackage{color}

\usepackage{exscale}

\usepackage{geometry}

\usepackage[final]{graphicx}

\newcommand{\href}[2]{#2}

\usepackage[latin9]{inputenc}

\usepackage{makeidx}

\usepackage{mathpple}

\usepackage{thumbpdf}

\makeindex

\index{gauge evolution condition|see{gauge condition}}
\index{gauge fixing condition|see{gauge condition}}

\newcommand{\sep}{\hspace{0pt}}

\newcommand{\trace}{\mathrm{trace}\,}

\begin{document}

% -*-LaTeX-*-
% $Header: /home/eschnett/cvs/diss/prefix.tex,v 1.6 2003/01/10 16:39:49 eschnett Exp $

%% \title{Gauge fixing for the simulation of black hole spacetimes}
%% \author{Erik Schnetter}
%% \date{Anno MMII}
%% \maketitle

\thispagestyle{empty}

\begin{centering}
\large

{\huge Gauge fixing for the simulation of black hole spacetimes}

\bigskip
Dissertation

zur Erlangung des Grades eines Doktors

der Naturwissenschaften

der Fakultät für Mathematik und Physik

der Eberhard--Karls--Universität zu Tübingen

\bigskip
vorgelegt von

Erik Schnetter

aus Letmathe

\bigskip
{\Large 2003}

\end{centering}

%% \noindent Gauge fixing for the simulation of black hole spacetimes
%% \\[3ex] Erik Schnetter
%% \\[1ex] Anno 2002
%% % \\[3ex] Theoretische Astrophysik
%% % \\ Auf der Morgenstelle
%% % \\ Universit\"at T\"ubingen
%% % \\ 72076 T\"ubingen
%% % \\ Germany
%% \\[3ex] Falkenweg 14
%% \\ 72076 T\"ubingen
%% \\ Germany
%% \\[1ex] Email: \href{mailto:schnetter@uni-tuebingen.de} {Erik
%% Schnetter \textless schnetter@uni-tuebingen.de\textgreater}
%% \\[1ex] Web: \href{http://www.tat.physik.uni-tuebingen.de/~schnette/}
%% {http://www.tat.physik.uni-tuebingen.de/\textasciitilde schnette/}

\newpage

\begin{tabular}{ll}
Tag der mündlichen Prüfung: & 13. Juni 2003 \\

Dekan: & Prof. Dr. Herbert Müther \\

1. Berichterstatter: & Prof. Dr. Hanns Ruder \\

2. Berichterstatter: & Priv.-Doz. Dr. Jörg Frauendiener
\end{tabular}

% -*-LaTeX-*-
% $Header: /home/eschnett/cvs/diss/abstract.tex,v 1.5 2003/01/10 16:39:49 eschnett Exp $

\chapter*{Abstract}
\label{abstract}
\addcontentsline{toc}{chapter}{Abstract}

%% I consider the initial-boundary-value-problem of fully nonlinear
%% general relativistic vacuum spacetimes, which today cannot yet be
%% evolved numerically in a satisfactory manner.  That is, I look at
%% gauge conditions, classifying them into gauge evolution conditions and
%% gauge fixing conditions.  I then present a system of evolution
%% equations containing a gauge fixing condition, and describe an
%% efficient numerical implementation.  I examine the behaviour of this
%% system for several test problems, such as Brill waves or black holes.

I consider the initial-boundary-value-problem of nonlinear general
relativistic vacuum spacetimes, which today cannot yet be evolved
numerically in a satisfactory manner.  Specifically, I look at gauge
conditions, classifying them into gauge evolution conditions and gauge
fixing conditions.  In this terminology, a gauge fixing condition is a
condition that removes all gauge degrees of freedom from a system,
whereas a gauge evolution condition determines only the time evolution
of the gauge condition, while the gauge condition itself remains
unspecified.  I find that most of today's gauge conditions are only
gauge evolution conditions.

I present a system of evolution equations containing a gauge fixing
condition, and describe an efficient numerical implementation using
constrained evolution.  I examine the numerical behaviour of this
system for several test problems, such as linear gravitational waves
or nonlinear gauge waves.  I find that the system is robustly stable
and second-order convergent.  I then apply it to more realistic
configurations, such as Brill waves or single black holes, where the
system is also stable and accurate.

% LocalWords:  eschnett Exp

% -*-LaTeX-*-
% $Header: /home/eschnett/cvs/diss/ack.tex,v 1.5 2003/01/10 16:39:49 eschnett Exp $

\chapter*{Acknowledgements}
\addcontentsline{toc}{chapter}{Acknowledgements}

I have greatly enjoyed all the research that has gone into this
thesis.  This enjoyment comes not least from the interaction that I
have had with my supervisors and colleagues, and the liberal and
inspiring atmospheres that they created in their departments.

It was certainly luck that brought me to the Institut f\"ur
Theo\-re\-ti\-sche As\-tro\-phy\-sik in T\"ubingen.  In countless
discussions with my colleagues I there learned the joy that lies in
academic discourse and in the discovery of new and old ideas.  My
supervisor Hanns Ruder has the rare ability of capturing his
listeners' whole attention by explaining physics in a most interesting
manner.  He also encouraged me to spend some time abroad.

The Deutsche Akademische Austauschdienst then granted me a stipend
that allowed me to spend one year with Pablo Laguna at the Department
of Astronomy in Penn State.  He introduced me to numerical relativity,
his research inspired the topic of this thesis, and his Latin
mentality taught me how to defend my ideas.  Mijan Huq and Deirdre
Shoemaker patiently answered my questions, and I had many interesting
discussions with my fellow students Bernard Kelly and Ken Smith,
especially on Thursdays.

After returning to T\"ubingen, Hanns Ruder gave me very generous
financial support, and also gave me the paternal nudging that is
necessary to finish a thesis.  I would also like to thank J\"org
Frau\-en\-die\-ner and Ute Kraus for their competent advice, and
Daniel Kobras and Stefan Kunze for making the important hardware in
our department work.

It is almost by their nature that these acknowledgements are
incomplete, and I apologise to the many people whom I have to thank in
person instead.

% LocalWords:  eschnett Exp

% -*-LaTeX-*-
% $Header: /home/eschnett/cvs/diss/toc.tex,v 1.1 2002/11/06 19:54:51 eschnett Exp $

\cleardoublepage
\addcontentsline{toc}{chapter}{Table of contents}
\tableofcontents

% -*-LaTeX-*-
% $Header: /home/eschnett/cvs/diss/intro.tex,v 1.13 2003/01/10 16:39:49 eschnett Exp $

\chapter{Introduction}
\label{intro}

I want to simulate spacetimes containing black holes.

I am not alone with this project; it has a long history.  This was the
goal of the Binary Black Hole Coalescence Grand Challenge in the USA,
and is also part of the programme of the SFB 382 in Germany which pays
my salary.  One wants to numerically simulate spacetimes that contain
singularities, where the spacetimes are time-dependent and have no
symmetries that reduce the dimensionality, and where no approximations
are possible to simplify the nonlinear equations.

Such calculations are often justified by referring to gravitational
wave detectors such as GEO 600, LIGO, or LISA, which will soon produce
data which then need to be compared to theoretical calculations.  I,
however, think that the ability to and the knowledge how to simulate
fully generic spacetimes have value in itself and need no further
justification.

The two major problems that one encounters today when simulating
spacetimes are the computational expense and numerical instabilities.
Three-\sep dimensional time-dependent simulations always produce much
data, and always need supercomputers to run on.  Supercomputers are
awkward to use, and it is time-consuming to experiment and try new
ideas.

The numerical instabilities that are encountered in spacetime
simulations are a problem of a different kind.  For many years now,
simulations have been plagued by instabilities of unknown origin.
Simulations run fine for some time, but in many cases they encounter
an unphysical and poorly understood unbounded growth.

Many approaches have been tried to find a cure for these
instabilities.  People have blamed the formulation of the equations,
the boundary conditions, the gauge conditions, and the discretisation
methods.  It is clear that errors in any of these can cause a
simulation to fail.  The only successful methods to remove the
instabilities known today rely on a combination of several steps that
seem to be random, looking like a magician's recipe, and can only be
justified after the fact.  While it is now possible e.g.\ to evolve a
static black hole for a long time, or to run a black hole collision
for a reasonable time, the overall situation is still far from
satisfactory (see e.g.\ \cite{stable-simulations}).

In this thesis I look at \index{gauge condition}gauge conditions,
which are one of the possible causes of instabilities mentioned above.
I distinguish between \index{gauge fixing condition}\emph{gauge
fixing} and \index{gauge evolution condition}\emph{gauge evolution}
conditions.  Gauge fixing conditions are ``real'' gauge conditions,
whereas the weaker gauge evolution conditions describe only the time
evolution of an (unspecified) gauge condition.  The majority of
today's gauge conditions are only gauge evolution conditions in this
terminology.  I propose a set of gauge fixing conditions, and show
that their additional power can lead to a stable evolution, even
without using any magic ingredients.

While experimenting with different formalisms, I am willing to
sacrifice speed.  I do believe that it is enough to have a formulation
that can at least in principle be implemented efficiently.  Computers
become faster every year, and run time on supercomputers is rather
easy to obtain for research purposes.  It is better to start with a
slow, but stable formulation, and optimise later on, than trying to
find stability among fast implementations.  It is also better to have
a simple scheme for doing things, than having to follow a long recipe
containing many steps, each of which is necessary.

This thesis is structured as follows.  I start with a general
introduction to simulating vacuum spacetimes and the problems
encountered therein (chapter \ref{sim}).  I then explain gauge fixing
conditions in general and the condition I chose in particular (chapter
\ref{gauge}), as well as a system of evolution equations that is
suited to this kind of condition (chapter \ref{system}).  After that I
discuss some boundary conditions (chapter \ref{boundary}) and initial
data (chapter \ref{initial}) that I use later on.  I briefly mention
some facts about the code that I developed (chapter \ref{code}) and
move on to present in some detail the test cases and their results
that demonstrate the properties of my gauge fixing conditions (chapter
\ref{results}).  I make some concluding remarks (chapter
\ref{conclusion}), and add some appendices containing equations and
explanations that are likely not of general interest.

% LocalWords:  eschnett Exp

% -*-LaTeX-*-
% $Header: /home/eschnett/cvs/diss/sim.tex,v 1.17 2003/01/10 16:39:49 eschnett Exp $

\chapter{Simulating spacetimes}
\label{sim}

A spacetime can be characterised by its four-metric $g_{\mu\nu}$,
which has to satisfy the Einstein equation $G_{\mu\nu} = 8\pi
T_{\mu\nu}$.  I will restrict myself in this thesis to the vacuum case
where $T_{\mu\nu}=0$.  (The notational conventions are explained in
appendix \ref{symbols}.)  The four-formalism couples the spatial and
temporal degrees of freedom.  The astrophysical interpretation,
however, calls for an initial--boundary--value formulation.

\section{The ADM formalism}

One commonly used way of transforming the Einstein equations into such
a system of time evolution equations is the so-called ADM formalism
\cite{adm}.  This is one form of a so-called $3+1$ split into $3$
spatial and $1$ temporal dimensions.  It leads to the primary
variables three-metric $\gamma_{ij}$ and extrinsic curvature $K_{ij}$.
The time evolution equations contain the freely specifiable quantities
lapse $\alpha$ and shift $\beta^i$.  The three-metric and extrinsic
curvature have to satisfy two constraints, the Hamiltonian constraint
$H$ and the momentum constraint $M_i$.

In order to evolve a spacetime in an ADM-like formalism, one needs
initial data, boundary conditions, and a choice of lapse and shift.
Additionally, one can choose to enforce the constraints or a gauge.
For reference, the ADM variables are given in appendix
\ref{adm-variables}, the ADM evolution equations in appendix
\ref{adm-time-evolution}, and the ADM constraints in appendix
\ref{adm-constraints}.

%% \section{Degrees of freedom}

At first sight, the primary variables of an ADM system (three-metric
$\gamma_{ij}$ and extrinsic curvature $K_{ij}$) seem to contain six
physical degrees of freedom, i.e.\ twelve independent quantities.
However, only four of those twelve have a true physical meaning.  Out
of the twelve apparent independent quantities, the constraints
eliminate four, and the choice of gauge eliminates another four.  Thus
there are only four independent quantities left, corresponding to the
two degrees of freedom for gravitational waves.

These days\footnote{I write this in 2002}, the original ADM system has
gone out of fashion in the numerical relativity community.  Instead,
one uses extended ADM-like systems, often called ctADM (``conformal
traceless ADM'') systems.  They have two more variables, a conformal
factor $\psi$ for the three-metric and the trace $K$ of the extrinsic
curvature.  The three-metric is replaced by a conformal three-metric
$\tilde \gamma_{ij}$, and the extrinsic curvature by a conformally
rescaled traceless extrinsic curvature $\tilde A_{ij}$.  These systems
also have two additional constraints, namely $\det \tilde \gamma_{ij}
= 1$ and $\trace \tilde A_{ij} = 0$.

Both the ADM and the ctADM systems exist in multiple variations.  The
constraint equations are of the form $C=0$, and it is therefore
possible to add arbitrary multiples of the constraints $C$ to the
right hand sides of the time evolution equations.  Doing this can
dramatically change the behaviour of a system.

The ctADM systems have the advantages that the new variables $\psi$
and $K$ have a direct interpretation as constraint and gauge
quantities.  $\psi$ is closely related to the Hamiltonian constraint
$H$, while $K$ is related to the lapse $\alpha$.  These relations make
it easier to influence the time evolution of these quantities.  It
also makes it potentially easier to enforce the constraints with the
York--Lichnerowicz method, which also relies on a conformal traceless
decomposition.

Both the Maya \cite{maya, ct-formulation, cure} and the AEI variants
of the BSSN systems \cite{nakamura1, nakamura2, shibata-nakamura,
baumgarte-shapiro} are variants of ctADM systems, where certain
derivatives of the conformal three-metric have additionally been made
independent variables, which leads to three additional constraints.
The additional variables capture gauge degrees of freedom that are
closely related to the shift, and thus allow further control over the
system.

%% A choice of lapse and shift is often called a gauge condition.  I call
%% it a \emph{gauge evolution condition} here, because choosing lapse
%% does not completely fix the gauge.  Choosing lapse and shift does not
%% determine the gauge freedom that one has in specifying the initial
%% data.  The gauge that is present in the initial data is evolved in
%% time by the choice of lapse in shift, hence the name gauge evolution
%% condition.
%% 
%% A \emph{gauge fixing condition} is stronger; it is a restriction on
%% the possible values of the primary variables, and thus also on the
%% initial data.  From each gauge fixing condition follows a
%% corresponding gauge evolution condition through the time evolution
%% equations.  The converse is not true; gauge evolution conditions do
%% not imply a gauge fixing condition.  One might call an evolution
%% scheme where only lapse and shift are specified ``evolving without
%% fixing the gauge''.
%% 
%% For example, geodesic slicing ($\alpha=1$) and normal coordinates
%% ($\beta^i=0$) are gauge evolution conditions.  Maximal slicing is a
%% gauge fixing condition, as it specifies $K=0$.  Minimal distortion is
%% a only a gauge evolution condition, because it only selects a shift,
%% and does not restrict the three-metric.  In fact, most gauge
%% conditions that are commonly used in nonlinear numerical relativity
%% are only gauge evolution conditions with this terminology.

\section{Constrained evolution}

Sometimes it is necessary or convenient to enforce the constraints
explicitly.  This can either be a part of a scheme to obtain initial
data, or it can be part of an evolution scheme, e.g.\ to prevent
deviation from the constraint hypersurface due to accumulated
numerical errors.  Enforcing the constraints can also help stability,
as e.g.\ described by Evans \cite{evans1989} or Marsha and Choptuik
\cite{marsa1996}.  The constraint equations contain no time
derivatives, and also do not contain the lapse or the shift.  They
connect the three-metric and extrinsic curvature only.

When spherical or axial symmetry is assumed, then some constraint
equations are algebraic conditions for some of the evolved variables.
That means that these constraints can be enforced rather easily.  This
is unfortunately not the case in three dimensions.  There, the
constraint equations are not suitable for enforcing in their ADM form.
However, they can be rewritten so that they are elliptic equations in
certain quantities that are connected to the three-metric and the
extrinsic curvature.  In the York--Lichnerowicz formalism \cite{york,
york2} \cite[section 2.2.1]{cook}, a ctADM system is extended to
include a vector potential $V_i$ for the conformal traceless extrinsic
curvature.  The constraints determine the conformal factor and this
vector potential.  This requires variable transformations before and
after solving.

As it is somewhat expensive to solve elliptic equations, it has become
customary to perform \emph{unconstrained evolution} for
three-dimensional simulations, i.e.\ to evolve without regard to the
constraints, monitoring the constraints only as a measure of quality
for the time evolution.  Given initial data that satisfy the
constraints, the Bianchi identities then guarantee that the
constraints will be satisfied at all times, of course only modulo
numerical errors.  Choptuik \cite{choptuik1991} argues that the order
of accuracy of an unconstrained evolution will not be worse than that
of a constrained evolution.

\section{Picturing spacetimes}
\label{picturing-spacetimes}

There are two fundamentally different points of view that one can take
to interpret a spacetime.  In the following, I use the terms
\emph{grid points} and \emph{coordinate system} in an equivalent
manner.  One can think of coordinate lines forming a grid, and the
intersections of these coordinate lines are then the grid points.

\begin{description}

   \item[The coordinate based view] treats the primary variables
   similar to hydrodynamic or electrodynamic field quantities.  One
   pictures a non-physical background grid, containing regularly
   spaced grid points forming the numerical grid.  On each grid point,
   or in each grid cell, live the primary variables, i.e.\ the
   three-metric and extrinsic curvature.  The background grid has
   nothing to do with the evolved spacetime, it exists only in a flat
   coordinate space.

   Time evolution consists of updating the evolved quantities through
   source and advection terms.  The boundary of the grid stays in a
   fixed place in the coordinate space.  This picture is the one that
   is most often used when quantities are plotted or displayed.
   Instabilities show up as certain metric coefficients becoming
   negative or some quantity growing without bound.  In this picture,
   the apparent horizon of a static black hole can move across the
   grid.

   \item[In the physically based view,] one looks at the location of
   the coordinate grid points in the physical spacetime.  It is not
   always easy to picture (let alone draw or plot) a curved spacetime,
   but much of general relativity has to do only with strange
   coordinate systems, and applies even in a flat spacetime.  It is
   instructive to use this picture to look at a time evolution of flat
   space in a nontrivial, time-dependent coordinate system.

   In this picture, time evolution consists of choosing the locations
   of the next layer of coordinate grid points.  That is, it is the
   grid that is constructed layer by layer, while the (unknown)
   spacetime is not evolved, but only discovered.  Instabilities show
   up as grid points getting too close or too far apart, grid cells
   becoming too elongated, or grid lines crossing.  In this picture,
   the apparent horizon of a static black hole stays fixed, while the
   grid points can move across it.

\end{description}

The coordinate based view is much easier to draw or display, and is
hence what is usually presented.  The physically based view is much
more difficult to display, and requires in general some (artificial)
embedding.  However, the physically based point of view makes it also
much easier to interpret gauge modes.  For gauge modes, the embedding
stays the same, only the grid points move.

\begin{figure}
\begin{tabular}{rr}
\includegraphics[width=0.45\textwidth]{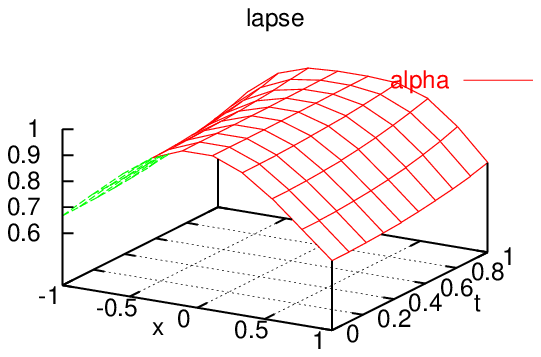} &
\includegraphics[width=0.45\textwidth]{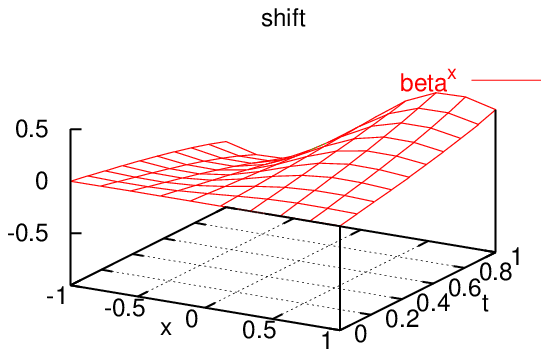} \\
\includegraphics[width=0.45\textwidth]{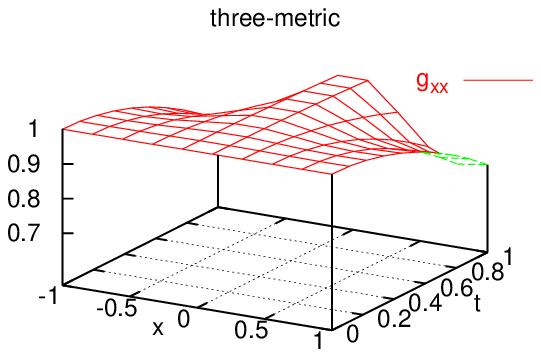} &
\includegraphics[width=0.45\textwidth]{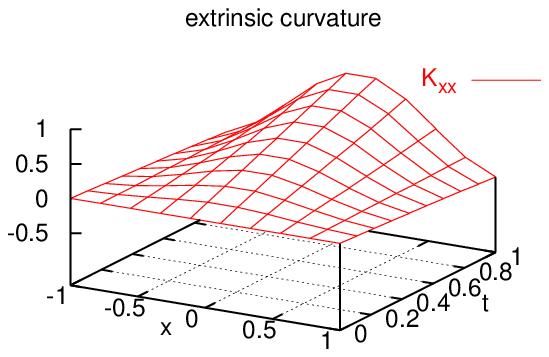}
\end{tabular}
\caption{\label{gauge-fields}Evolved field quantities vs.\ time and
space}
\end{figure}

As an example, consider the time-dependent metric
\begin{eqnarray}
\label{strange-metric}
   ds^2 & = & - \frac{1}{(1+Ax^2)^2}\; dt^2 + 2\, \frac{2 A t x}
   {(1+Ax^2)^3}\; dt\, dx
\\\nonumber
   & & + \left[ 1 - \frac{(2 A t x)^2} {(1+Ax^2)^4} \right]\, dx^2 +
   dy^2 + dz^2
\end{eqnarray}
with the free parameter $A$.  For $A=0$, this metric is identical to
the Minkowski metric.  This metric happens to describe a flat
spacetime for all values of $A$, which is not obvious to see.  Figure
\ref{gauge-fields} shows the graphs of the three-metric, extrinsic
curvature, lapse, and shift for the case $A=1/2$.  This figure uses
the coordinate based view.

Assume now that this deviation from the Minkowski metric is due to an
unwanted gauge mode.  It is possible to arrive at the conclusion that
this metric is identical to the Minkowski metric after a gauge
transformation by looking at the field quantities shown in figure
\ref{gauge-fields}.  Yet it is not trivial to do so, nor can the
nature of the gauge transformation (i.e.\ the gauge mode) easily be
found out.  However, one wants to know what gauge transformation was
applied so that the gauge mode can be counteracted in a numerical
evolution.

\begin{figure}
\begin{tabular}{rr}
\includegraphics[width=0.45\textwidth]{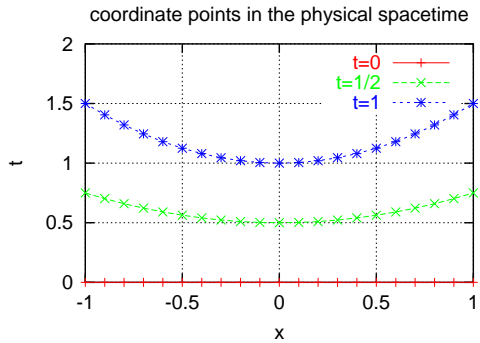}
\end{tabular}
\caption{\label{gauge-coords}Location of the coordinate grid points in
the (flat) physical spacetime}
\end{figure}

Figure \ref{gauge-coords} shows the same metric, but now uses the
physically based view.  That is, shown are the grid points of the
coordinate system described by the metric (\ref{strange-metric}),
using a Cartesian coordinate system (i.e.\ the Minkowski metric) as
reference.  This graph is only possible because the spacetime is flat;
otherwise one would need an embedding.  It is obvious from this graph
that the gauge transformation is a very simple one, namely just $t' =
t\, (1+Ax^2)$ with $A=1/2$.  Knowing this, one could e.g.\ choose a
different lapse to prevent the time slices from curling up.  While
this can in principle also be deduced by looking only at the primary
field variables, it is certainly easier in the physically based view.

\section{Interpreting the quantities}
\label{interpreting}

It is quite useful to have a mental picture of what the quantities in
an ADM or ctADM system mean:
\begin{itemize}

   \item The conformal factor determines the volume that is contained
   in a grid cell; the volume is given by $\sqrt{\det \gamma_{ij}} =
   \psi^6$.

   \item The diagonal elements of the three-metric determine the
   physical length of the coordinate lines between grid points within
   a time slice.  The off-diagonal elements determine the angles
   between the coordinate lines.

   \item The trace of the extrinsic curvature determines the mean
   curvature of the layer of grid points forming the current time
   slice.  It is closely related to the time derivative of the
   conformal factor.  It is also closely connected to the lapse, as it
   is mostly the lapse that determines the change of the mean
   curvature in time.

   \item I have found no immediately useful interpretation of the
   extrinsic curvature tensor.  One can view it as the time derivative
   of the three-metric.

   \item The lapse is the physical distance between two time slices,
   measured in proper time of an observer that is at rest in those
   slices.  It is thus the ratio between proper and coordinate time
   for such an observer.  When the lapse is zero, coordinate time
   increases, but no proper time elapses.

   \item The shift determines the difference between the spatial
   coordinate systems of two neighbouring time slices.  It is the
   velocity of the coordinate system with regard to a resting
   observer.

\end{itemize}

Given the three-metric and extrinsic curvature of a time slice, there
is no direct way of knowing just where in a given spacetime the
current time slice is located.  One has to work out how the time slice
is embedded in the spacetime, and then compare the resulting
three-metrics and extrinsic curvatures.  This is in general a
difficult matching process, and I know of no attempts to try this in
four dimensions (but see the Lazarus project \cite{lazarus}).  An
efficient algorithm for this problem would surely be very useful for
comparing results of runs with different formulations, or of runs with
different coordinate conditions \cite{mexico-unam, mexico-scott,
appleswithapples}.

\section{Instabilities}

Three-dimensional numerical simulations of nonlinear general
relativistic spacetimes have been plagued by instabilities.  To date,
it is still safe to say that these instabilities are not well
understood.  There are many ideas about on how to remove these
instabilities, but none have truly succeeded so far.

I use the term \emph{instability} here in a rather loose manner.  I
take it to mean that a certain quantity grows exponentially (or
faster) and without bound.  I want to use this term here not for a
theoretical argument, but only to describe empirical results from
simulation runs.

I do not want to use the definition of instability from the theory of
hyperbolic systems.  I think that this definition is not too useful in
this case, as it considers an exponential growth to be stable.  (On
the other hand, it introduces the concept of resolution independence,
which is crucial for numerical simulations.)

Jeffrey Winicour has introduced the very practical definition of
\emph{robust stability} \cite{robust-stability, robust-stability-2},
which allows for at most polynomial growth even when arbitrary noise
is continuously injected into the system.  This can easily be tested
in a numerical implementation.

Apart from being stable, a system must also be convergent.  That is,
it must be able to track a given solution arbitrarily closely, where
the deviation should depend on the resolution e.g.\ in a quadratic
manner.  I will concentrate on stability, rather than convergence, in
the following section.  That does, of course, not mean that I believe
convergence to be unimportant.

%% \section{Classifying instabilities}

Let me try to classify the possible instabilities of a system
implementing the Einstein equations in a $3+1$ split:
\begin{description}

   \item[Physical instabilities] have their cause in the spacetime
   itself.  A spacetime can contain singularities, and if the part of
   the spacetime that is described by the numerical metric gets close
   to or contains such a singularity, there will be a physical
   instability.  There is not much one can do against such a
   singularity.  Possible approaches include choosing a slicing that
   does not intersect the singularity, or excising that part of the
   spacetime, or trying to ``subtract'' the singularity and treating
   it analytically.

   \item[Constraint-violating instabilities] occur because the
   constraints are not exactly satisfied in a numerical simulation.
   Many theorems about general relativity hold only when the
   constraints are satisfied.  It is not known what happens when the
   constraints are not satisfied, and it is quite possible that such a
   constraint violation would grow exponentially.  Constraint
   violating modes may also propagate faster than light; they do not
   have to follow causality.  The constraint violation can be easily
   monitored.  It is possible to enforce the constraints at any time
   by e.g.\ solving elliptic equations, but this is expensive.

   \item[Gauge instabilities] result from a bad choice of coordinates.
   With a gauge evolution condition, i.e.\ a choice of lapse and
   shift, one does not choose the coordinate system directly, but
   rather only its time evolution.  Especially when lapse and shift
   are themselves the result of a time evolution equation with its
   associated boundary conditions, the relationship between them and
   the quality of the coordinate system becomes complicated.  Gauge
   instabilities can theoretically always be cured by a different
   choice of coordinates, but this is difficult in practice.  There is
   no clear-cut distinction between physical and gauge degrees of
   freedom, which makes it difficult to separate between physical and
   gauge instabilities.

\end{description}

Instead of classifying instabilities by the affected quantities, one
can also try to classify them by their sources:
\begin{description}

   \item[Instabilities in the formulation] are another name for an
   ill-posed problem.  The prototype of an ill-posed problem is the
   backward heat equation $\dot u = - u''$.  While the (forward) heat
   equation has a smoothing property, the backward heat equation
   amplifies any perturbation exponentially.  The amplification time
   scale depends on the length scale of the perturbation, where
   smaller perturbations are amplified more quickly.

   \item[Instabilities due to boundaries] are caused by inappropriate
   boundary conditions.  The advection equation $\dot u = u'$ on the
   unit interval $[0; 1]$ needs a boundary condition at $x=1$, but
   cannot have one arbitrarily imposed at $x=0$.  This follows from
   the characteristic of the equation.  However, characteristics for
   more complicated, nonlinear equations that are not given in a
   first-order form are difficult to find.  The characteristics also
   depend on the formulation and the gauge conditions, so that
   expressing the same physical system with different variables or
   with a different gauge will lead to different characteristics.

   \item[Instabilities in the discretisation] are introduced by an
   unsuitable discretisation scheme.  A prototypical example of such
   an unsuitable scheme for e.g.\ the advection equation $\dot u = u'$
   is combining centred differencing with an explicit first order time
   integrator.  Suitable schemes for this equation are e.g.\ using
   upwind differencing, adding artificial diffusion, using an implicit
   time integrator, or using interpolation instead of finite
   differencing and time integration.

\end{description}

\section{My goal in this thesis}

In this thesis, I am not attempting to find a well-posed formulation
for the Einstein equations.  I am instead using a simple,
straightforward formulation that might well not be well-posed
(although I do not know this).  I am adding artificial diffusion to
counteract problems caused by this.  This artificial diffusion cures
the discretisation instabilities.

Also, in this thesis I am not trying to find an especially suited
discretisation scheme for the Einstein equations.  I am using centred
differencing and a second order explicit time integrator, which are
among the most simple tools that can be used for this.  First order
artificial diffusion makes my overall discretisation scheme first
order accurate only.

Instead, in this thesis, I use the first classification of
instabilities, and discuss gauge modes.
%% I believe that gauge modes have up to now not received the attention
%% that they deserve.  One is easily lead to believe that gauge modes
%% cannot pose serious problems, because one can ``just choose a
%% different gauge'' if so.  In my opinion this is just the very reason
%% that they cause problems, namely because there is no accepted
%% prescription to determine what even constitutes or how to measure a
%% gauge mode.  Below
I introduce generic gauge fixing conditions which apply in ways
equivalent to the constraints.  That allows gauge modes and gauge
instabilities to be identified, and also to be removed by enforcing
the gauge.

% LocalWords:  eschnett Exp ij ctADM pics betax gxx kxx appleswithapples

% -*-LaTeX-*-
% $Header: /home/eschnett/cvs/diss/gauge.tex,v 1.18 2003/01/10 16:39:49 eschnett Exp $

\chapter{Gauge conditions}
\label{gauge}

The ADM evolution equations contain the free quantities lapse $\alpha$
and shift $\beta^i$, which need to be specified prior to a time
evolution.  Lapse and shift determine how the coordinate system in the
spacetime is constructed, starting from the current time slice.  Such
a choice of lapse and shift is also commonly called a \emph{gauge
condition}.  Unfortunately, the simplest gauge choices will in general
lead to coordinate singularities, so that more involved conditions are
necessary.

The gauge choice also influences the properties of the system of
evolution equations.  Hyperbolicity, and the stability properties in
general, do crucially depend on the gauge condition.  The search for
good gauge choices forms an important part of the current efforts to
obtain stable evolutions.

Below, I list several gauge choices that are or were commonly used,
and mention some of the classifications for gauge conditions that have
proven useful in numerical relativity.  I discuss the difference
between \emph{gauge evolution conditions} and \emph{gauge fixing
conditions}, and point out the properties that a good gauge fixing
condition needs to have.

I then describe such a gauge fixing condition that has these
properties, and that I will use in the following chapters.  I also
examine the behaviour of its induced coordinate systems and compare it
to existing gauge conditions.

\section{Common gauge conditions}

\label{geodesic-slicing}
\label{normal-coordinates}
Clearly the easiest way to specify lapse and shift is by prescribing
them as functions of space and time, i.e.\ as functions of the
coordinates $x^i$ and $t$.  Doing so might be called using an
\emph{exact gauge condition}\footnote{I think that this expression was
coined by Manuel Tiglio.}.  Again the simplest possible way to do this
is to prescribe $\alpha=1$ and $\beta^i=0$ everywhere, which is called
geodesic slicing and normal coordinates, respectively.
%% Both together are also called ``temporal gauge''.
For other exact gauge conditions, one often uses a known solution of
Einstein's equations to obtain expressions for the lapse and shift.
As mentioned above, such a simple gauge choice usually leads to
unstable evolutions.  The intuitive explanation for these
instabilities is that numerical errors accumulate, creating gauge
modes or a gauge drift, while the gauge condition does not compensate.

An only slightly more complicated way to specify lapse and shift is by
taking the values of the primary variables into account.  This is
often called an \emph{analytic gauge condition}.  In an ADM-like
system, one specifies lapse and shift as functions of the coordinates
$x^i$ and $t$, the three-metric $\gamma_{ij}$, and the extrinsic
curvature $K_{ij}$.  Thus analytic gauges are a superset of exact
gauges.  The additional freedom in choosing lapse and shift can be
used to counteract instabilities.  It is also possible to solve time
evolution equations for lapse and shift.  For example, harmonic
slicing \cite{harmonic-slicing} can be implemented by setting
$\partial_t \alpha = - \alpha^2 K$.  Harmonic slicing has singularity
avoiding properties similar to maximal slicing (see below).
%% A related condition is $1+\log$ slicing, which sets $\partial_t \alpha
%% = -2 \alpha K$.

\label{maximal-slicing}
The most expensive gauge conditions are \emph{elliptic gauge
conditions}, consisting of elliptic equations for lapse or shift.
They are a subset of analytic gauge conditions.  The prototype for
this is maximal slicing \cite{maximal-minimal}.  This gauge condition
prescribes $K=0$, which leads to the condition $\triangle \alpha =
\alpha R$ for the lapse.  Other well-known elliptic gauge conditions
are minimal strain and minimal distortion \cite{maximal-minimal}, and
the Gamma-freezing shift derived from $\partial_t \tilde \Gamma^i = 0$
\cite{gamma-freezing} in the BSSN system.

The gauge conditions currently en vogue in the community are analytic
gauges with time evolution equations.  Elliptic conditions have
desirable properties, but are considered to be too expensive.
Harmonic slicing and variations thereof can be seen as \emph{driver
conditions}.  Instead of solving an elliptic equation for lapse or
shift, one transforms the elliptic equation into a parabolic or
hyperbolic time evolution equation.  These evolution equations drive
the lapse or shift towards the desired value.  A parabolic equation
acts as a diffusive mechanism that disperses the difference, and a
hyperbolic equation acts as a (damped) radiative system that
transports the difference away.

Hyperbolic driver gauge conditions have been successful to some extent
for the BSSN system, and also for the KST system \cite{kst} with a
densitised lapse.

As mentioned in the introduction, I do not share the belief that
elliptic gauge equations are too expensive.  While it is obviously
true that they are inconveniently expensive, it is often possible to
optimise them for speed, such as e.g.\ solving only every $n$ time
steps, and using some approximation in between.  Furthermore,
stability is more important than speed.

\section{Gauge evolution and gauge fixing}
\label{gauge-fixing}

A gauge condition is in four-space a condition that is applied to the
four-metric $g_{\mu\nu}$.  A choice of lapse and shift corresponds to
a specific choice of $g_{00}$ and $g_{0i}$ and is thus a true gauge
condition there.

However, in an ADM-like system, the primary variables are not the
four-metric, but rather the three-metric and the extrinsic curvature.
A gauge condition thus has to be applied to these quantities.  A
choice of lapse and shift does not restrict the primary variables, and
should therefore not quite be called a gauge condition.  Because I do
not want to break with tradition too much, I call it a \emph{gauge
evolution condition} in this text.  While a choice of lapse and shift
does not eliminate the gauge freedom present in the initial data, it
does determine how the gauge condition is evolved in time.
%% Lapse and shift transport the gauge from one time slice to the next.
A gauge evolution condition can be obtained from an evolution equation
of a primary variable, e.g.\ from $\partial_t K = \cdots$ or
$\partial_t \tilde \Gamma^i = \cdots$.

A ``true'' gauge condition is what I call a \emph{gauge fixing
condition} here.  A gauge fixing condition is a direct restriction on
the possible values of the primary variables.  This will implicitly
also lead to a choice for lapse and shift.  Maximal slicing ($K=0$) is
an example of a gauge fixing condition; it not only selects a lapse,
but also restricts the possible sets of values for the primary
variables.

(Gauge fixing is not a new idea.  In fact, gauge fixing is the way in
which gauge conditions are applied in other fields.  Calling a choice
of lapse and shift a gauge condition seems to be a relatively new
convention in general relativity, which was used by Smarr and York in
\cite{maximal-minimal}.)

Most of the gauge conditions mentioned above and used today are only
gauge evolution conditions.  This fact does not make them inferior per
se, but using them can easily lead to (coordinate) singularities,
which show up as instabilities.  Let me consider, as an example,
geodesic slicing ($\alpha=1$) in a flat spacetime, and let me
disregard the shift for now.  In Minkowski coordinates, geodesic
slicing works fine.  With different initial data for a flat spacetime,
geodesic slicing can lead to coordinate singularities, while of course
the spacetime is still flat.  This is numerically a problem when one
starts out close to Minkowski coordinates, with the difference being
only due to numerical errors.

Unexpected gauge singularities, i.e.\ gauge instabilities, cannot
happen with a gauge fixing condition.  A gauge fixing condition
imposes a condition on a primary variable, e.g.\ specifies the values
for $K$, and then selects a lapse so that this condition continues to
hold.  Because $K$ is specified, it is impossible to accidentally
select a lapse that makes $K$ diverge, as in the example above.  The
only way to produce a gauge singularity is by specifying a diverging
$K$ to begin with.

This does not prevent accumulated numerical errors, which could still
make $K$ diverge.  But the gauge fixing condition can be monitored
during the evolution, so that one can judge the quality of the
simulation.  This is equivalent to the way one handles the
constraints.  Furthermore, a gauge fixing condition can also be
directly enforced when numerical errors have accumulated, eliminating
gauge instabilities altogether.

The interplay between lapse and shift and gauge singularities is
further complicated by the boundary conditions.  One needs boundary
conditions for the primary variables, and also for lapse and shift if
they are evolved in time.  Without a proper gauge fixing condition to
start from it is prohibitively difficult to ensure that a certain
choice of lapse and shift does not, under the influence of numerical
errors, lead to a gauge singularity at the boundary.  With a gauge
fixing condition, such instabilities are not possible, or can at least
be easily detected.

The only gauge fixing condition that is in use today in nonlinear
three-dimensional numerical relativity is maximal slicing.

A close candidate is AEI's Gamma freezing, which sets $\partial_t
\tilde \Gamma^i = 0$ and derives an elliptic shift condition from
that.  Placing a condition on a time derivative is not directly a
gauge fixing condition, which requires placing a condition on the
variable itself.  But setting a time derivative to zero can be viewed
as fixing the variable at its initial value, and I therefore want to
count it as gauge fixing.  However, Gamma freezing does not try to
keep the value of $\tilde \Gamma^i$ consistent with $\tilde
\gamma_{ij}$, i.e.\ the new constraints arising from introducing
$\tilde \Gamma^i$ as additional primary variable
\cite{baumgarte-shapiro} are not controlled.  (These constraints seem
to be important, because they seem to be growing without bound when
they are not controlled.)  Re-calculating $\tilde \Gamma^i$ from
$\tilde \gamma_{ij}$ destroys the gauge fixing properties of Gamma
freezing.

Minimal distortion is not a gauge fixing condition.  This condition
minimises the distortion, which is the \emph{change} of shape during a
time step.  It does not try to attain any specific coordinate shape
(e.g.\ for a sphere), and is therefore not a gauge fixing condition.
It restricts only the time derivative of the three-metric, and not the
three-metric itself.

%% \todo{distortion tensor: $\Sigma^{ij} = \frac{1}{2} \psi^4 \partial_t
%% \tilde \gamma^{ij}$}
%% 
%% \todo{minimal distortion: $\nabla_j \Sigma^{ij} = 0$}

One interesting case of a gauge fixing condition is area locking.
Area locking was introduced in spherically symmetric situations, and
it leads to very stable evolutions in this case.  It sets $\partial_t
\gamma_{\theta\theta} = 0$, which leads to a condition for the shift
component $\beta^r$.  As above, I count this as gauge fixing.
Unfortunately, area locking was never successfully generalised to
three dimensions.

\section{Good gauge fixing conditions}

A gauge fixing condition is a condition that acts directly on the
primary variables, i.e.\ the three-metric and the extrinsic curvature
for an ADM-like system.  It has thus properties similar to the
constraints.  Gauge fixing conditions are a superset of gauge
evolution conditions which select a lapse and a shift only.

A good gauge fixing condition has to have several properties:
\begin{enumerate}

   \item Of course, it has to leave enough degrees of freedom so as to
   not constrain any physics, even in the full nonlinear case.  No
   physical degree of freedom must be removed from the system.

   \item It has to lead to a unique prescription for lapse and shift,
   so that the gauge condition will also hold for the neighbouring
   time slices.

   \item It should be easy to evaluate, so that one knows whether the
   gauge condition still holds, or by how far one has left this gauge.

   \item There should be a reasonable way to enforce this gauge, so
   that one can counteract numerical errors that have accumulated over
   time, or create initial data in this gauge.

   \item It should be independent of the constraints.  That is, the
   condition should be independent of the variables that are changed
   when the constaints are enforced, so that one can at the same time
   enforce the gauge condition and the constraints.

   \item It should have an understandable physical interpretation, so
   that the physicist can apply his or her intuition.

   \item It should be general enough so that many kinds of initial
   data can be prepared with this gauge, and many spacetimes can be
   described in it.

\end{enumerate}

I do count solving elliptic equations as being ``reasonable'' in the
context of the above.  There are kinds of equations that are much more
unwieldy than elliptic ones.  And there exist efficient methods for
solving elliptic equations with a cost that grows only linearly with
the problem size.  I do not strive for the fastest possible
implementation here, I only want something ``reasonable''.

Maximal slicing almost satisfies these properties, but it is not a
complete gauge choice.  It selects a slicing, but does not determine
the three-coordinate system.  Additionally, the condition $K=0$ is too
constraining in practice --- it does e.g.\ not permit a
horizon-penetrating static slicing of a static black hole (see e.g.\
\cite[section 3.1.1]{cook}).

\section{My gauge fixing condition}
\label{my-gauge-condition}

My goal in this thesis is to present a gauge fixing condition that can
be used to simulate black hole spacetimes.  This condition has to have
the properties that I list above.  As mentioned in the introduction, I
also want a simple system.  The currently promising systems (e.g.\ the
BSSN variants) all contain many subtle rules that have to be followed
closely.  The gauge condition should be conceptually as simple as
possible, even if it is expensive to use.

For my gauge condition, I choose as gauge variables the trace $K$ of
the extrinsic curvature and certain combinations of partial
derivatives of the conformal three-metric, which I call $F_i$ after
Shibata and Nakamura \cite{shibata-nakamura}.  My gauge condition then
consists of prescribing, in advance, values for these quantities as
functions of space and time, i.e.\ depending on $x^i$ and $t$ only.

Prescribing the trace of the extrinsic curvature leads to an elliptic
equation for the lapse (see appendix \ref{ctadm-time-evol}).  Given
that, it is clear that a choice of $K$ has all of the above
properties.  Interpreting $K$ as a gauge variable is a common choice.
It is a straightforward generalisation of maximal slicing
\cite{maximal-minimal}, and I think that this gauge variable needs no
further justification.

My gauge condition acting on the metric is a bit more involved.  I
first define (for technical reasons) the traceless conformal
three-metric $\bar h_{ij}$ as
\begin{eqnarray}
   \bar h_{ij} & = & \tilde \gamma_{ij} - \frac{1}{3} \delta_{ij}
   \delta^{kl} \tilde \gamma_{kl} \quad \mathrm{.}
\end{eqnarray}
The bar $\bar\cdot$ indicates here that the indices are to be raised
and lowered with the coordinate metric $\delta_{ij}$.  My gauge
variable $F_i$ is then defined by
\begin{eqnarray}
   F_i & = & \bar h_{ij,j} \quad \mathrm{.}
\end{eqnarray}
That is, my gauge quantity $F_i$ is closely related to the divergence
of the conformal three-metric $\tilde \gamma_{ij}$.  The fact that I
use the traceless part only has only technical reasons, having to do
with the method used to enforce the gauge condition, which is
explained further down.

This gauge condition is similar to the transverse-traceless gauge used
in linearised general relativity, where one often requires of the
metric perturbation the condition $h_{ij,j} = 0$.  (Here $h_{ij}$ is a
three-metric perturbation, which is different from the variable $\bar
h_{ij}$ above.)  However, I do not require that $F_i=0$, but only that
it be prescribed in advance.

This gauge variable is also very similar to, but slightly different
from the auxiliary variable $F_i$ introduced by Shibata and Nakamura
\cite{shibata-nakamura} as $F_i = \tilde \gamma_{ij,j}$.   The BSSN
system has instead $\tilde \Gamma^i = \tilde \gamma^{jk}\, \tilde
\Gamma^i_{jk}$ \cite{baumgarte-shapiro}, which is also closely related
to the Shibata--Nakamura $F_i$.

It is by design that the gauge condition is not covariant.  The gauge
defines the coordinate system, and the use of a specific coordinate
system is what makes a calculation be not covariant.  That means that
it does not make sense to speak of a covariant gauge condition ---
such a gauge condition could not single out a specific coordinate
system.

The choice of the gauge variable $F_i$ constrains the possible shapes
of the three-coordinate system.  The gauge condition judges the
quality of the current coordinate system, and prevents a deterioration
by placing the grid points appropriately.

As described in section \ref{interpreting}, the metric components
$\gamma_{ii}$ (no summation implied) describe the physical distance
between grid points in the $i$ direction, while the components
$\gamma_{ij}$ (with $i \ne j$) describe the physical angle between the
$i$ and $j$ coordinate directions.  The direct metric divergence
$\gamma_{ij,j}$ hence describes the spatial change of the behaviour of
the grid lines in the $i$ direction --- the change in length, plus the
change in the angles with the other grid lines.

The physical meaning of the components of the conformal traceless
metric $\bar h_{ij}$ is more complex, but will be similar in nature.
Hence enforcing a certain $\bar h_{ij,j}$ means to enforce a certain
spatial rate of change of the coordinate system.  Enforcing the gauge
changes the metric, and therefore effectively moves the grid points
%% within the time slice
until they have the ``right'' location with respect to their
neighbours.

%% \todo{example pictures for Minkowski, Kerr--Schild, etc.}

As one can set the quantity $F_i$ to an arbitrary function, this gauge
choice can be used for all initial data.  Fixing $F_i$ leads, via the
time evolution equation of $\tilde \gamma_{ij}$, to an elliptic
equation defining the shift (see appendix
\ref{coordinate-conditions}), so that using this shift preserves the
gauge condition.  That means that this gauge condition is generic, can
be used for all spacetimes, and does not constrain any physical
degrees of freedom.\footnote{In this I assume that this elliptic
equation has a solution.  This depends e.g.\ also on the boundary
conditions.  For the cases I am interested in, one can assume that the
domain is simply connected and has Dirichlet boundaries.  I have not
proven that a solution exists, but numerical tests support my
assumption.}

\label{coord-boundary}
The resulting elliptic equations for the lapse $\alpha$ and the shift
$\beta^i$ also require boundary conditions.  These boundary conditions
are part of the gauge condition.  One can use the boundary conditions
for the lapse to slow down or speed up the time evolution, or to tilt
the time slice.  The boundary conditions for the shift can be used to
move the boundary further in or out, or to move or rotate the whole
simulation domain.  This can be used to set up co-rotating coordinate
systems, for example.

By introducing a vector potential $W_i$ for the traceless conformal
metric $\bar h_{ij}$, one can enforce this gauge condition in a manner
equivalent to enforcing the momentum constraint on the extrinsic
curvature.\footnote{The existence and uniqueness properties of the
York--Lichnerowicz method thus are trivially valid for this method as
well.}  One can reconstruct the conformal metric $\tilde \gamma_{ij}$
from its traceless part through the condition $\det \tilde \gamma_{ij}
= 1$, which determines its trace.  Furthermore, as enforcing the
constraints in the York--Lichnerowicz formalism (see section
\ref{enforce-contraints}) does not change the conformal metric or the
trace of the extrinsic curvature, it preserves this gauge conditions.

This means that this gauge has all the properties (1) -- (7) above for
a good gauge fixing condition.  I will use it in the following
chapters and examine its numerical properties.

\bigskip

Another, potentially more elegant gauge condition on the metric could
be $F_i = \tilde \gamma_{ij,j}$.  Another, potentially more elegant
way of enforcing the gauge condition could be to apply a
four-dimensional coordinate transformation.  A coordinate
transformation would have the advantage that the constraints are not
influenced.  I did not investigate these alternatives.

% LocalWords:  eschnett Exp ij gv symmetrizable Commun Phys const cgc bbhgc hy
% LocalWords:  shibata bo lic

% -*-LaTeX-*-
% $Header: /home/eschnett/cvs/diss/system.tex,v 1.16 2003/01/10 16:39:49 eschnett Exp $

\chapter{The system of equations}
\label{system}

\section{The variables}

I chose not to use the standard ADM system to represent a spacelike
hypersurface, but rather a variant of a conformal traceless ADM
(ctADM) system.  Such a system can more easily be used to enforce the
constraints, and furthermore it also already makes the gauge degrees
of freedom more explicit.

My primary variables are
\begin{itemize}

   \item the conformal factor $\psi$, with $\psi^{12} = \det \gamma$

   \item the conformal three-metric $\tilde \gamma_{ij}$, with $\psi^4
   \tilde \gamma_{ij} = \gamma_{ij}$

   \item the trace $K$ of the extrinsic curvature, with $K =
   \gamma^{ij} K_{ij}$

   \item the traceless conformal extrinsic curvature $\tilde A_{ij}$,
   with $\psi^{-2} \tilde A_{ij} = A_{ij}$.  Here $A_{ij}$ is the
   traceless extrinsic curvature, i.e.\ $A_{ij} = K_{ij} - \frac{1}{3}
   \gamma_{ij} K$

   \item the divergence $F_i$ of the traceless conformal three-metric,
   with $F_i = \bar h_{ij,j}$.  Here $\bar h_{ij}$ is the traceless
   conformal three-metric, i.e.\ $\bar h_{ij} = \tilde \gamma_{ij} -
   \frac{1}{3} \delta_{ij} \delta^{kl} \tilde \gamma_{kl}$

\end{itemize}
In the above, $\gamma_{ij}$ and $K_{ij}$ are the three-metric and
extrinsic curvature of the standard ADM system.  This system
implicitly fulfils the constraints $\det \tilde \gamma_{ij} = 1$ and
$\mathrm{trace}\, \tilde A_{ij} = \tilde \gamma^{ij} \tilde A_{ij} =
0$.  I use a tilde $\tilde \cdot$ for quantities that are connected to
the conformal metric $\tilde \gamma_{ij}$ rather than the physical
metric $\gamma_{ij}$, and a bar $\bar \cdot$ for quantities that are
connected to the coordinate metric $\delta_{ij}$.  The time evolution
equations for this system are derived in appendix
\ref{ctadm-time-evol}, the constraints in these variables in appendix
\ref{ctadm-constraints}.

This system, which I call the \emph{TGR system}, is similar to the
BSSN \cite{nakamura1, nakamura2, shibata-nakamura, baumgarte-shapiro}
system.  The main differences are that it has the conformal factor
$\psi$ instead of its logarithm $\phi = \log \psi$ as primary
variable, and that the traceless conformal extrinsic curvature is
scaled with a different power of the conformal factor; while the TGR
system uses the scaling factor $\psi^{-2}$, the BSSN system uses the
same factor $\psi^4$ as for the conformal metric.  It also has $F_i$
instead of $\tilde \Gamma^i$, which is similar but different.

Compared to the original ADM system, the TGR system has the advantage
that it can more easily be used to enforce the constraints.  Enforcing
the constraints using the York--Lichnerowicz formalism \cite{york,
york2} \cite[section 2.2.1]{cook} requires a decomposition into a
system with a conformal metric and a traceless conformal extrinsic
curvature.  With the standard ADM system, this requires variable
transformations before and after each enforcing, which leads to
additional discretisation errors.  In fact, the very system used by
Cook in \cite[section 2.2.1]{cook} motivated my choice of variables.
The BSSN system would also suitable for solving the constraint
equations, although only with a slightly different method.

\section{Enforcing the gauge and the constraints}

\label{enforce-contraints}
In order to enforce the constraints with the York--Lichnerowicz
method, one introduces a vector potential $V_i$ for the traceless
conformal extrinsic curvature.  This transforms the constraint
equations into a set of coupled nonlinear elliptic equations for the
conformal factor and this vector potential, which can then be solved.
This is described in appendix \ref{ctadm-enforce-constraints}.  The
gauge condition can be enforced in a similar way by introducing a
vector potential $W_i$ for the traceless conformal metric.  This is
described in appendix \ref{ctadm-enforce-gauge}.

With given initial data, boundary, and gauge conditions, the time
evolution of a physical system is defined by the time evolution
equations for the above system.  The constraint and the gauge
equations can also be used to enforce the contraints and the gauge,
i.e.\ to create initial data, or to counteract numerical errors during
the time evolution.

Enforcing the constraints and the gauge conditions has to happen in a
certain order.  This order is necessary, because e.g.\ enforcing the
gauge condition for $F_i$ on $\tilde \gamma_{ij}$ changes the
constraints.  Therefore, the gauge condition for $F_i$ has to be
enforced before the constraints are enforced.  Lapse and shift are
calculated after all enforcing.  One possible order is the following:

\begin{enumerate}
   \item Enforce gauge, part I: set $K = K^*$

   \item $\det \tilde \gamma_{ij}$: Rescale $\tilde \gamma_{ij}$ such
   that $\det \tilde \gamma_{ij} = 1$

   \item Enforce gauge, part II: change $\tilde \gamma_{ij}$ such that
   $F_i = F_i^*$, preserving $\det \tilde \gamma_{ij}$

   \item $\trace \tilde A_{ij}$: Change $\tilde A_{ij}$ such that
   $\trace \tilde A_{ij} = 0$ (depends on $\tilde \gamma_{ij}$)

   \item Enforce constraints: change $\psi$ and $\tilde A_{ij}$ such
   that $H=0$ and $M_i=0$ (depends on $\tilde \gamma_{ij}$, $K$, and
   $\tilde A_{ij}$)

   \item Calculate lapse $\alpha$ and shift $\beta^i$ (depends on
   $\psi$, $\tilde \gamma_{ij}$, $K$, and $\tilde A_{ij}$)
\end{enumerate}
In the above, values marked with an asterisk $\cdot^*$ are values
prescribed by the gauge condition.

In principle, enforcing the gauge and the constraints is not necessary
for a perfect time evolution.  In practice, it is necessary or
advantageous to counteract numerical errors.  It might be possible to
change these elliptic equations into hyperbolic ones which would be
much cheaper to solve; I have not tried this.

% LocalWords:  eschnett Exp ij kl

% -*-LaTeX-*-
% $Header: /home/eschnett/cvs/diss/boundary.tex,v 1.20 2003/01/10 16:39:49 eschnett Exp $

\chapter{Boundary conditions}
\label{boundary}

Boundary conditions are necessary while integrating the evolution
equations in time, and for solving the elliptic gauge and constraint
equations.  They are needed at the outer boundary, and if excision
(see below) is used, also at the excision boundary.

\section{Outer boundary}

Most astrophysically interesting spacetimes are asymptotically flat
\cite[chapter 11]{wald}.  That means that at large distances from the
origin, the spacetime is flat plus some perturbation that falls off at
least with $1/r$.  Of the spacetimes considered in this thesis, only
some numerical test problems are not asymptotically flat.

For single black hole and binary black hole collision simulations, the
outer boundary should ideally be located at spatial or null infinity.
In that case, no gravitational radiation will enter or leave the
simulation domain, which makes the boundary condition easy to
implement.

\subsection{Location of the outer boundary}

In a numerical simulation, the outer boundary will be at a finite
location (and not at spatial infinity), where the four-metric is not
flat.\footnote{It is possible to put the outer boundary at null
infinity or beyond by using a conformal approach
\cite{conformal-approach}.  I will not consider this here.}  It can
usually be assumed to consist of a known background metric that is
superposed with small perturbations.  These perturbations can be
gravitational waves, or can be gauge modes, or can be constraint
violating modes that arise from numerical errors.

It is theoretically possible to put the outer boundary indeed at
spatial infinity by choosing a suitable coordinate system.  However,
one would have to get there with a finite number of grid points, and
the gravitational waves would be backscattered by the change in
resolution.  One would have to make sure that the gravitational waves
have been absorbed or dissipated before they reach a region where they
cannot be resolved any more.  Given this, there seems to be no real
advantage in putting the outer boundary at spatial infinity.

\subsection{Physical boundary conditions}

Astrophysically, one wants an outgoing radiation boundary condition at
the outer boundary.  All gravitational radiation that reaches the
outer boundary should be let out, and no gravitational radiation
should be sent in or reflected back.  This means that the outer
boundary has to be in a region where one can meaningfully speak of
gravitational radiation ``on top of something else'', i.e.\ one needs
to be close to a known background solution, and the radiation
amplitude has to be small.

A true radiative boundary condition would require splitting the
quantities at the boundary into a background solution containing the
total mass and spin, plus a linearised perturbation representing the
gravitational waves, separated into their incoming and outgoing modes.
This would be a rather complex and time consuming process.  People
have instead contemplated matching a perturbative formulation to the
nonlinear formulation at the outer boundary \cite{ccm}.  This should
in principle lead to a very clean boundary condition, but is hampered
in practice by the coordinate transformations necessary to match the
formulations.  I know of no stable implementation using this approach.

One approximation is to use some kind of sponge layer near the outer
boundary.  Within this layer, one introduces an increasing absorption
coefficient.  It emits no radiation by itself, absorbs the waves
reaching the layer, and reflects almost nothing back.  On the outer
side of this sponge layer one imposes a Dirichlet boundary condition.
Penn State uses a blending zone \cite{grazing}, which is similar to
this approach.

\label{radiative-boundary}
One other approximation is to assume that the waves are spherical
about the origin.  AEI uses conditions like this.  Such radiative
boundary conditions assume that the evolved quantity is an outgoing
radial wave, satisfying $f(t,r) = f_{\infty} + u(r-t)/r$, where the
constant $f_{\infty}$ is given, and the function $u$ is determined
implicitly at each boundary point.

\label{robin-boundary}
For elliptic equations, the usual boundary condition is a Robin
boundary condition, which enforces a certain falloff towards infinity,
i.e.\ a condition $f(r) = f_{\infty} + C/r^n$ with given $f_{\infty}$
and $n$.  Here $f_{\infty}$ is the desired value at infinity, and $n$
is the falloff power.  The constant $C$ is determined implicitly at
each boundary point.

If the solution at the outer boundary is known, which is usually not
the case, then one can use Dirichlet boundary conditions.  For waves,
a constant-in-time Dirichlet boundary is a reflective boundary, so
that waves cannot leave the simulation domain.  This makes Dirichlet
boundaries unsuitable for realistic applications, but also well suited
for test problems, e.g.\ to study the stability of a formulation.

\subsection{Periodicity}

It is, especially for test problems, sometimes convenient to assume
periodicity.  Periodicity means that there are no real boundary
conditions; this allows one to examine the system of evolution
equations independent of any problems that might be caused by the
boundary conditions.  For example, the first two stages of Winicour's
robust stability test \cite{robust-stability, robust-stability-2}
involve periodic boundaries.  Additionally, periodicity allows one to
more easily examine the long-term behaviour of dynamic systems,
because the interesting features cannot leave the simulation domain.

Unfortunately, periodicity poses a severe problem for elliptic
equations.  Their solution is determined by the boundary condition,
and without a boundary their solution is not unique any more, or might
cease to exist.  If the solution is not unique any more, then one can
impose a pseudo ``boundary condition'' at a single point in the domain
to select a solution.

The elliptic equations arising in the TGR system seem in fact to be
ill-posed when given periodic boundaries.  That is, they do not admit
a solution any more.  This can already be seen at the example of the
Poisson equation $\Delta \phi=\rho$ with periodic boundaries.  For
$\rho=0$, one needs to define $\phi$ at one point to make the solution
unique.  For $\rho \ne 0$, the shape of $\phi$ is a parabola, and is
therefore clearly not periodic.  Similarly, the Hamiltonian constraint
equation in the York--Lichnerowicz formalism with periodic boundaries
seems to be ill-posed when $K_{ij}$ is not zero.  For $K_{ij}=0$, a
solution exists, but is not unique.

This prevents many interesting test cases from being run while
enforcing the gauge conditions and constraints.  It is in some cases
possible to convert these test cases into almost equivalent ones with
Dirichlet boundaries.  A travelling wave with periodic boundaries can
e.g.\ be changed into a standing wave with periodic boundaries, and
that system is then very similar to a standing wave with Dirichlet
boundaries at the wave knots.  And fortunately, test cases with
Dirichlet boundaries are often considered to be more difficult to pass
than those with periodic boundaries, i.e.\ they form stronger test
cases.

\subsection{Gauge and constraint boundary conditions}

A physical boundary condition does not define how to handle the gauge
and constraint violating degrees of freedom.  Physically, the
constraints have to be satisfied.  Numerically, there will always be
errors.  One has to try to reduce the violation of the constraints in
some way, and this mechanism will interact with the boundary
condition.

Similarly, after fixing the gauge, the gauge will numerically only be
satisfied approximately.  The boundary conditions have to deal with
violations of the gauge and the constraints, meaning that there also
has to be a boundary condition for the gauge and constraint
violations.  Unfortunately it is not clearly defined just which
(combination) of the evolved or solved-for quantites contain these
gauge and constraint violating degrees of freedom.

In my system, the variables $\psi$, $\tilde \gamma_{ij}$, and $\tilde
A_{ij}$, are evolved in time, while $K$ and $F_i$ are prescribed in
advance by the gauge condition.  It is not necessary to evolve the
conformal factor $\psi$ in time, as it can also be calculated from the
Hamiltonian constraint.  I do evolve it only to have a better initial
guess for enforcing the Hamiltonian constraint, and in order to have
more freedom in choosing the boundary conditions.  Depending on the
problem, I use either Dirichlet or radiative boundary conditions for
the time evolution of these evolved variables.

It is clear that a Dirichlet boundary condition, even if it is
consistent with the equations, is not suitable for an astrophysical
problem.  The radiation that is reflected at the outer boundary makes
the result unphysical.  But as such reflected radiation is considered
to introduce instablities, a Dirichlet boundary is actually a stronger
stability test.  I would use a sponge layer or a radiative boundary
condition if I were to run a simulation with a domain large enough and
a resolution fine enough to permit meaningful extraction of
gravitational radiation.

The variables $W_i$, $\psi$, and $V_i$ need boundary conditions when
the gauge and the constraints are enforced by solving elliptic
equations.  I use either Dirichlet or Robin boundary conditions for
them.  I impose $W_i=0$ and $V_i=0$ on the vector potentials.  For the
conformal factor $\psi$ I distinguish between initial data and time
evolution.  For initial data I use either a Dirichlet or a Robin
boundary condition.
%% with $\lim_{r\to\infty} \psi(r) = 1 + C/r$.
During time evolution, I use a Dirichlet or a radiative boundary
condition.

I also apply Dirichlet or Robin boundary conditions to the lapse
$\alpha$ and the shift $\beta^i$.  When using a Dirichlet boundary
condition, I retain the boundary values from the initial data.

The lapse and the shift determine the coordinate system that is
constructed for the spacetime during time evolution, and therefore
their boundary values are important and have a clear meaning.  Their
boundary condition is part of the gauge condition, as described in
section
%% \ref{coord-boundary}%
\ref{my-gauge-condition}%
.

\section{Excision boundary}
\label{excision}

Black hole spacetimes contain singularities.  These singularities can,
of course, not be directly treated numerically.  Several methods have
been used to treat them in a $3+1$ split.

The likely oldest and mathematically most elegant ansatz is
\emph{maximal slicing} \cite{maximal-minimal}.  For this, one chooses
the coordinate time in such a way that it reaches the singularity only
for $t \to \infty$.  The relation between the coordinate time $t$ and
the proper time $\tau$ of an observer can be chosen almost arbitrarily
through a suitable lapse function.  With maximal slicing it is
nevertheless the case that the whole spacetime is eventually covered
by the time coordinate.

This elegant ansatz has the disadvantage that the time coordinate in
the neighbourhood of the singularity proceeds more and more slowly,
which eventually leads to a distortion of the coordinate system
(``grid stretching'').  Although the coordinate system stays well
defined all the time, certain metric components start to grow without
bound.  This can in the end not be treated numerically any more.
Maximal slicing can thus describe a black hole only for a certain
amount of coordinate time before a code crashes for numerical reasons.
There are also different slicings, such as harmonic
\cite{harmonic-slicing} or $1+\log$ \cite{1+log} slicings, with
similar singularity--avoiding properties, and which also have similar
disadvantages.

%% \todo{remove these two paragraphs?:}
A different ansatz was developed at the AEI and is called
\emph{punctures} \cite{punctures}.  At the singularity, certain metric
components are infinite.  This behaviour can be described by
decomposing the three-metric $\gamma_{ij}$ into a conformal factor
$\psi$ and a conformal three-metric $\tilde \gamma_{ij}$.  For certain
initial data (such as e.g.\ Brill--Lindquist black holes; see section
\ref{brill-lindquist}), the conformal three-metric remains finite over
the singularity, and only the conformal factor diverges.  The
conformal factor for the initial data is known analytically, and by
certain gauge choices (maximal slicing ($K=0$) and normal coordinates
($\beta^i=0$) at the punctures), the time derivative of the conformal
factor can be set to zero.  Additionally, one chooses a numerical grid
such that the singularity does not coincide with a grid point.
Through that, the code never has to calculate any diverging
quantities. The conformal factor and its spatial derivatives are known
analytically and can be provided with arbitrarily high accuracy.

However, the restriction on the gauge conditions has certain
disadvantages; e.g.\ the position of the singularity cannot change in
time.  When two black holes coalesce, their physical distance
decreases, while the coordinate distance stays constant with this
method.  This leads to similar problems as with maximal slicing.  But
a single black hole, which is an important test problem in numerical
relativity, can in principle be described without problems with
punctures.

Another way to treat singularities is \emph{excision}.  It consists of
removing a certain region around the singularity from the simulation
domain.  Because there is, under certain reasonable assumptions,
always an event horizon enclosing a singularity, this excising is
possible without influencing the part of the domain outside the
horizon.  In the language of hydrodynamics, the excision or inner
boundary is a supersonic outflow boundary for the physical degrees of
freedom.  Outflow means here ``out of the simulation domain'', i.e.\
in the direction towards the singularity, further into the black hole.

\begin{figure}
\begin{tabular}{rr}
\includegraphics[width=0.45\textwidth]{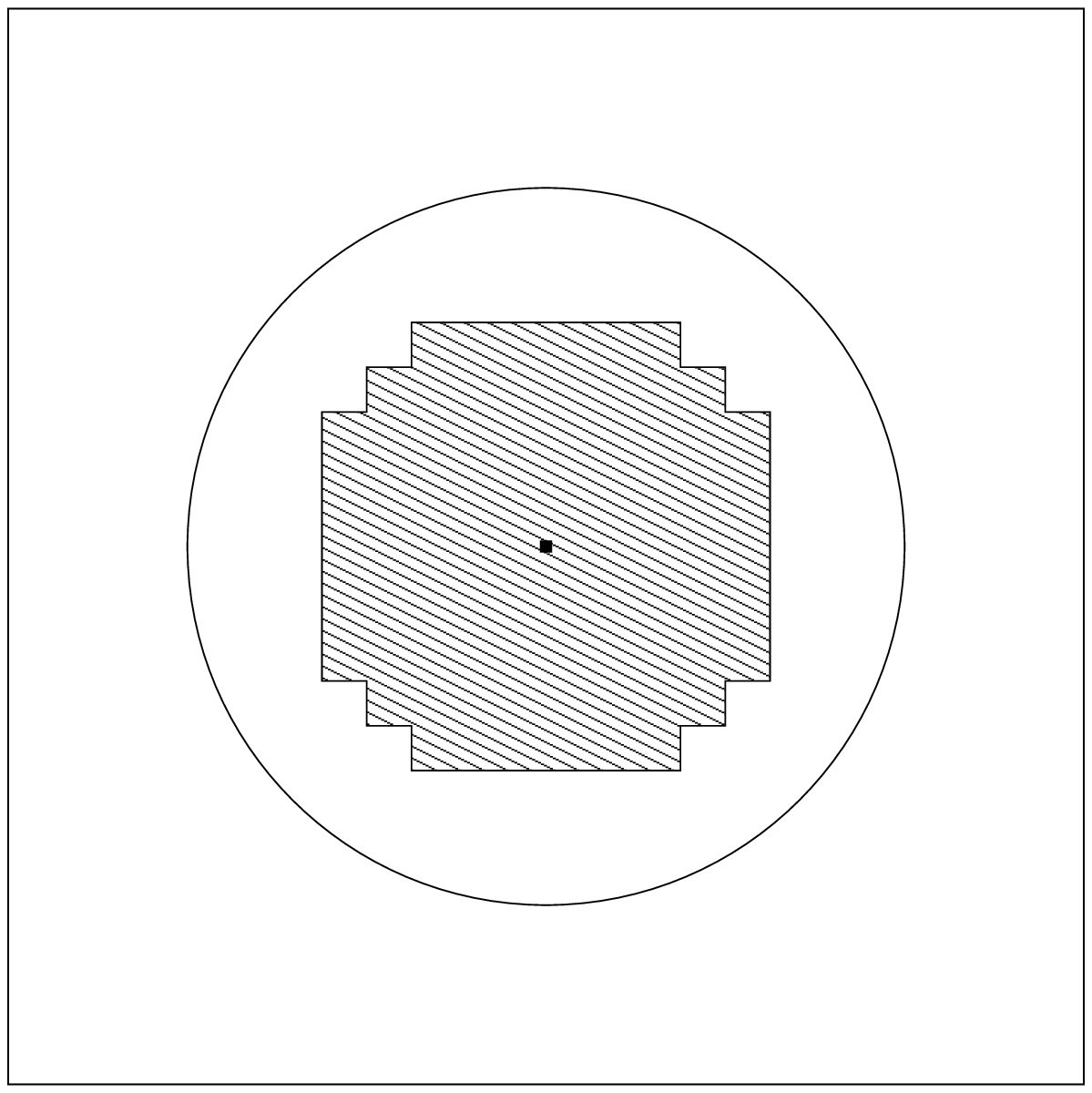}&
\includegraphics[width=0.65\textwidth]{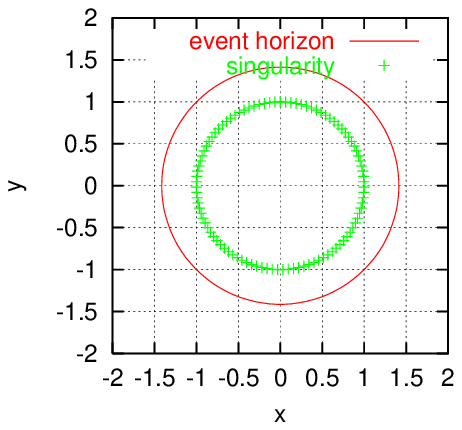}
\end{tabular}
\caption{\label{excision-figure}Left: Schematic picture of an excision
region.  The circle depicts the apparent horizon.  The grey grid cells
are excised.  The excision boundary has to follow the grid cell
boundaries.  Right: A cut through the equatorial plane for an $M=1$,
$a=1$ black hole in Kerr--Schild coordinates.  There is not much
coordinate space between the event horizon and the ring-shaped
singularity.}
\end{figure}

In spherical coordinates, excision is quite customary, and it poses no
problems, as the boundary can be put at an $r = \mathrm{const}$
coordinate plane.  Unfortunately, spherical coordinates are not
suitable for describing two coalescing black holes, which is why one
usually uses Cartesian coordinates for that.  The excised region on a
Cartesian grid can not be spherical any more, but will rather look
like a sphere that has been built from Lego blocks (see figure
\ref{excision-figure}).  This irregular shape leads to problems when
implementing the excision boundary condition.

It is generally assumed that excision is the most promising way to
treat singularities, and that it will be widely used as soon as good
boundary conditions are known for the excision region.

I will only consider excision boundaries in the remainder of this
chapter.  I will not consider punctures at all, and in order to use
maximal slicing, one needs of course no special treatment of the
singularity.

\subsection{Location and shape of the excision boundary}

In order to excise a region of space, one generally has to find a
closed two-surface that is guaranteed to be both within the event
horizon, and away from the singularity.  One surface that is
guaranteed to be within or on an event horizon is an apparent horizon
(see section \ref{ahfinding}).  For numerical reasons, one wants to
stay several layers of grid points away from the event horizon, and
also sufficiently far (as far as possible) from the singularity.  It
is thus necessary to locate apparent horizons to define or at least
check the location of the excision boundary.

Choosing the right shape for the boundary is another rather difficult
issue.  Spherical or elliptic shapes are difficult to handle, because
the do not have a normal direction that is aligned with the grid.
Such normal directions are necessary for extrapolations.  On the other
hand, using a cubic boundary is not always possible.  For example, the
singularity of a rotating black hole in Kerr--Schild coordinates (see
section \ref{kerr-schild}) is ring-shaped, and the space available
between the ring-shaped singularity and the elliptic horizon shape is
too small to permit using a cubic boundary (see figure
\ref{excision-figure}).  One thus has to deal with boundary shapes
that are not aligned with the computational grid.  Some possible
shapes include boxes ``with the corners cut off'', which are polyhedra
with 18 or 26 faces.\footnote{I think this was first suggested by
Miguel Alcubierre in 2000.}

\subsection{Boundary conditions}

Excision boundaries are very different from outer boundaries.  Apart
from the fact that they are more difficult to handle numerically
because of their irregular shape, they are located in the strong field
regime, and hence one cannot assume that any approximation holds.
Luckily, there is an event horizon present, and the fact that no
physical information can escape from the horizon means that one has
some freedom in managing this boundary.  As long as one stays within
the limits of physics, one can choose any boundary condition without
influencing the physics of the spacetime outside the horizon.

Yet there are things that do escape from the horizon.  Information
about the gauge and about constraint violations does not carry
physical information.  It can propagate faster than light, and can
also leave the horizon.  There is \emph{a priori} no reason to assume
that it would not do so.  Care has to be taken in separating the
individual quantities, which is much more difficult than at the outer
boundary, where everything can be assumed to be a small perturbation
of a known background.  The inner boundary is a place where a
hyperbolic formulation would definitively be of help.

While solving elliptic equations, it is consistent to use (arbitrary,
constant in time) Dirichlet boundary conditions for the quantities
$V_i$, $W_i$, $\alpha$, and $\beta^i$.  Here the boundary values of
the shift have to prevent the grid boundary from falling further into
the black hole and getting too close to the singularity, or from
rising out of it and leaving the event horizon (see also section
\ref{picturing-spacetimes}).  The latter would invalidate the
assumption underlying excision and must therefore not be permitted.
%% \todo{what about $\psi$?}

%% This paragraph still has a big question mark associated with it.
While integrating in time, the boundary value for the conformal factor
$\psi$ has to be determined by a Dirichlet boundary condition from a
known solution, or by extrapolation.  The quantities $\tilde
\gamma_{ij}$ and $\tilde A_{ij}$ carry a mixture of physical, gauge,
and constraint information.  The physical information has to be
advected out of the simulation domain.  Using a Dirichlet boundary
condition for the physical degrees of freedom is consistent only if
one uses the correct values.  Of course, the correct values are only
known if one tests the code with an analytic solution, so this is not
possible for real-world cases.

%% \todo{the following happens to not be the case:}
%% The gauge and constraint information should have boundary conditions
%% imposed that is consistent with the choices $W_i=0$ and $V_i=0$, i.e.\
%% that do not let the boundary float freely.  Using e.g.\ an
%% extrapolation boundary condition for the gauge degrees of freedom does
%% not really fix the gauge, and hence would allow gauge modes.

A supersonic outflow boundary condition can be realised by
extrapolation.  Unfortunately, extrapolating the conformal metric
causes problems, because the extrapolated conformal metric will
normally not satisfy the constraint $\det \tilde \gamma_{ij} = 1$ any
more.  It is not a good idea to enforce this constraint by rescaling
the whole metric, because this will change certain gauge degrees of
freedom in unpredictable ways.  Allowing for arbitrary gauge changes
at the excision boundary might allow gauge modes to appear.

Instead, the gauge degrees of freedom have to be fixed, according to
the plan I set out in section \ref{gauge-fixing}.  The obvious gauge
degree of freedom to fix at the inner boundary is the location of the
boundary itself.  That is, I want to clearly define how the inner
boundary moves inward or outward (again, see also section
\ref{picturing-spacetimes}).  Therefore I decompose the metric
components $\tilde \gamma_{ij}$ into those parallel to and those
normal to the boundary.  I enforce the constraint $\det \tilde
\gamma_{ij} = 1$ by rescaling only those components that are normal to
the boundary.  This is also consistent with the fact that
gravitational waves cause transverse length changes only, because
these lengths will not be changed by the rescaling.

The traceless conformal extrinsic curvature happens to be less
critical.  It does not contain any gauge degrees of freedom.  By
extrapolating $\tilde A_{ij}$ one violates the constraint $\tilde
A_i^i=0$, especially because the conformal metric used to calculate
that trace is also extrapolated, and hence changes as well.  Both the
extrapolation of $\tilde A_{ij}$ and enforcing $\tilde A_i^i=0$ later
on will change the boundary condition for the constraints.  However,
the constraints can be enforced no matter what the boundary values.
Numerical experiments show that the handling of the traceless
conformal extrinsic curvature is not critical and does not lead to
instabilities.

I assume that the constraint violation error that is introduced by
extrapolating $\tilde A_{ij}$ is less malevolent than the gauge error
introduced by extrapolating $\tilde \gamma_{ij}$.  This gauge error
has the ability to change the extent of the simulation domain, and
that is not acceptable.  On the other hand, the constraint error could
lead to certain components of $\tilde A_{ij}$ growing without bound,
although the constraints would stay satisfied.  Luckily, that seems to
not happen.

With the the above boundary conditions for the excision boundary, the
time evolution of the TGR system is consistent, and all degrees of
freedom are fixed.

% LocalWords:  eschnett Exp ij const

% -*-LaTeX-*-
% $Header: /home/eschnett/cvs/diss/initial.tex,v 1.16 2003/01/10 16:39:49 eschnett Exp $

\chapter{Initial data}
\label{initial}

It is a nontrivial task to generate initial data for a black hole
simulation.  There are many analytic solutions for single black holes,
but the proposed methods to construct spacetimes with two or more
black holes either restrict the possible configurations, or require
solving elliptic equations.  Multiple black hole initial data usually
need to be interpreted in terms of their ADM mass and spin
\cite[chapter 6.6]{weinberg}, and the apparent horizons (see chapter
\ref{ahfinding}) that can be found.

Initial data are usually presented in the ADM variables $\gamma_{ij}$
and $K_{ij}$, even if they are actually calculated in other variables.
This has the advantage that the form of the initial data is
independent of the formulation used in the evolution, and facilitates
exchanging initial data between different formulations.  The
transformation between the ADM and other variables is usually
straightforward.

I present three major kinds of initial data in this chapter.  I start
with data without black holes.  These data often have an analytic
solution for all times and can thus be easily used as test cases.  I
then present various coordinate systems for static or stationary
single black holes.  They are also mostly test cases of little
astrophysical relevance.  Last I present initial data for multiple
black holes, for which no closed form solution exists for later times.

I follow the convention of giving names not to spacetimes, but rather
to coordinate systems.  This is motivated by the fact that different
coordinate systems behave very differently numerically.  On the other
hand, the fact that two different coordinate systems might describe
the same physical spacetime is largely irrelevant for me, because it
cannot easily be tested.  I also hope that this helps to prevent
confusion.  I thus distinguish between flat space (a spacetime) and a
Minkowski metric (one possible coordinate system for it).  Similarly,
there are static black holes, of which Schwarzschild is just one
possible metric, with zero-spin Kerr--Schild and Painlev\'e--Gullstand
being others.

\section{Data without black holes}

\subsection{Minkowski metric (flat space)}
\label{minkowski}

The Minkowski metric is the simplest case.  It describes flat space in
Cartesian coordinates.  It has $g_{\mu\nu} = \eta_{\mu\nu}$, or
$\gamma_{ij} = \delta_{ij}$, $K_{ij} = 0$, $\alpha = 1$, and $\beta^i
= 0$.

\subsection{Weak Bondi wave (linear planar wave)}
\label{weak-bondi-wave}

A weak Bondi wave is a planar gravitational wave with a small
amplitude.  This is a solution to the linearised Einstein equations,
presented e.g.\ in \cite[chapter 35.9, eqn.\ (35.32)]{mtw}.  Assuming
that the wave propagates in the $z$ direction, the ADM quantities are
given by
\begin{eqnarray}
   \gamma_{xx} & = & 1 + b \\
   \gamma_{yy} & = & 1 - b \\
   \gamma_{zz} & = & 1 \\
   K_{xx} & = & - \frac{1}{2} \dot b \\
   K_{yy} & = & \frac{1}{2} \dot b \\
   K_{zz} & = & 0
\end{eqnarray}
where all other components of $\gamma_{ij}$ and $K_{ij}$ are zero, and
$\alpha=1$ and $\beta^i=0$.  The free parameter function $b(t,z)$ can
e.g.\ be chosen as
\begin{eqnarray}
   b & = & A \sin k\, (z-t) \\
   \dot b = \partial_t b & = & - k\, A \cos k\, (z-t)
\end{eqnarray}
for a wave propagating in the $z$ direction, or as
\begin{eqnarray}
   b & = & A\; \sin k z\; \cos k t \\
   \dot b = \partial_t b & = & - k\; A\; \sin k z\; \sin k t
\end{eqnarray}
for a standing wave, where in all cases $|b| \ll 1$ has to hold.  The
parameters $A$ and $k$ determine the amplitude and wave number of the
weak Bondi wave.  This metric is a solution of the Einstein equations
only for $|A| \ll 1$.

\subsection{Gauge pulse (nonlinear planar gauge wave)}
\label{gauge-pulse}

A gauge pulse is a planar nonlinear gauge wave.  This is a solution to
the full vacuum Einstein equations which does not contain
gravitational waves.  It represents a flat spacetime in a strange
coordinate system.  Assuming that the wave propagates in the $z$
direction, the ADM quantities are given by
\begin{eqnarray}
   \gamma_{xx} & = & 1 \\
   \gamma_{yy} & = & 1 \\
   \gamma_{zz} & = & \exp\left\{ b \right\} \\
   K_{zz} & = & - \frac{1}{2} \exp\left\{ \frac{b}{2} \right\}\; \dot
   b \\
   \alpha & = & \exp\left\{ \frac{b}{2} \right\}
\end{eqnarray}
where all other components of $\gamma_{ij}$, $K_{ij}$, and $\beta^i$
are zero.  The free parameter function $b(t,z)$ can e.g.\ be chosen as
\begin{eqnarray}
   b & = & A \sin k\, (z-t) \\
   \dot b = \partial_t b & = & - k\, A \cos k\, (z-t)
\end{eqnarray}
for a wave propagating in the $z$ direction, or as
\begin{eqnarray}
   b & = & A\; \sin k z\; \cos k t \\
   \dot b = \partial_t b & = & - k\; A\; \sin k z\; \sin k t
\end{eqnarray}
for a standing wave.  The parameters $A$ and $k$ determine the
amplitude and wave number of the gauge pulse.  The amplitude $A$ can
be arbitrarily large.

\subsection{Brill wave}
\label{brill-wave}

A Brill wave is a nonlinear, axially symmetric gravitational wave
\cite{brill}.  A Brill wave can be strong enough to form a black hole.
This is an interesting test case, insofar as a black hole can form
from scratch, without a singularity or matter being present in the
beginning.

In contrast to the solutions presented above, which are valid for all
times, the Brill wave metric given below describes only initial data
at $t=0$.  The metric can be written as
\begin{eqnarray}
   ds^2 = \psi^4 \left[ e^{2q} (d\rho^2 + dz^2) + \rho^2 d\phi^2
   \right]
\end{eqnarray}
with the free function $q(\rho,z)$ than can e.g.\ be chosen as
\begin{eqnarray}
%%    q = A \rho^2 e^{-r^2} \left[ 1 + C \frac{\rho^2}{1+\rho^2}
%%    \cos^2(\omega\phi) \right]
   q = A\, \rho^2\, e^{-r^2}
\end{eqnarray}
with the amplitude $A$.
%% the angular modulation amplitude $C$, and the angular modulation
%% frequency $\omega$.
$\rho$ is the cylindrical radial coordinate with $\rho^2 = x^2 + y^2$.

These initial data are then chosen to be time-symmetric with
$\partial_t \gamma_{ij}=0$.  Together with the gauge choice
$\beta^i=0$ this leads to the extrinsic curvature $K_{ij}=0$,
satisfying the momentum constraint identically.  The conformal factor
$\psi$ has to be chosen (solved for) so that the Hamiltonian
constraint is satisfied.  One usually uses a Robin type boundary
condition
%% with $\lim_{r\to\infty} \psi(r) = 1 + C/r$
for this.

One usually uses maximal slicing ($K=0$) and normal coordinates
($\beta^i=0$) when simulating a Brill wave.  One also usually uses a
Robin boundary condition for the lapse.
%% with $\lim_{r\to\infty} \alpha(r) = 1 + C/r$.

\section{Single black hole data}

There are many coordinate systems that describe single black holes.  I
chose to test my formulation with three, namely Kerr--Schild
coordinates, Painlev\'e--Gullstrand coordinates, and harmonic
coordinates.  The original Schwarzschild coordinates are not suitable
here, because they do not penetrate the horizon.  That would not allow
me to use an excision boundary condition (see section \ref{excision}),
which has to be applied inside the event horizon.

%% The original Schwarzschild\footnote{Please allow me to remark that the
%% name ``Schwarzschild'' should be pronounced and hyphenated as
%% Schwarz-schild --- it has nothing to do with the English word
%% ``child''.} coordinates are not suitable for a numerical simulation,
%% as they do not cover all of the spacetime.

\subsection{Kerr--Schild coordinates}
\label{kerr-schild}

For single black hole test runs I prefer the Kerr--Schild
coordinates\footnote{For some historic reason, these coordinates are
sometimes also called ``ingoing Eddington--Finkelstein coordinates''
in the community, although the real iEF coordinates are different.}.
This is a static (or stationary, when there is a nonzero spin)
solution with a spin that can be specified freely.  The slicing does
intersect the singularity, and the singularity in the time slice is a
point for a non-spinning, and is ring-shaped for a spinning black
hole.  The singularity diverges with $1/r^2$ only, which still allows
reasonable numerical resolutions to be used.

Kerr--Schild coordinates \cite[section 3.3.1]{cook} are usually
expressed in an elliptic coordinate system, where the $r$ coordinate
is given by
\begin{eqnarray}
   \rho^2 & = & r^2 + a^2 \left( 1 - \frac{z^2}{r^2} \right)
%
%% r^4 & = & r^2 \left( \rho^2 - a^2 \right) + a^2 z^2
\end{eqnarray}
with the usual radial coordinate $\rho$, i.e.\ $\rho^2 = x^2 + y^2 +
z^2$.  The four-metric can be written as
\begin{eqnarray}
   \label{kerr-schild-metric} g_{\mu\nu} & = & \eta_{\mu\nu} + 2 H
   l_\mu l_\nu
\end{eqnarray}
where $\eta_{\mu\nu}$ is the Minkowski metric, $H$ is a scalar, and
$l_\mu$ is a null vector, which are given by
\begin{eqnarray}
   H & = & \frac{M r^3}{r^4 + a^2 z^2} \\
   l_t & = & 1 \\
   l_x & = & \frac{r x + a y}{r^2 + a^2} \\
   l_y & = & \frac{r y - a x}{r^2 + a^2} \\
   l_z & = & \frac{z}{r}
\end{eqnarray}
where $M$ is the black hole mass, and $a$ is its spin about the $z$
axis.  The three-metric, lapse, and shift then follow as
\begin{eqnarray}
   \gamma_{rr} & = & 1 + \frac{2Mr}{\rho^2} \\
   \gamma_{r\phi} & = & -a \left[ 1 + \frac{2Mr}{\rho^2} \right]
   \sin^2 \theta \\
   \gamma_{\theta\theta} & = & \rho^2 \\
   \gamma_{\phi\phi} & = & \left[ r^2 + a^2 + \frac{2Mr}{\rho^2} a^2
   \sin^2 \theta \right] \sin^2 \theta \\
   \alpha & = & \frac{1}{\sqrt{1 + \frac{2Mr}{\rho^2}}} \\
   \beta^r & = & \alpha^2 \frac{2Mr}{\rho^2} \quad.
\end{eqnarray}
The event horizon is located at $r = M + \sqrt{M^2-a^2}$.

The extrinsic curvature is defined implicitly via $\gamma_{ij}$,
$\alpha$, $\beta^i$, and their derivatives through the ADM time
evolution equation for $\gamma_{ij}$ (see eqn.\ (\ref{evol-metric})).

\subsection{Painlev\'e--Gullstrand coordinates}
\label{painleve-gullstrand}

Painlev\'e--Gullstrand coordinates \cite[section 3.3.3]{cook} are
interesting because they have a flat three-metric $\gamma_{ij} =
\delta_{ij}$.  However, it seems empirically that their time evolution
leads to higher discretisation errors.  The spacetime is described by
\begin{eqnarray}
   \gamma_{ij} & = & \delta_{ij} \\
   K_{ij} & = & \sqrt{\frac{2M}{\rho^3}} \left[ \delta_{ij} -
   \frac{3}{2} q^i q^j \right] \\
   \alpha & = & 1 \\
   \beta^i & = & \alpha^2 \sqrt{\frac{2M}{\rho}} q^i
\end{eqnarray}
where the parameter $M$ is the mass of the black hole.  The auxiliary
quantity $q^i$ is defined as $q^i = x^i/\rho$.  I do not allow for
spin.  $\rho$ is the usual radial coordinate with $\rho^2 = x^2 + y^2
+ z^2$.

\subsection{Harmonic coordinates}
\label{harmonic}

Harmonic coordinates follow from the coordinate conditions $\square
x^{(\mu)} = 0$ \cite[section 3.3.2]{cook}.  A black hole in harmonic
coordinates is described by
\begin{eqnarray}
   \gamma_{ij} & = & \delta_{ij} + \left( v + v^2 + v^3 \right) q^i
   q^j \\
   \alpha & = & \frac{1}{\sqrt{1 + v + v^2 + v^3}} \\
   \beta^i & = & \alpha^2 v^2 q^i
\end{eqnarray}
where $v = 2M/\rho$, and $q^i = x^i/\rho$.  $\rho$ is the usual radial
coordinate with $\rho^2 = x^2 + y^2 + z^2$, and $M$ is the mass of the
black hole.  I do not allow for spin.  The extrinsic curvature is
defined implicitly via $\gamma_{ij}$, $\alpha$, $\beta^i$, and their
derivatives through the ADM time evolution equation for $\gamma_{ij}$
(see eqn.\ (\ref{evol-metric})).

\subsection{Coordinate transformations}
\label{coordinate-transformation}

In order to have access to a larger class of initial data and analytic
solutions, I allow for a generic coordinate transformation to be
applied to the initial data.\footnote{Currently I implemented this
only for the Kerr--Schild solution.}  This transformation is applied
to the solution's four-metric, making it automatically covariant.  The
transformation can e.g.\ be used to rotate a black hole to point the
spin into any direction, boost the black hole, deform the black hole
in various more or less useful ways, and change to a moving coordinate
system.

Such a generic coordinate transformation is defined by the usual
$x^\mu = T^\mu_\nu \hat x^\nu$ with the transformation tensor $T^\mu_\nu =
\partial x^\mu / \partial \hat x^\nu$.  I also allow for a coordinate
translation $x^\mu = \hat x^\mu + C^\mu$.  This makes for altogether
$16+4=20$ free parameters.  In order to allow the coordinate change to
be prescribed conveniently, I decompose the coordinate transformation
into
\begin{eqnarray}
   \mathsf{T} & = & \mathsf{Rot} \circ \mathsf{Slow} \circ
   \mathsf{Dfrm} \circ \mathsf{Cshn} \circ \mathsf{Shear} \circ
   \mathsf{Bst} \circ \mathsf{Vel}
\end{eqnarray}
where
%
% 3 1 3 3 3 3 3
%
\begin{description}
\item[$\mathsf{Rot}$] is a rotation of the spatial components,
\item[$\mathsf{Slow}$] is a slowdown, i.e.\ a scaling of the lapse,
\item[$\mathsf{Dfrm}$] is a deformation of the spatial components, i.e.\ a
rescaling of the coordinate axes,
\item[$\mathsf{Cshn}$] is a cushion deformation of the spatial components,
\item[$\mathsf{Shear}$] is a shear transformation of the spatial components,
\item[$\mathsf{Bst}$] is a Lorentz boost, and
\item[$\mathsf{Vel}$] is a change to a moving coordinate system.
\end{description}
Of the above, $\mathsf{Bst}$ is not really necessary and could be
replaced by a combination of the other transformations, but I keep it
for convenience.  The individual transformation operators are defined
as
\begin{eqnarray}
   \mathtt{Rot}^i_j & = & \delta^i_j - \Omega^i_k \Omega^k_j\; \left(
   \cos 2\pi\alpha - 1 \right) \\
   & & \quad {} + \Omega^i_j\; \sin 2\pi\alpha \\
   & & \textrm{with} \quad \Omega^i_j = \epsilon_{ijk}\;
   \mathtt{rot}^k / \alpha \\
   & & \textrm{and} \quad \alpha = |\mathbf{rot}| \\
   \mathtt{Slow}^0_0 & = & \mathtt{slow} \\
   \mathtt{Dfrm}^i_j & = & \mathtt{diag}\left( \mathtt{dfrm^k} \right)
   \\
   \mathtt{Cshn}^i_j & = & \delta^i_j + |\epsilon_{ijk}|\;
   \mathtt{cshn}^k \\
   \mathtt{Shear}^i_j & = & \delta^i_j + \epsilon_{ijk}\;
   \mathtt{shear}^k \\
   \mathtt{Bst}^0_0 & = & \gamma \\
   \mathtt{Bst}^i_0 = \mathtt{Bst}^0_i & = & \mathtt{bst}^i\; \gamma
   \\
   \mathtt{Bst}^i_j & = & \delta^i_j + \frac{\gamma^2}{\gamma+1}\;
   \mathtt{bst}^i\; \mathtt{bst}^j \\
   & & \textrm{with} \quad \gamma = \frac{1}{\sqrt{1 -
   |\mathbf{bst}|^2}} \\
   \mathtt{Vel}^i_0 & = & \mathtt{vel}^i
\end{eqnarray}
where the remaining components of these four-tensors are set to
$\delta^\mu_\nu$.  The lower-case three-vectors $\mathtt{rot}^i$,
$\mathtt{dfrm}^i$, $\mathtt{cshn}^i$, $\mathtt{shear}^i$,
$\mathtt{bst}^i$, $\mathtt{vel}^i$, and the scalar $\mathtt{slow}$
parameterise these transformations with 19 (instead of the necessary
16) free parameters.  Additionally there are the 4 translation
parameters $C^\mu$.

A rotation is described by a three-vector $\mathtt{rot}^i$.  Its
direction is the axis of rotation, and its length the angle, where a
length of $1$ means one full rotation.  A cushion transformation
changes a square into a cushion- or a barrel-shaped object, where the
components of the three-vector $\mathtt{cshn}^k$ describe the
distortion factors along the diagonals.  A shear transformation
changes a square into a rhomboid, where the components of the
three-vector $\mathtt{shear}^k$ describe the shear factor along the
corresponding coordinate axis.

Rotations are useful e.g.\ to point the spin of an object in an
arbitrary direction.  Deformations and shears can be used e.g.\ to
create coordinates where the contravariant and covariant three-vectors
differ (for testing purposes).  A velocity transformation can be used
to cancel effects of a boost, e.g.\ to create a stationary boosted
black hole (which differs from a non-boosted black hole).  Rotations
and boosts together form the proper Lorentz transforms.

From the transformed four-metric I then calculate the ADM quantities,
i.e.\ the three-metric, the extrinsic curvature, lapse, and shift.
The extrinsic curvature is defined via the time derivative of the
three-metric using equation (\ref{evol-metric}).  The necessary
partial derivatives can easily be calculated numerically to any given
accuracy, much more accurately than to the grid spacing accuracy.

When lapse and shift are later, during the time evolution, determined
through elliptic equations, one needs boundary conditions for them.  I
often use the boundary values from the initial data as Dirichlet
boundary conditions.  This makes the method used to initially
calculate lapse and shift actually important.

\section{Multiple black hole data}

\subsection{Brill--Lindquist data}
\label{brill-lindquist}

Brill--Lindquist data \cite[section 3.1.2]{cook} can contain
arbitrarily many black holes.  The black holes are described by their
coordinate location and by a mass parameter.  This solution is also
only valid on a single time slice.  It is time-symmetric and
conformally flat:
\begin{eqnarray}
   \psi & = & 1 + \sum_n \frac{\mu_n}{2|\mathbf{x}-\mathbf{x}_n|} \\
   \gamma_{ij} & = & \psi^4 \delta_{ij} \\
   K_{ij} & = & 0 \\
   \alpha & = & 1 \\
   \beta^i & = & 0
\end{eqnarray}
where $\mathbf{x}_n$ are the positions of the black holes, and $\mu_n$
are their mass parameters.

If there is only one black hole, then the convention to write $\mu/2$
instead of $2\mu$ means that the event horizon is at $r=\mu/2$ instead
of at $r=2\mu$.  In this case, the black hole still has the mass
$M=\mu$.

\subsection{Superposed Kerr--Schild data}
\label{superposed-kerr-schild}

For multiple black hole runs I prefer superposed Kerr--Schild data.
They were to my knowledge first proposed by Mazner et al.\
\cite{humash} and have recently been refined by Moreno et al.\
\cite{humash2}.  They follow a rather intuitive approach.  One views a
single black hole as the sum of a flat space metric and a black hole
metric contribution.  This view is motivated by the form of equation
(\ref{kerr-schild-metric}) in which the Kerr--Schild four-metric can
be written.  One then combines two black holes by adding two different
black hole contributions to a flat space metric.  This combined metric
does, however, not satisfy the constraints any more, so that a
constraint solving step has to follow.

The black holes can be combined in two ways.  In the first way, one
combines the three-metrics and the extrinsic curvatures; in the second
way, one combines the four-metrics and their time derivatives.  Both
ways are equally ``valid'' in principle.  Note, however, that the
second way breaks down when the black holes are close to each
other.\footnote{The combined four-metric then does not have a
$(-,+,+,+)$ signature any more.  This cannot happen when superposing
the three-metrics and extrinsic curvatures.}  It can be convenient to
attenuate the combined solutions as one gets close to one of the black
holes, regaining a single black hole solution near the individual
holes.

When superposing the three-metrics and the extrinsic curvatures of $n$
black holes, one uses
\begin{eqnarray}
   \gamma_{ij} & = & \delta_{ij} + \sum_n \left[\gamma^{(n)}_{ij} -
   \delta_{ij} \right] \\
   K_{ij} & = & \sum_n K^{(n)}_{ij} \\
   \alpha & = & 1 + \sum_n \left[ \alpha^{(n)} - 1 \right] \\
   \beta^i & = & \sum_n \beta^{(n)}{}^i
\end{eqnarray}
The quantities $\cdot^{(n)}$ are the corresponding quantities from the
to-be-superposed black holes.  Note that the way in which the
extrinsic curvatures are superposed means that the superposition of
two black holes at the same position does not give a single black hole
with the combined mass, and does not satisfy the constraint equations
any more.

When superposing the four-metric and its time derivative, one uses
\begin{eqnarray}
   g_{\mu\nu} & = & \eta_{\mu\nu} + \sum_n \left[ g^{(n)}_{\mu\nu} -
   \eta_{\mu\nu} \right]
\end{eqnarray}
and then calculates three-metric, extrinsic curvature, lapse, and
shift from this four-metric and its time derivatives.  In both cases,
one can also use a different lapse and shift than calculated here, but
this will lead to a different extrinsic curvature with the second
method.  When attenuating, one attenuates the contributions from the
individual black holes.

As mentioned above, the resulting spacetime will in general not
satisfy the constraints.  One has to use the combined metric as
background metric and initial guess to explicitly solve the constraint
equations (see chapter \ref{system}).  I usually use the conformal
factor $\psi$ resulting from the superposition and vector potential
$V_i=0$ as boundary conditions for this.  It is also possible to use a
Robin boundary condition
%% with $\lim_{r\to\infty} \psi(r) = 1 + C/r$ for that.
for $\psi$.  In order to verify that this procedure does indeed lead
to black holes, one has to locate the apparent horizons (see section
\ref{ahfinding}).

% LocalWords:  eschnett Exp ij Bondi Painlev Gullstrand schild iEF eqn Mazner
% LocalWords:  Moreno

% -*-LaTeX-*-
% $Header: /home/eschnett/cvs/diss/code.tex,v 1.16 2003/01/10 16:39:49 eschnett Exp $

\chapter{Code}
\label{code}

The computer code that is used by us numerical physicists is of utmost
importance to us.  It is our experimental setup, the testbed for our
ideas, the centre about which our group assembles in the morning, and
the equipment that we check upon even on weekends.  Small codes might
be written on rainy afternoons or on lazy weekends; they come a dime a
dozen and are quickly forgotten.  The large codes take several people
and several months or years to assemble and fine-tune, and
correspondingly one has to put a lot of effort into keeping a code
maintainable and understandable to others.

Yet a code is never, at least not to us numerical physicists, an end
in itself.  It is a tool; it is an important one, but yet only a tool.

\section{The Tiger code}
\label{tiger}

I created and used the \emph{Tiger code}\footnote{TGR, pronounced
``Tiger'': the T\"ubingen General Relativity Code.  (Codes apparently
have to have names.)  The pronunciation ``Tiger'' was suggested by
Gabrielle Allen.} to test the ideas presented in the previous
chapters.  It descended from the Maya code \cite{maya}, written in
2000 at Penn State under Pablo Laguna, which in turn descended from
the Agave code \cite{agave}, which in turn was one of the results of
the Binary Black Hole Coalescence Grand Challenge \cite{bbhgc}.

The Tiger code is written in the Cactus framework
\cite{cactus-grid, cactus-tools, cactus-webpages}, which relieves
the programmer of many merely soft\-ware-en\-gi\-nee\-ring related
problems, and which is also supposed to make easier the sharing of
code modules.  At the time of this writing, the sharing still has to
happen.  The main obstacles seem to be rather mundane things such as
e.g.\ differing variable names.  A promising standardisation effort
for ADM-like evolution codes was started in the spring of 2002 by the
home institution of Cactus, the Albert--Einstein--Institut in Golm
near Potsdam.

When writing a numerical code for a complicated system of equations,
there is always the issue of how to translate the equations into code.
One can either create the code semi-automatically with a symbolic
algebra package such as Maple \cite{maple}, or one can code the
equations by hand.  Both approaches have their advantages.  I chose to
code the equations by hand.  In doing that, I strove for clarity,
relying on the compiler to produce efficient code.  Appendix
\ref{coding-equations} shows some example code.

The Tiger code has the standard structure found in time evolution
codes.  (This structure is also mandated by Cactus.)  It consists of
five large components: initial data, constraint solving, time
evolution, boundary conditions, and analysis routines.  Each of these
components is described below.

\section{Initial data}

In order to facilitate exchanging initial data routines and initial
data themselves, it is customary in numerical general relativity to
create initial data in the ADM variables (see chapter \ref{initial}),
i.e.\ for the three-metric $\gamma_{ij}$ and the extrinsic curvature
$K_{ij}$.  While these are sufficient as initial data, it is also
often customary to additionally provide the initial lapse $\alpha$ and
shift $\beta^i$.  With lapse and shift included, the whole four-metric
$g_{\mu\nu}$ can be reconstructed on the initial time slice.

The Tiger code contains initial data routines for several analytic
solutions and background data.  Among these are the Minkowski
spacetime (flat space, see section \ref{minkowski}), weak Bondi waves
(linear planar gravitational waves, see section
\ref{weak-bondi-wave}), the background data for Brill waves (see
section \ref{brill-wave}), and black holes in Kerr--Schild coordinates
(see section \ref{kerr-schild}), Pain\-le\-v\'e--Gull\-strand
coordinates (see section \ref{painleve-gullstrand}), and harmonic
coordinates (see section \ref{harmonic}), as well as
Brill--Lind\-quist black holes (see section \ref{brill-lindquist}).
The Kerr--Schild data can have generic coordinate transformations
applied (see section \ref{coordinate-transformation}), and they can
also be used as background data to superpose black holes (see section
\ref{superposed-kerr-schild}).

The background data mentioned above do not satisfy the constraints.
They can be used as background and as initial guess to solve the
constraint equations numerically (see below), which is necessary to
obtain a solution to Einstein's equations from them.

Sometimes it is convenient to create initial data e.g.\ in spherical
or cylindrical coordinates, and then transform these into Cartesian
coordinates.  For this, the initial data are still calculated at the
grid points forming the Cartesian grid, but the tensors have their
components in another coordinate system.  A coordinate transformation
into Cartesian tensor components is then a linear transformation
according to the usual $T^i = (\partial x^i / \partial x^j) T^j$.

After creating initial data in the ADM variables, these are converted
to the TGR variables (see chapter \ref{system}), which form the
primary variables in the Tiger code.

\section{Constraint solvers}

After creating initial data in the TGR variables, and after every time
step, these data are used as background and as initial guess to
enforce the gauge and the constraints, and to calculate the lapse and
shift.  This is the part that distinguishes the Tiger code from other
codes that evolve unconstrained and without fixing the gauge.

\label{using-petsc}
Enforcing the gauge condition for $F_i$, enforcing the constraints,
and calculating lapse and shift requires solving nonlinear coupled
elliptic equations.  This is currently done using the thorn
TATelliptic in Cactus.  This thorn is an interface to generic
nonlinear elliptic solvers.  The only currently available reasonable
solvers are TATPETSc, which is an interface to the PETSc library
\cite{petsc-home-page, petsc-efficient, petsc-manual}, and TATMG,
which is a full approximation storage multigrid solver
\cite{wesseling} that is currently being written by me.

Enforcing the gauge condition $F_i$ yields the traceless conformal
metric $\bar h_{ij}$.  In order to calculate $\tilde \gamma_{ij}$ from
this, one has to choose $\trace \tilde \gamma_{ij}$ such that $\det
\tilde \gamma_{ij} = 1$.  This is a nonlinear equation that does not
involve derivatives.  I solve it with the \texttt{zriddr} routine of
Numerical Recipes \cite{nr}.

\section{Time integration}

The evolution equations are the equations for the time derivatives of
the primary variables.  In the Tiger code, these are $\psi$, $\tilde
\gamma_{ij}$, and $\tilde A_{ij}$ (see chapter \ref{system}).
Evaluating these equations is straightforward but tedious (see
appendix \ref{ctadm-time-evol}, eqns.\ (\ref{evol-conffact}),
(\ref{evol-confmetric}), and (\ref{evol-confextcurv})).  One usually
introduces the Christoffel symbols $\Gamma^i_{jk}$ and/or $\tilde
\Gamma^i_{jk}$ and the Ricci tensor $R_{ij}$ as explicit intermediate
quantities.

The time evolution equation for $\psi$ is not strictly necessary
during time evolution, but is there nonetheless for analysis purposes.
The Tiger code contains additionally the time evolution equations for
the gauge variables $K$ and $F_i$, which are needed to determine the
lapse $\alpha$ and the shift $\beta^i$ (see appendix
\ref{ctadm-time-evol}).

The time evolution equations for the Hamiltonian and momentum
constraints are not implemented.  One could theoretically use them to
verify the Bianchi identities at run time, and thus gather an
additional measure for discretisation errors.

%% Some people prefer to derive the evolution equations using a symbolic
%% algebra package (such as Maple or Mathematica) and have the results
%% converted to C or Fortran code automatically.  These packages usually
%% do a bad job at introducing suitable intermediate quantities, and
%% hence the resulting code is very large and very unreadable, and often
%% exceeds internal limits in compilers (or the symbolic algebra packages
%% themselves).
%% 
%% In the Tiger code, the evolution equations are directly written in
%% Fortran code.  This leads to code that is readable, but might be
%% wrong.

The time integrator proper uses the iterative Crank--Nicholson scheme
\cite{icn} (see appendix \ref{icn}).  This is an explicit second-order
scheme that can be used to introduce some diffusion in order to obtain
a stable discretisation for advection terms.  As the name suggests,
this diffusion is introduced via additional iterations.  I often run
with just a single iteration, so that the scheme is identical to the
midpoint rule (see appendix \ref{midpoint}) and does not add any
artificial diffusion.  Instead, I add explicit artificial diffusion
terms to the time evolution equations (see appendix \ref{artvisc}).

\section{Boundary conditions}

Boundary conditions are necessary for enforcing the constraints and
gauge conditions, and for the time evolution.

\subsection{Outer boundary}

For the constraint, gauge condition, and coordinate condition solvers,
there are Dirichlet and Robin boundary conditions available on the
outer boundary.  Dirichlet boundaries are kept constant while solving,
but they do not have to be constant in space.  Robin boundary
conditions enforce a certain falloff towards infinity (see section
\ref{robin-boundary}).  They are readily availabe in Cactus
\cite{cactus-webpages}.

For the time evolution, there are Dirichlet and radiative boundary
conditions available.  Dirichlet boundary conditions are kept constant
in time, but they do not have to be constant in space.  Radiative
boundary conditions assume that the evolved quantity is an outgoing
spherical radial wave (see section \ref{radiative-boundary}).  They,
too, are availabe in Cactus.

Additionally, certain symmetry conditions can be enforced at the
boundaries, e.g.\ to restrict the simulation domain to a quadrant or
an octant of space, or to enforce periodicity.

For the robust stability tests (see section
\ref{robust-stability-test}), noise can be introduced at the boundary.

\subsection{Excision boundary}

For the constraint, gauge condition, and coordinate condition solvers,
there are only Dirichlet boundary conditions available on the excision
boundary.  These Dirichlet boundaries are kept constant while solving,
but may vary in space.

For the time evolution, there are Dirichlet and extrapolation boundary
conditions available on the excision boundary.  Dirichlet boundaries
are kept constant in time, but may vary in space.  The extrapolation
boundaries apply an $n$-th order polynomial extrapolation ($n \in [0
\ldots 3]$) towards the excision boundary.  They are applied along the
normals of the excision boundary.  The high order of extrapolation is
necessary because it is applied close to the singularity, where the
shape of the extrapolated functions is not anywhere near flat.  It is
quite possible that a non-polynomial extrapolation (e.g.\ in terms of
$1/r$) would perform better.

%% It is important to select consistent boundary conditions, and to not
%% let some part of the boundary condition unspecified.  \todo{hyperbolic
%% would solve this.}  \todo{put this someplace else} \todo{(ref)}
%% Boundaries for elliptic equations should usually be Dirichlet, while
%% boundaries for the evolution of $\tilde \gamma_{ij}$ and $\tilde
%% A_{ij}$ should be extrapolated.  The boundary conditions also interact
%% with enforcing the gauge and constraints; the extrapolated values for
%% $\tilde \gamma_{ij}$ and $\tilde A_{ij}$ will in general not satisfy
%% the constraints and gauge conditions any more.  This is explained in
%% more detail in chapter \ref{boundary}.

\section{Analysis routines}

After every time step, the ADM variables are calculated from the TGR
variables.  The ADM variables by themselves are important analysis
quantities, because they allow comparisons to analytic solutions, or
to results from other codes.  However, because the coordinate system
that is used to represent the ADM variables is generally not known,
the three-metric and the extrinsic curvature alone do not provide much
insight into the result of a simulation.  While they are, together
with the lapse and shift, sufficient to extract the four-metric and
therefore theoretically all information about the spacetime, these
quantities are basically impossible to interpret when visualised on
their own.

%% \footnote{There is no known good way to visualise a second-rank
%% tensor.  Colleagues in the SFB 382 are currently experimenting.  One
%% idea is to use ellipsoids for that. \todo{not true, this is an old
%% idea.}}

%% \todo{coordinate transformations?}

%% One has to analyse a spacetime in order to make sense of it.  And of
%% course it is convenient to be able to analyse a simulation while it is
%% still running.  This can take much tension out of the already
%% difficult juggling that is necessary to coerce today's high
%% performance computers into doing physics.

The Tiger code provides, as analysis quantities, also all the
intermediate quantities that are calculated during the time evolution.
These are the three-Ricci tensor, the time derivatives of the TGR
variables, the constraints, the gauge variables and their time
derivatives, and the gauge violations.  It also provides the time
derivatives of the ADM variables, and the constraints as calculated
from the ADM variables.  Additionally it calculates the four-Weyl
tensor, which can be used to extract information about gravitational
radiation from a simulation, although I did not attempt this in my
test runs.

\label{ahfinding}
Apparent horizons \cite[chapter 4]{thornburg-thesis},
\cite{ah-finding, shibata1997, shibata2000, huq-finder, flow-finder,
ah} are the only locally (in time) detectable structures that are
present in vacuum spacetimes.  Event horizons are global structures
and cannot be detected during a time evolution.  The Tiger code can
locate and track multiple apparent horizons.  It reports the area and
irreducible mass of the horizon.  If the horizon is isolated
\cite{ih}, it calculates also the spin and total mass.

% LocalWords:  eschnett Exp ij jk Mathematica SFB

% -*-LaTeX-*-
% $Header: /home/eschnett/cvs/diss/results.tex,v 1.20 2003/01/10 16:39:49 eschnett Exp $

\chapter{Results}
\label{results}

In this chapter I present results from example runs with the Tiger
code, using the formalism laid out in the previous chapters.  I begin
with some convergence tests demonstrating that the Tiger code is
second-order convergent apart from artificial diffusion, and also
showing the typical magnitude of the discretisation errors.  A section
about runs with added noise then supports my claim that the formalism
and its implementation are robustly stable in the sense of Szil\'agyi
et al.\ \cite{robust-stability, robust-stability-2}.  Two sections on
Brill waves and Kerr--Schild black holes show that the code works well
in the presence of strong field dynamics and black holes.

%% The last section shows some runs for spacetimes containing two black
%% holes.  I do not claim, however, that the runs I present there have a
%% large enough domain, enough accuracy, reasonable initial data, or
%% suitable boundary conditions to make any physical prediction about the
%% fate of two black holes.

\section{Convergence tests}

\subsection{Static and stationary tests: black holes}
\label{convtest-static}

Here I present a series of convergence tests for highly nonlinear but
static or stationary spacetimes.  By calculating the time derivatives
of several quantities, essentially the whole right hand side
evaluation subsystem of the code is tested.  The tests here contain no
time evolution (but all the following tests do).  All time derivatives
should converge to zero to second order as the numerical resolution is
increased.

As test cases I use black holes in several coordinates, namely
Kerr--Schild (see section \ref{kerr-schild}), Painlev\'e--Gullstrand
(see section \ref{painleve-gullstrand}), and harmonic coordinates (see
section \ref{harmonic}).  In all cases, the mass of the black hole is
$M=1$.  For Kerr--Schild coordinates, the spin is $a=0.9$, and for the
other two cases, the spin is zero.  These are stationary or static
solutions, meaning that the analytic values of the time derivatives
are all zero.  The time derivatives of certain primary quantities are
shown, as well as some constraints.  All shown quantities should be
zero.  The absolute values of the numerical values along the $x$ axis
are plotted, where the results from the finer resolutions have been
scaled by factors of 4 and 16, respectively.  The dips in the graphs
indicate places where the numerical error vanishes because it changes
sign.

\begin{figure}
\begin{tabular}{rr}
\includegraphics[width=0.45\textwidth]{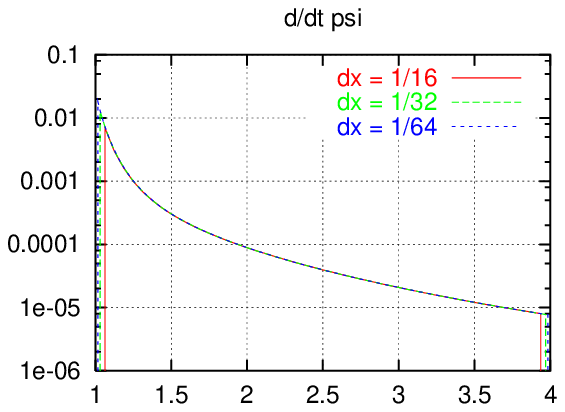} &
\includegraphics[width=0.45\textwidth]{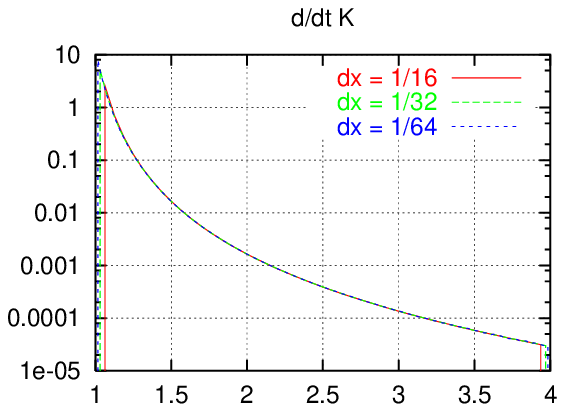} \\
\includegraphics[width=0.45\textwidth]{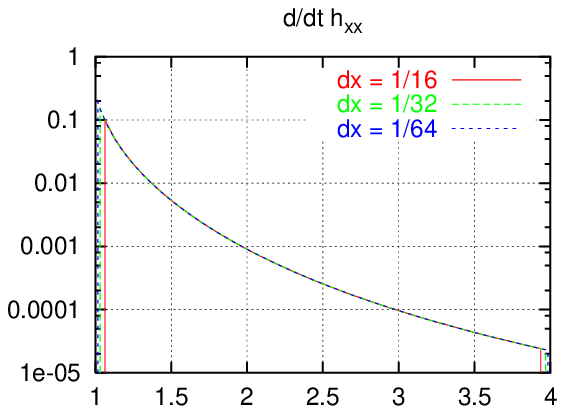} &
\includegraphics[width=0.45\textwidth]{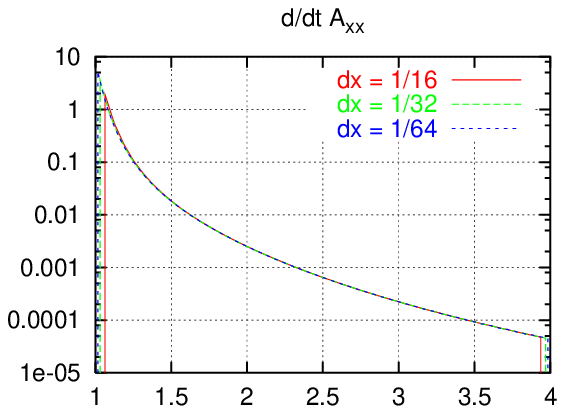} \\
\includegraphics[width=0.45\textwidth]{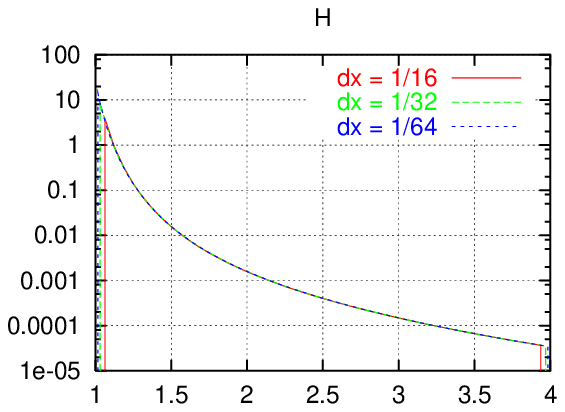} &
\includegraphics[width=0.45\textwidth]{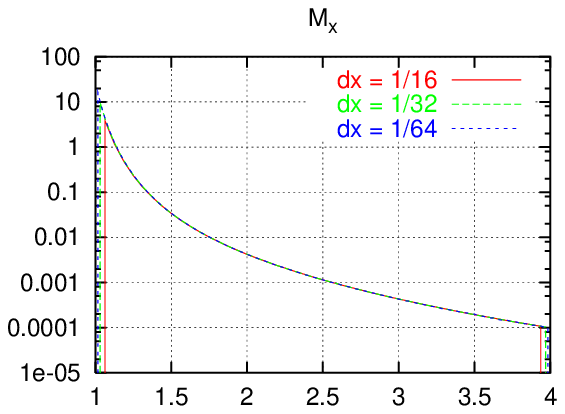}
\end{tabular}
\caption{\label{convtest-kerr-schild}Convergence test with
Kerr--Schild coordinates.  The graphs show the errors in the time
derivatives of various quantities for three resolutions.  The errors
for the finer resolutions have been scaled by the factors 4 and 16,
respectively.  The coinciding graphs show that there is second order
convergence towards zero.}
\end{figure}

\begin{figure}
\begin{tabular}{rr}
\includegraphics[width=0.45\textwidth]{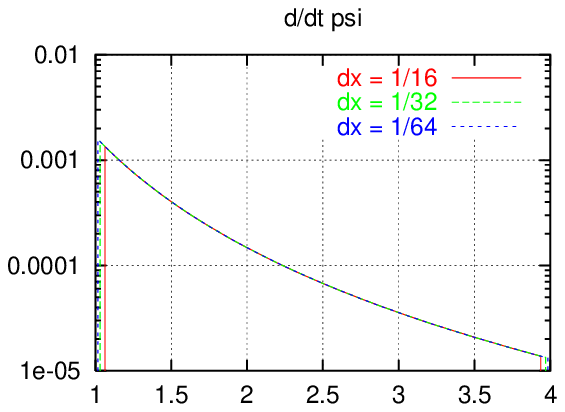} &
\includegraphics[width=0.45\textwidth]{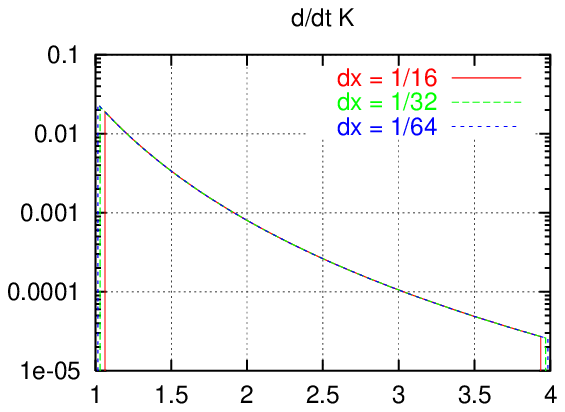} \\
\includegraphics[width=0.45\textwidth]{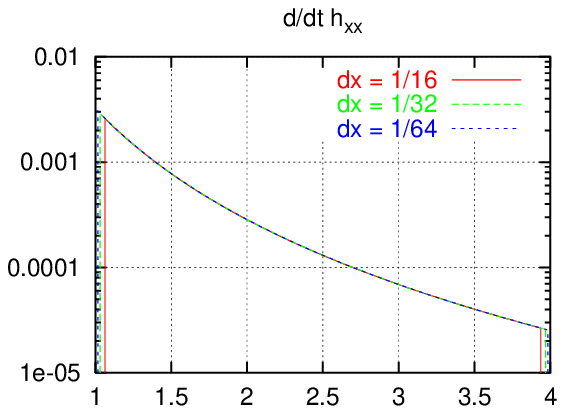} &
\includegraphics[width=0.45\textwidth]{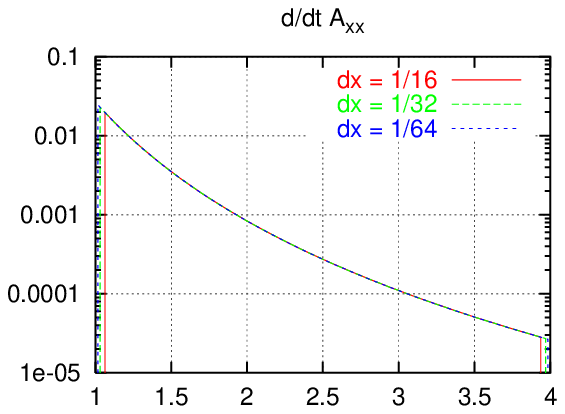} \\
\raisebox{2cm}{($H=0$ everywhere; not shown)} &
\includegraphics[width=0.45\textwidth]{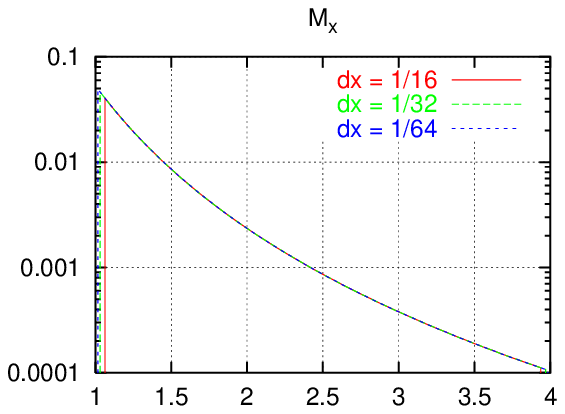}
\end{tabular}
\caption{\label{convtest-painleve-gullstrand}Convergence test with
Painlev\'e--Gullstrand coordinates.  The graphs show the errors in the
time derivatives of various quantities for three resolutions.  The
errors for the finer resolutions have been scaled by the factors 4 and
16, respectively.  The coinciding graphs show that there is second
order convergence towards zero.}
\end{figure}

\begin{figure}
\begin{tabular}{rr}
\includegraphics[width=0.45\textwidth]{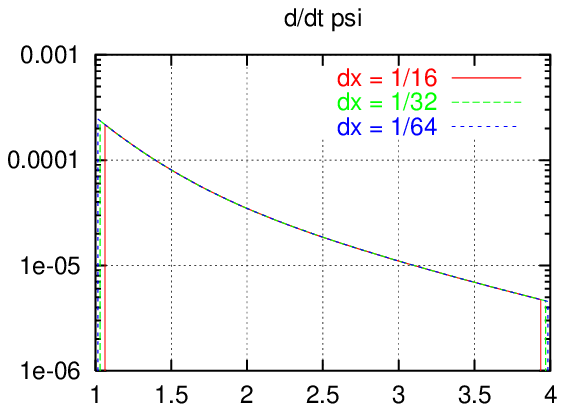} &
\includegraphics[width=0.45\textwidth]{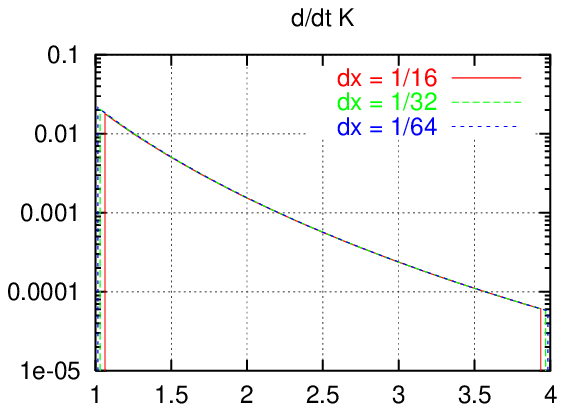} \\
\includegraphics[width=0.45\textwidth]{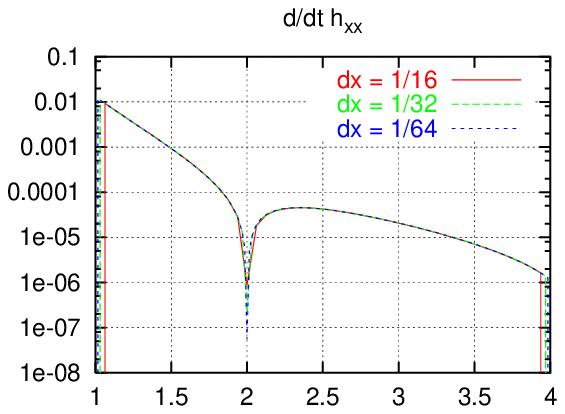} &
\includegraphics[width=0.45\textwidth]{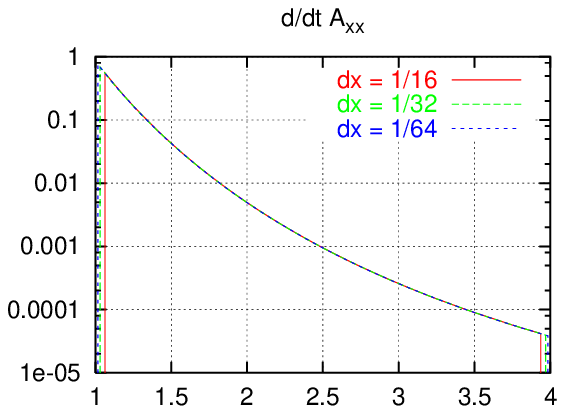} \\
\includegraphics[width=0.45\textwidth]{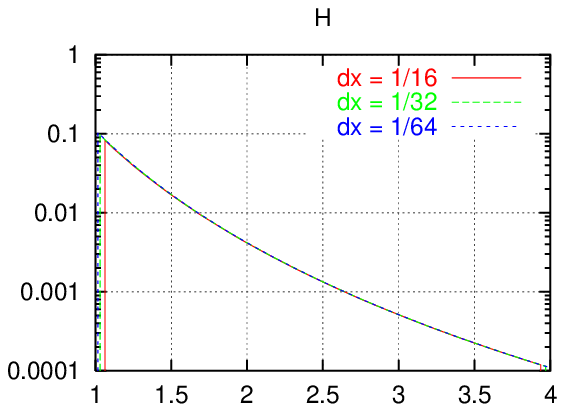} &
\includegraphics[width=0.45\textwidth]{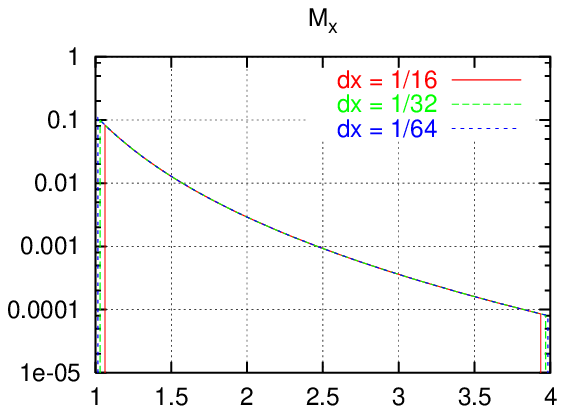}
\end{tabular}
\caption{\label{convtest-harmonic}Convergence test with harmonic
coordinates.  The graphs show the errors in the time derivatives of
various quantities for three resolutions.  The errors for the finer
resolutions have been scaled by the factors 4 and 16, respectively.
The coinciding graphs show that there is second order convergence
towards zero.}
\end{figure}

The fact that the graphs for the different resolutions in figures
\ref{convtest-kerr-schild}, \ref{convtest-painleve-gullstrand}, and
\ref{convtest-harmonic} do overlap indicates second order convergence.
Nevertheless, the errors of some of these quantities are surprisingly
large.  Note that the coarsest resolution that I present here is
$dx=1/16$, which is already considered to be a rather fine resolution
for a typical black hole collision simulation.

The $L_2$ norms of the convergence factors for the resolutions
$dx=1/32$ and $dx=1/64$ and for the three coordinate systems (cs)
Kerr--Schild (KS), Painlev\'e--Gullstrand (PG), and harmonic
coordinates (hc) are given below.  The fact that the factors are so
close to 4 indicates near-perfect second order convergence of the
code:

\smallskip
\begin{tabular}{l|llllll}
   cs & $\partial_t \psi$ & $\partial_t K$ & $\partial_t h_{xx}$ &
   $\partial_t A_{xx}$ & $H$ & $M_x$
\\\hline
   KS & 3.99554 & 4.00827 & 4.00078 & 4.00763 & 3.99803 & 3.99002
\\
   PG & 3.99958 & 4.00089 & 3.99852 & 4.0006 & (n/a) & 3.99987
\\
   hc & 3.99976 & 3.99453 & (4.59489) & 4.00214 & 3.99408 & 3.99923
\end{tabular}
\smallskip

The convergence factor for the Hamiltonian constraint $H$ for
Painlev\'e--Gullstrand coordinates is ill-defined because the metric
is flat, and hence the Hamiltonian constraint is exactly zero
everywhere.  The convergence factor for $\partial_t h_{xx}$ for
harmonic coordinates is larger than 4 because the discretisation error
changes sign at $x=2$ and hence the convergence factor is ill-defined
there.  Excluding this grid point, the $L_2$ norm of this convergence
factor is 4.00058.

%% flush figure floats
\clearpage

\subsection{Dynamic linear test: weak Bondi wave}

As a test of the dynamic behaviour in the linear regime I use weak
Bondi waves (see section \ref{weak-bondi-wave}).  Because a
periodicity boundary condition is not possible when solving the
elliptic constraint and gauge equations, I decided to use standing
waves with Dirichlet boundary conditions instead.  It is also possible
to use Dirichlet boundaries for a travelling wave, but they would
continuously inject the analytic solution into the simulation domain,
which I want to avoid.  By putting the Dirichlet boundaries at nodes
of the standing wave, the boundary values are constant in time, which
is a stronger test of the code.

I use $K=0$ and $F_i=0$ as gauge conditions.  This is consistent with
the analytic solution.

I use an effectively one-dimensional simulation domain, which
corresponds to having a translational symmetry in two directions.  My
simulation domain extends only into the $z$ direction with $z \in
[-0.5; +0.5]$.  I use a wave length $L=1$ and an amplitude of
$A=10^{-6}$.  I choose a resolution of $dx=1/100$ and add artificial
diffusion with a coefficient $C_{\mathrm{SM}}=1/4$ (see appendix
\ref{artvisc}).  I use the midpoint rule (see appendix \ref{midpoint})
for time integration.

Figure \ref{wbw-run} shows the result of this simulation.  The phase
of the wave stays correct, which is to be expected for a standing
wave.  The amplitude decreases with time, which is also to be expected
due to numerical and artificial diffusion.  The quantities $g_{zz}$,
$F_z$, and the constraints $H$ and $M_z$ are all correct up to
floating point accuracy.

\begin{figure}
\begin{tabular}{rr}
\includegraphics[width=0.45\textwidth]{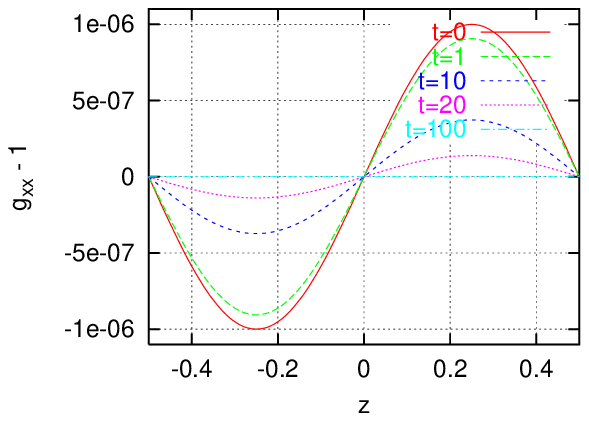}&
\includegraphics[width=0.45\textwidth]{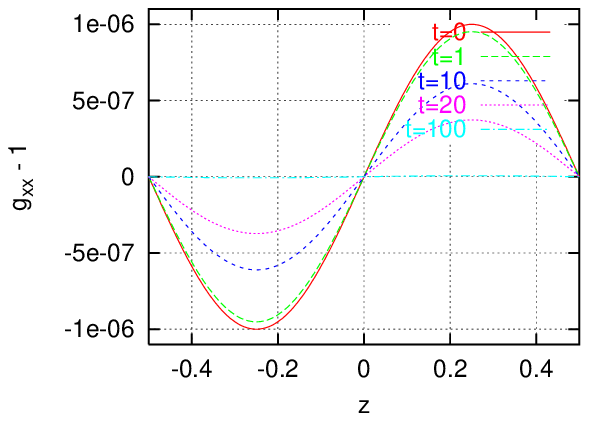}\\
\includegraphics[width=0.45\textwidth]{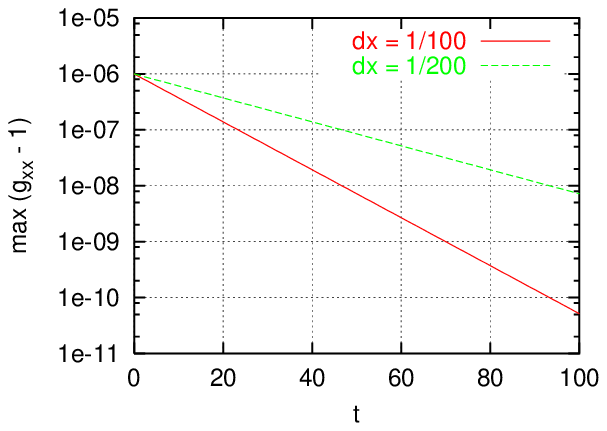}
\end{tabular}
\caption{\label{wbw-run}Performance test with a weak Bondi wave for
the resolutions $dx=1/100$ (top left) and $dx=1/200$ (top right).
Shown are the wave forms at certain times during the evolution.  The
bottom graph compares the amplitude decay for the two resolutions.}
\end{figure}

The amplitude decay is exponential in time.  It can be empirically
described by the expression $A(t) = A(0)\, e^{-\tau\, t}$ with a
resolution and wave length dependent decay rate $\tau$.  The table
below shows decay rates for several example resolutions and wave
lengths:

\begin{center}
\begin{tabular}{ll|lll}
$dx$	& $L$	& $A(t=1) / A(t=0)$	& $\tau$		& $\tau/\tau_0$	\\\hline
$1/100$	& $1$	& 0.906			& 0.0987 ($=:\tau_0$)	& 1		\\
$1/200$	& $1$	& 0.952			& 0.0493		& 0.500		\\
$1/100$	& $1/2$	& 0.673			& 0.3964		& 4.017		\\
$1/100$	& $1/4$	& 0.207			& 1.5764		& 15.98		
\end{tabular}
\end{center}

This table indicates that the decay rate $\tau$ is proportional to the
resolution $dx$ and inversely proportional to the square of the wave
length $L$.  

This rate can be explained by the kind of artifical diffusion that I
use: Given initial data with $A \ll 1$ as described in section
\ref{weak-bondi-wave}, gauge conditions as described above, and
including artificial viscosity with $C_{\mathrm{SM}} \ne 0$ as
described in section \ref{artvisc}, the set of equations governing the
time evolution of $g_{xx}$ is
\begin{eqnarray}
   \partial_t g_{xx} & = & -2\, K_{xx} + C_{\mathrm{SM}}\, dx\,
   \partial_{zz} g_{xx}
\\
   \partial_t K_{xx} & = & -\frac{1}{2}\, \partial_{zz} g_{xx} +
   C_{\mathrm{SM}}\, dx\, \partial_{zz} K_{xx} + O(A^2) \quad
   \textrm{.}
\end{eqnarray}
Neglecting the terms in $O(A^2)$, these equations can be combined into
\begin{eqnarray}
   \partial_{tt} g_{xx} & = & \partial_{zz} g_{xx} + 2\,
   C_{\mathrm{SM}}\, dx\, \partial_{tzz} g_{xx} + O(dx^2)
   \quad \textrm{.}
\end{eqnarray}
Neglecting again the terms in $O(dx^2)$, the ansatz
\begin{eqnarray}
   g_{xx}(t,z) & = & \exp(- k^2\, C_{\mathrm{SM}}\, dx\, t) \sin(kt)
   \cos(kz) \quad \textrm{,}
\end{eqnarray}
which is motivated by the form of the initial data, solves this
equation.  That means that the decay rate $\tau$ caused by the
artificial diffusion should be
\begin{eqnarray}
   \tau & = & \frac{C_{\mathrm{SM}}\; dx}{L^2}
\end{eqnarray}
with $L=1/k$, and this is indeed the case in the table above.  It is
therefore reasonable to assume that the amplitude decay is due to the
artificial diffusion.

%% Analytic solution:\\
%% $f = g_{xx}$, $g = -2K_{xx}$, $D = C_{\mathrm{SM}} h$, assume $k\; dx \ll 1$\\
%% $\partial_t f = g + D f_{,zz}$, $\partial_t g = f_{,zz} + Dg_{,zz}$\\
%% $\partial_{tt} f = f_{,xx} + 2 D f_{,txx} + O(D^2)$\\
%% $f(t,x) = \exp(-k^2Dt) \sin(kt) \cos(kx)$\\
%% $\tau = k^2 C_{\mathrm{SM}} dx$
%% 
%% Example: $dx=1/100$, $k=1$: $e^{-\tau} = 0.9003351238722$.\\
%% Experimental value is $1.006336058626$ times that.
%% 
%% $k=1$, $dx=200$: $0.9488599074006$.\\
%% $k=2$, $dx=100$: $0.6570777671638$.\\
%% $k=4$, $dx=100$: $0.1864090918806$.

Although an amplitude loss of 10\% per crossing time seems large, it
is actually acceptable for a binary black hole collision simulation.
Scaling the resolution to typical values for such a run, I arrive at a
simulation domain with $x^i \in [0; 20]$, a resolution of $dx=1/5$,
and a wave length of $L=20$.  This wave length is close to the
quasi-normal mode of a black hole with a mass of $M=2$, the result of
a collision of two $M=1$ black holes.  In this case, it takes a time
$T=20$ for the numerical amplitude to decrease to 90\% of its real,
physical value.  This is about the time scale in which the wave
reaches the outer boundary.  That means that the code is not perfect,
but the performance\footnote{I use the term ``performance'' as
measuring the quality of the numerical result.  This performance
depends on the magnitude of the numerical errors, not on the speed of
an implementation on some hardware.} is acceptable for a realistic
run.

%% flush figure floats
\clearpage

\subsection{Dynamic nonlinear test: gauge pulse}

As a test of the dynamic behaviour in the nonlinear regime I present a
nonlinear gauge pulse (see section \ref{gauge-pulse}).  For the same
reason as in the previous section, a periodicity boundary condition is
not possible.  Again, I therefore use a standing gauge pulse with
Dirichlet boundary conditions that are constant in time.

I use the analytically known (time and space dependent) values of $K$
and $F_i$ as gauge conditions.  This is permissible, because these are
gauge quantities only, and one is free to specify the gauge condition
in advance --- indeed, one has to.  It is certainly possible to use a
different gauge condition, but one then cannot compare the result to
this analytic solution any more.

I use an effectively one-dimensional simulation domain, which
corresponds to a translational symmetry in two directions.  My
simulation domain extends only into the $z$ direction with $z \in
[-0.5; +0.5]$.  I use a wave length $L=1$ and an amplitude of $A=1$.
I choose a resolution of $dx=1/100$ and add artifical diffusion with a
coefficient $C_{\mathrm{SM}}=1/4$ (see appendix \ref{artvisc}).  I use
the midpoint rule (see appendix \ref{midpoint}) for time integration.
Except for the wave amplitude, these parameters are the same as in the
previous section.

\begin{figure}
\begin{tabular}{rr}
\includegraphics[width=0.45\textwidth]{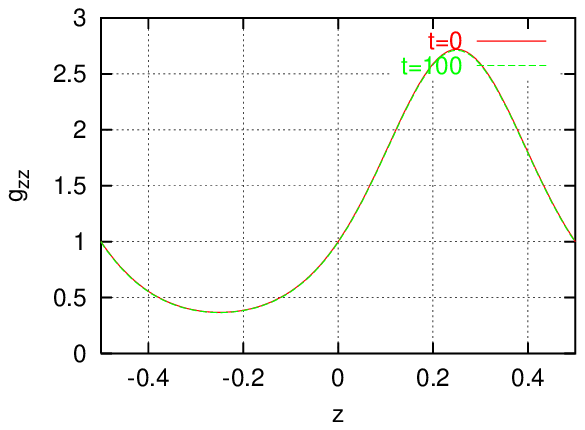}\\
\includegraphics[width=0.45\textwidth]{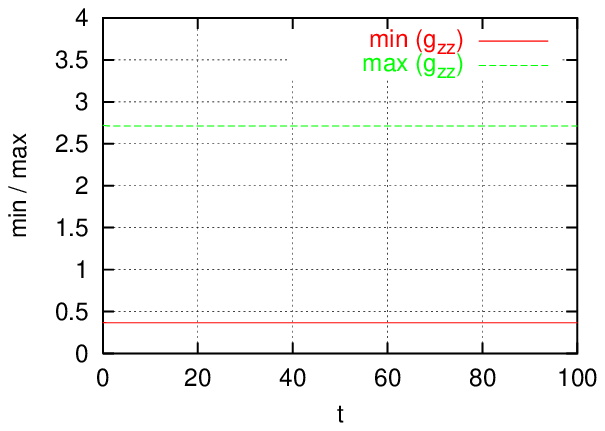}&
\includegraphics[width=0.45\textwidth]{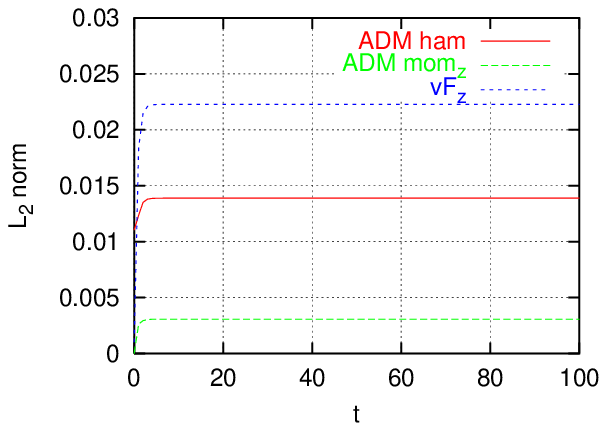}
\end{tabular}
\caption{\label{gp-run}Performance test with a gauge pulse.  The top
graph compares the initial wave form and the wave form after 100
crossing times.  The bottom graphs show the wave amplitude (left) and
constraint and gauge violations (right) vs.\ time.}
\end{figure}

Figure \ref{gp-run} shows the result of this simulation.  The phase
stays correct, which is to be expected for a standing wave.  The
amplitude does not decrease, it stays constant in time.  The
simulation result does not deviate from the analytic solution in any
significant manner.

This test case contains a pure gauge wave, and is therefore a
pathologically well suited\footnote{The term ``pathologically well
suited'' was coined by Ed Seidel in spring 2002 in Mexico.} case for
my formulation.  Because I explicitly enforce the gauge condition at
every time step, the nontrivial part of the evolution for these data
is prescribed.  The gravitational wave degrees of freedom are not
prescribed, but those stay zero all the time.

%% t=0 gzz=2.7208287934834 A=1.000936537382
%% t=1 gzz=2.7139802345828 A=0.9984162786237
%% A_1 / A_0 = 0.9974820993498

%% \todo{Compare to analytic solution?}

%% \todo{Creation of spurious gravitational waves?}

%% flush figure floats
\clearpage

\section{Stability test}
\label{robust-stability-test}

Szil\'agyi et al.\ \cite{robust-stability, robust-stability-2} define
the notion of \emph{robust stability} of a code.  This definition is
not an analytic, but rather an experimental one, which means that it
can be tested rather easily.  The basic idea is to add noise to the
initial data, and also to add noise to the values provided by the
boundary conditions.  A code that behaves gracefully, i.e.\ does not
show exponential growth, is called \emph{robustly stable}.

The robust stability test, as proposed by Szil\'agyi et al., comes in
four stages of increasing difficulty.  Stages I and II include
periodicity, which is not a valid boundary condition for my elliptic
equations.  I therefore concentrate on stage III, which uses a
rectangular domain without periodicity.  Stage IV requires a spherical
outer boundary, which I did not test.

For this test, I use Minkowski flat space (see section
\ref{minkowski}) as background to which the noise is added.  I use
$K=0$ and $F_i=0$ as gauge conditions and Dirichlet boundary
conditions.

I run this test in two configurations with different resolutions
$dx=1/4$ and $dx=1/8$.  I use a cubic box with a length of $L=4$, and
run the code up to $T=400$, which is equivalent to $T/L=100$ crossing
times.  I use a noise amplitude of $A=0.1$ on the initial data and on
the boundaries.  I add an artificial diffusion with a coefficient of
$C_{\mathrm{SM}} = 0.1$ (see appendix \ref{artvisc}).  I use the
midpoint rule (see appendix \ref{midpoint}) for time integration.

\begin{figure}
\begin{tabular}{rr}
\includegraphics[width=0.45\textwidth]{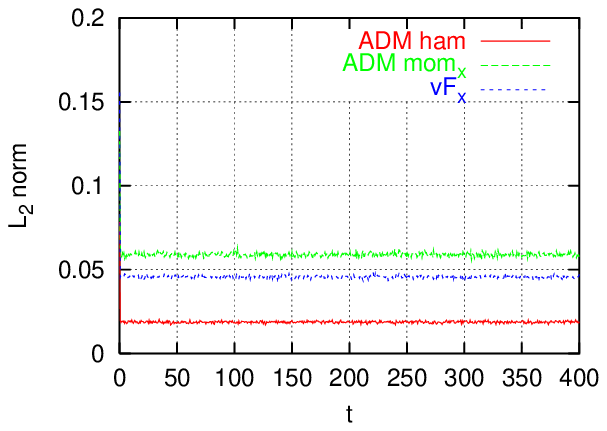} &
\includegraphics[width=0.45\textwidth]{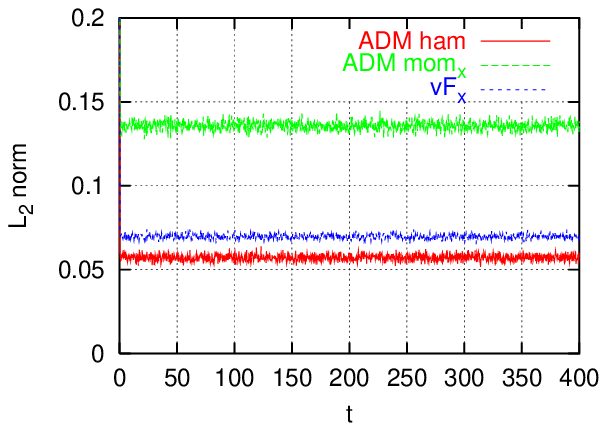}
\end{tabular}
\caption{\label{stability-test}Robust stability test.  Plotted are the
violations of the constraints and of the gauge condition versus time
for the resolutions $dx=1/4$ (left) and $dx=1/8$ (right).}
\end{figure}

The result of this test is that the formulation of Einstein's
equations as implemented in the Tiger code is robustly
stable.\footnote{A longer run time would have been preferable, but was
not possible because of the large computational resource
requirements.}  The constraints and the gauge violations, which are
shown in figure \ref{stability-test}, do not increase with time after
an initial transient.  Similarly, all components of the three-metric
and of the extrinsic curvature stay bounded.

I assume that the artificial diffusion is the reason for the fact that
there seems to be no growth at all.  I assume that the levels at which
the individual metric and extrinsic curvature components remain are
defined by the noise on the boundary, which inserts energy into the
system, and the artificial viscosity, which removes energy.

Note that the finer run is not just a higher-resolution version of the
coarser run.  The two spacetimes are different, in that the finer run
contains noise with a higher spatial frequency.  For that reason the
errors in the fine run cannot be expected to be smaller than those in
the coarse run.  The two runs do not form a convergence test.

%% flush figure floats
\clearpage

\section{Brill wave}

A Brill wave (see section \ref{brill-wave}) tests the dynamic
nonlinear behaviour in a realistic situation.  Although the initial
data that I use are axially symmetric, I perform a full
three-dimensional evolution.  Due to the Cartesian grid in the
simulation domain the setup loses its axial symmetry.

There is no complete analytic solution for the Brill wave initial
data; the conformal factor has to be determined numerically from the
Hamiltonian constraint.  In addition, there is no analytic solution at
all for the evolution of Brill waves at times $t \ne 0$.  It is thus
difficult to gauge the correctness and performance of a code.

The Albert--Einstein--Institut in Golm also has a code, called
\emph{Einstein code}, that solves Einstein's equations.  This code is
publicly available from the Cactus web pages \cite{cactus-webpages}.
I used both their and the Tiger code, and tried to compare the results
for Brill waves.  This is unfortunately very difficult, because the
codes use different gauge conditions during the evolution.  While both
codes can use maximal slicing, i.e.\ $K=0$, they differ in their shift
conditions.  The Tiger code uses a metric gauge condition
$F_i=\mathrm{const}$ or $F_i=0$, and derives a shift condition from
that.  The Einstein code cannot impose a gauge condition on the
metric, and one has to specify the shift condition directly.  For
Brill waves one usually uses normal coordinates, i.e.\ $\beta^i=0$.
This difference leads to very different time evolutions, although the
spacetimes should (barring implementation errors) be identical up to
the discretisation error.

I simulate Brill waves with initial data as described in section
\ref{brill-wave}.  I use an amplitude $A=4$, which makes the Brill
wave highly nonlinear, but still subcritical, i.e.\ it does not form a
black hole.  I also use $C=0$ and $\omega=1$.  I use maximal slicing,
i.e.\ the gauge condition $K=0$ on the extrinsic curvature.  I use an
iterative Crank--Nicholson time integrator with 2 iterations after the
initial Euler step (see appendix \ref{icn}).

I run the Tiger code with two different gauge conditions.  The first,
which I call here ``$F_i=\mathrm{const}$'', keeps the gauge variable
$F_i$ constant in time at the values from the initial data.  The
second, called ``$F_i=0$'', sets $F_i$ to zero at all times.  Thus
only the case $F_i=\mathrm{const}$ uses the same initial data as the
Einstein code.  The case $F_i=0$ evolves a different (but closely
related) spacetime.  This gauge condition has the advantage that it,
together with $K=0$, enforces a Minkowski metric when the spacetime is
flat (and with suitable boundary conditions).

\begin{figure}
\begin{tabular}{rr}
\includegraphics[width=0.45\textwidth]{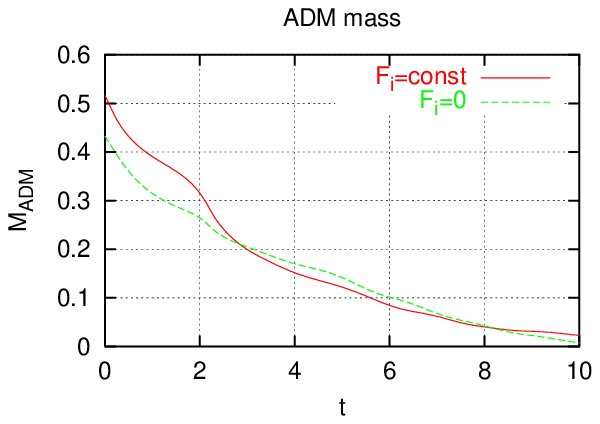}&
\includegraphics[width=0.45\textwidth]{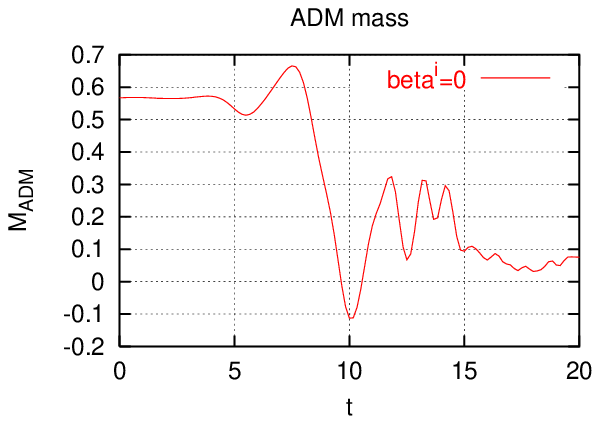}
\end{tabular}
\caption{\label{bw-runs-adm-m-vs-t}Brill wave runs: ADM mass vs.\ time
for different gauge and coordinate conditions.  Note the different
time scales.}
\end{figure}

It turns out that the cases $F_i=\mathrm{const}$ and $F_i=0$ show very
similar behaviour.  This indicates that my way of enforcing the gauge
leaves the physical degrees of freedom mostly unchanged.  This is also
confirmed by the time evolution of the ADM mass, as shown in figure
\ref{bw-runs-adm-m-vs-t}.  Strangely, the behaviour of the Einstein
code using a zero shift differs greatly, as can also be seen in the
following.

Traditionally, one of the most interesting quantities to look at in
maximal slicing is the lapse $\alpha$, or its minimum.  With maximal
slicing, the lapse drops to zero near a singularity; thus the minimum
of the lapse indicates heuristically whether a singularity is forming.
The Brill wave that I look at is subcritical, i.e.\ there is no
singularity in the spacetime.  Consequently, although the lapse drops
initially, it later ``recovers'' and ends up as $\alpha=1$ everywhere
at late times.

\begin{figure}
\begin{tabular}{rr}
\includegraphics[width=0.45\textwidth]{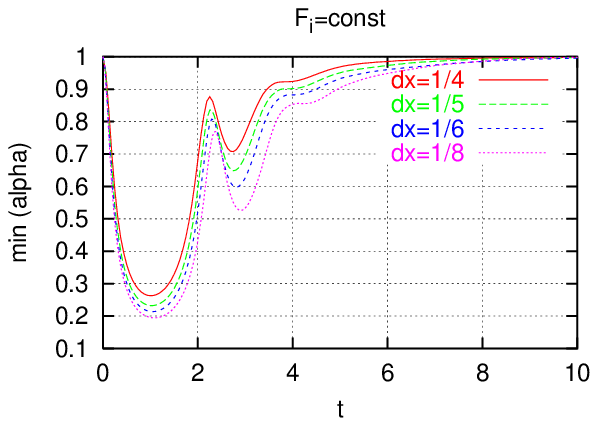}&
\includegraphics[width=0.45\textwidth]{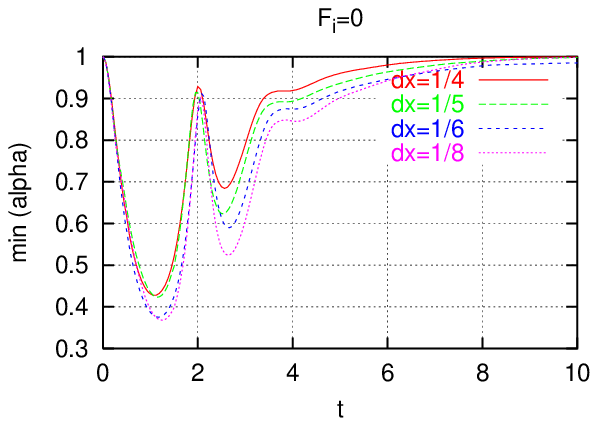}\\
&
\includegraphics[width=0.45\textwidth]{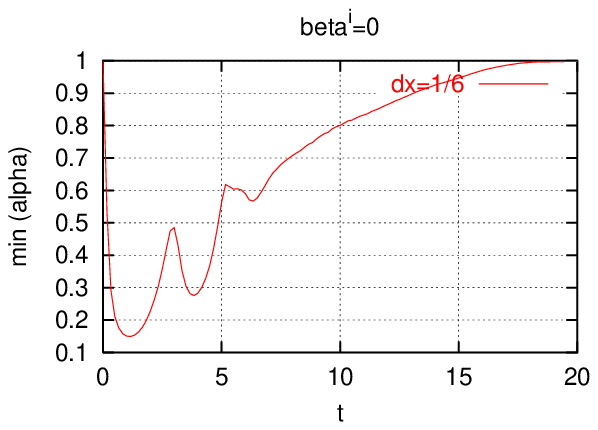}\\
\end{tabular}
\caption{\label{bw-runs-alpha-vs-t}Brill wave runs: Minimum of the
lapse $\alpha$ vs.\ time for different gauge and coordinate
conditions.  (This is always the value of the lapse at the origin.)
All cases use maximal slicing, i.e.\ $K=0$ everywhere.  Note the
different time scales.}
\end{figure}

\begin{figure}
\includegraphics[height=0.45\textheight]{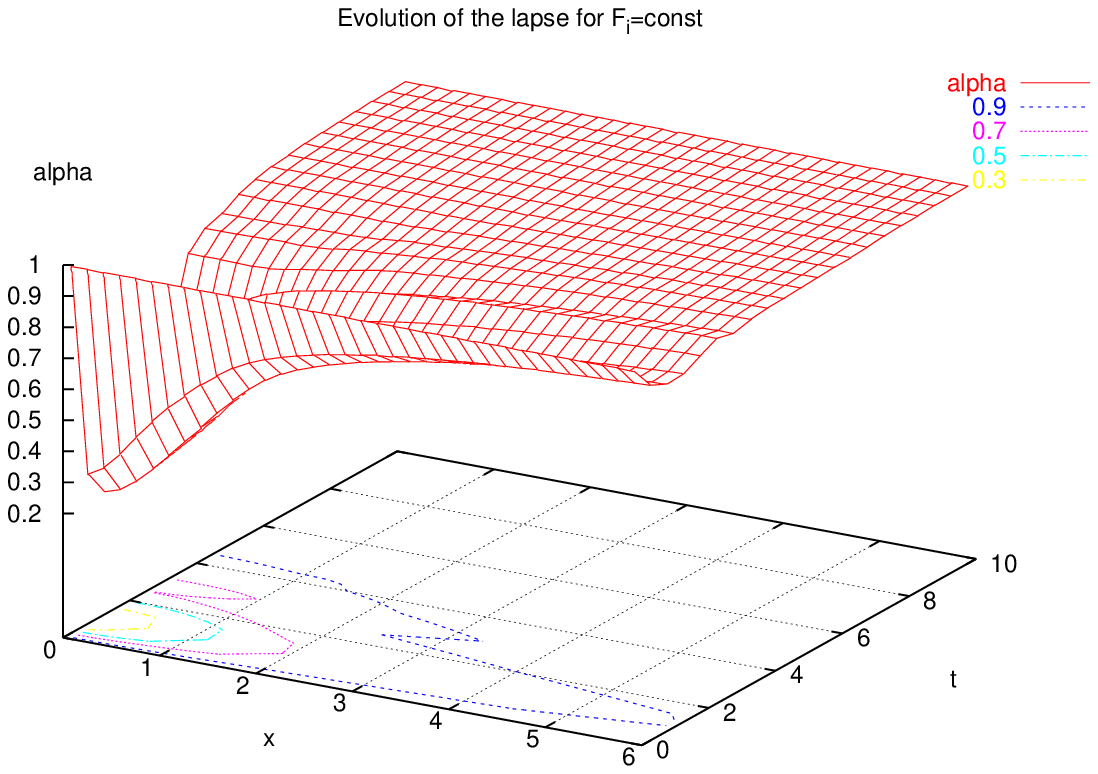}\\
\includegraphics[height=0.45\textheight]{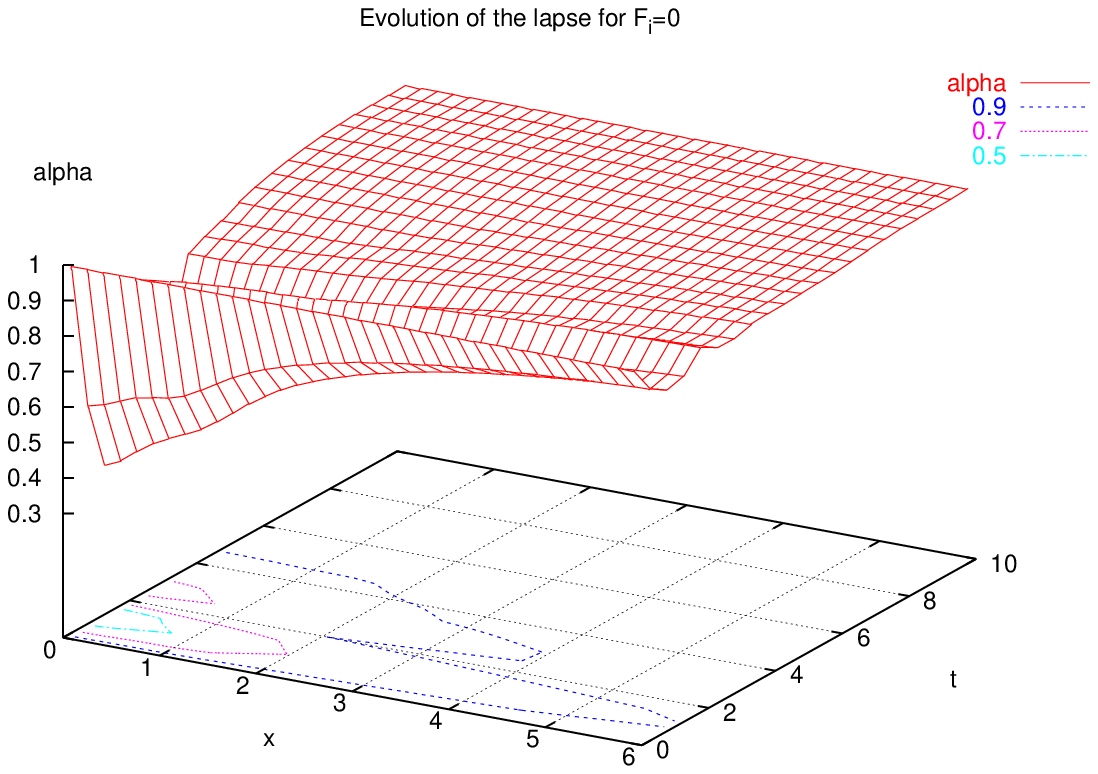}
\caption{\label{bw-runs-alpha-vs-x-t}Brill wave runs: Lapse $\alpha$
vs.\ radius (along the $x$ axis) and time for different gauge
conditions.  Note that also $K=0$ everywhere.}
\end{figure}

\begin{figure}
\includegraphics[height=0.45\textheight]{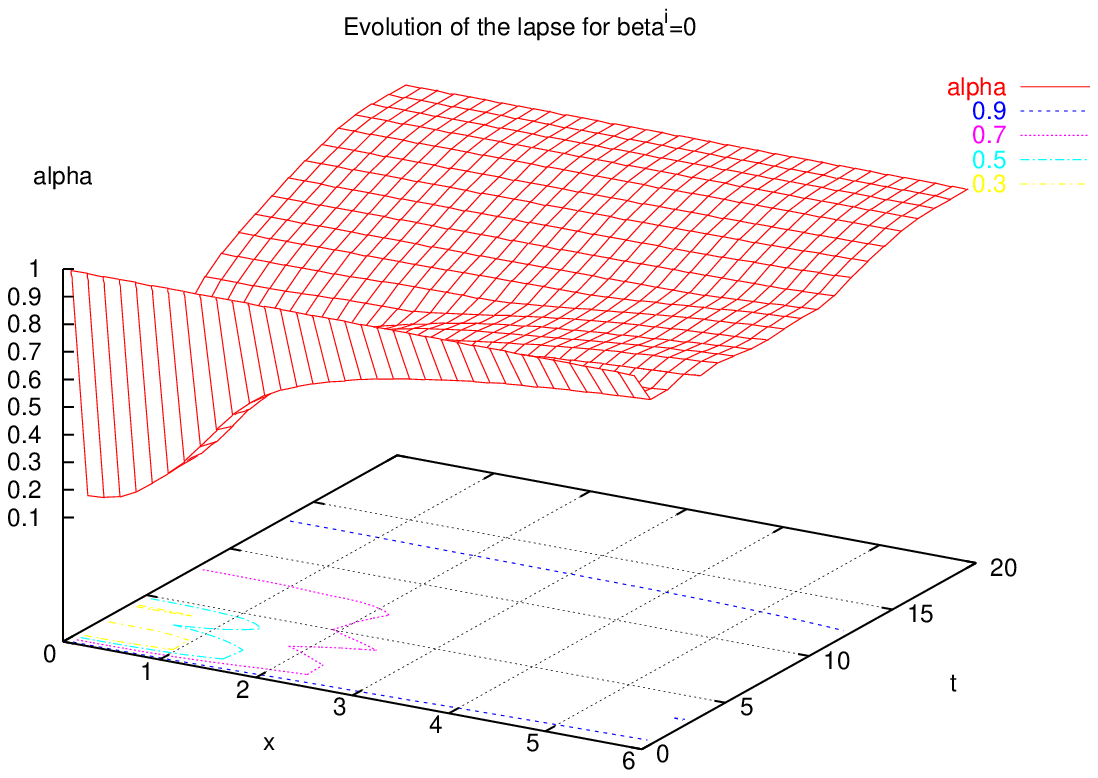}
\caption{\label{bw-runs-alpha-vs-x-t-2}Brill wave runs: Lapse $\alpha$
vs.\ radius (along the $x$ axis) and time for zero shift.  Note that
also $K=0$ everywhere.}
\end{figure}

These time evolutions of the minimum of the lapse are shown in figure
\ref{bw-runs-alpha-vs-t} for  four resolutions from $dx=1/4$ to
$dx=1/8$, run with the Tiger code.  Finer resolutions would have been
desirable, but would have required much longer run times.  (The PETSc
elliptic solver (see section \ref{using-petsc}) that I used does not
scale well on multiple processors.)  As the resolutions are, they form
only a weak convergence test.  The same figure also shows a run with
the Einstein code with a resolution of $dx=1/6$.  Figures
\ref{bw-runs-alpha-vs-x-t} and \ref{bw-runs-alpha-vs-x-t-2} show the
time evolution of the lapse along the $x$ axis.

All three evolutions show qualitatively the same features: the lapse
first drops sharply down from $\alpha=1$, bounces back up two times,
and then drifts back towards $\alpha=1$.  At late times, the Brill
wave has radiated away, and the spacetime becomes flat.  The largest
and maybe most puzzling difference is that it takes about twice as
much coordinate time until the lapse has recovered for the zero shift
coordinate condition than for the two other gauge conditions.

\begin{figure}
\includegraphics[height=0.45\textheight]{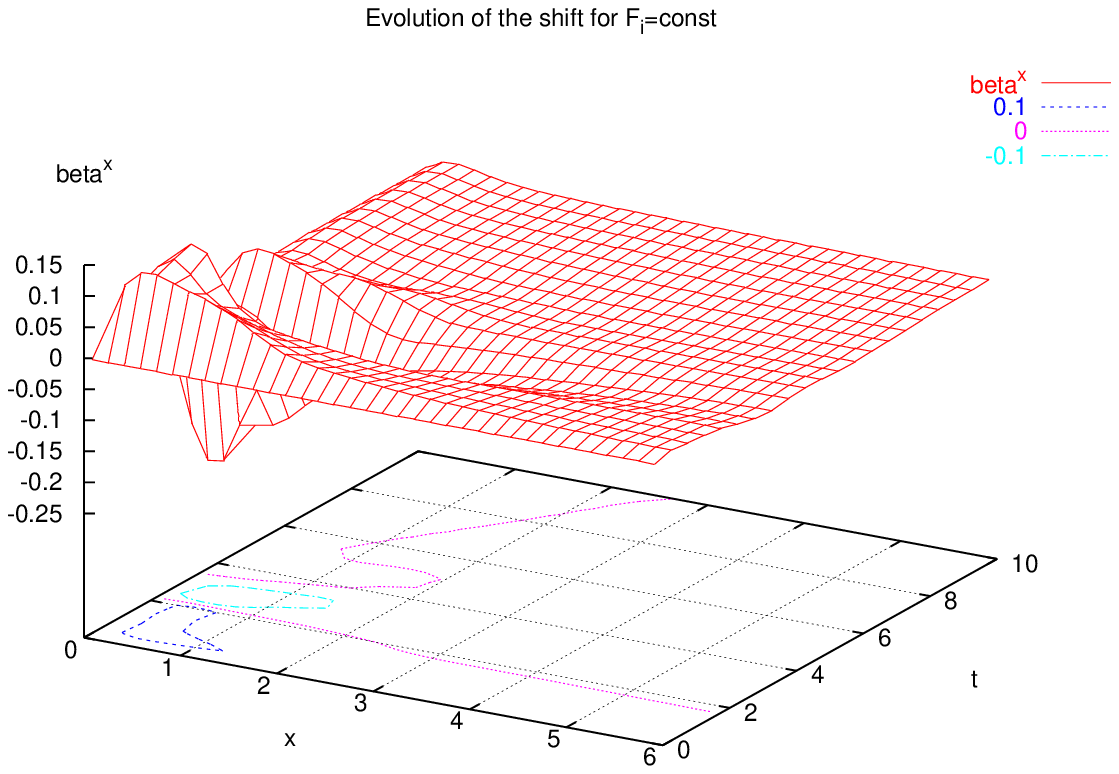}\\
\includegraphics[height=0.45\textheight]{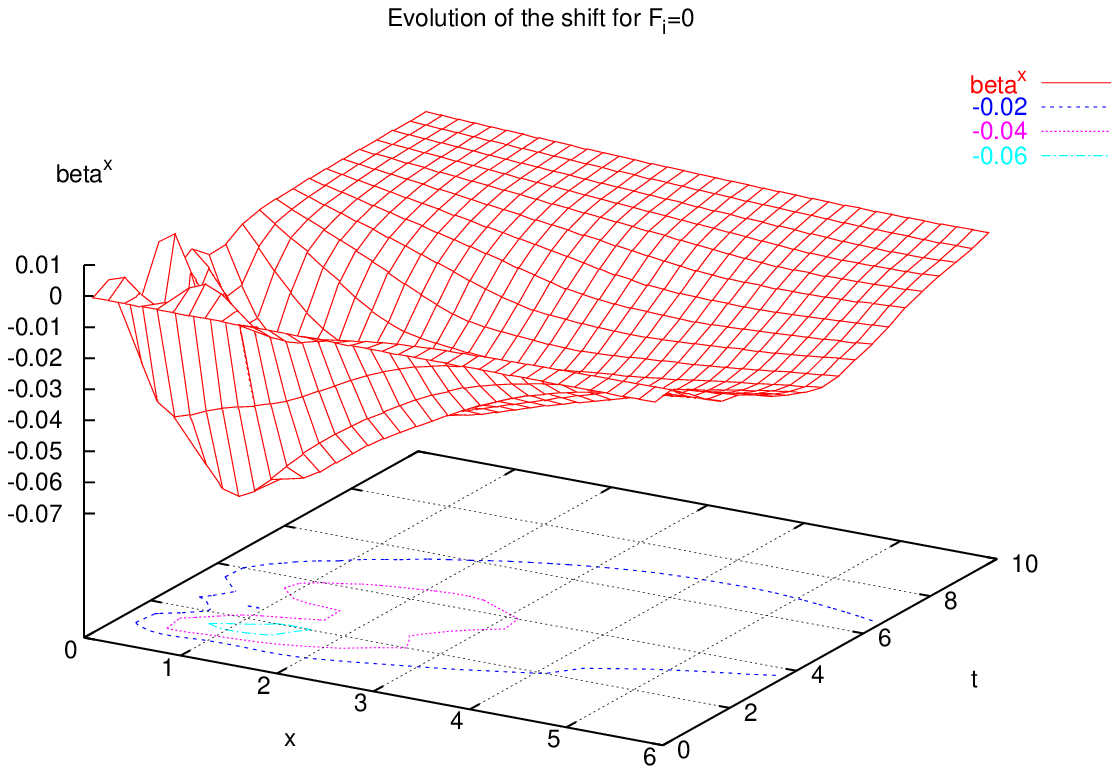}
\caption{\label{bw-runs-betax-vs-x-t}Brill wave runs: Shift component
$\beta^x$ vs.\ radius (along the $x$ axis) and time for different
gauge conditions.  Note that also $K=0$ everywhere.}
\end{figure}

\begin{figure}
\begin{tabular}{rr}
\includegraphics[width=0.45\textwidth]{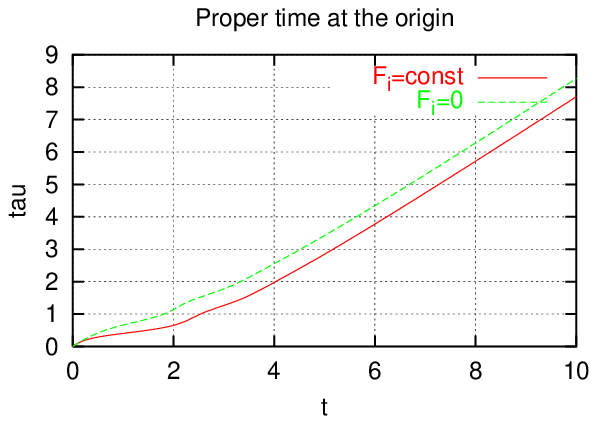}&
\includegraphics[width=0.45\textwidth]{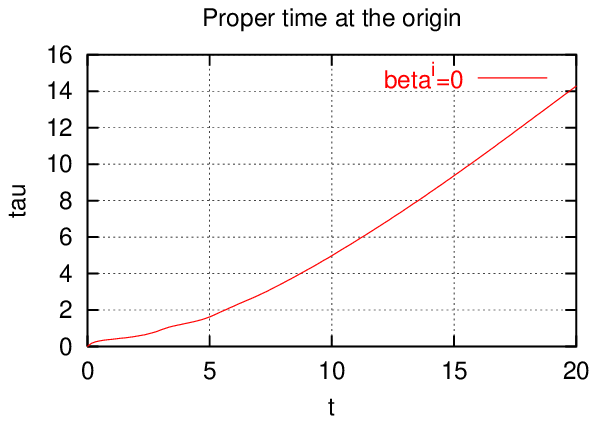}
\end{tabular}
\caption{\label{bw-runs-proper-time-vs-t}Brill wave runs: Proper time
vs.\ coordinate time at the origin for different gauge and coordinate
conditions.}
\end{figure}

This difference in coordinate time seems to be caused by a combination
of two effects.  First, due to the different shift values, the lapse
near the origin has different values.  This is because the shift
enters into the $\partial_t K$ equation (\ref{evol-tracek}) that
determines the lapse.  (The shifts are shown in figure
\ref{bw-runs-betax-vs-x-t}.)  A smaller lapse leads to a larger ratio
between coordinate time and proper time.  However, this explains only
a part of the difference, as can be seen when integrating proper time
along the geodesic formed by the origin (see figure
\ref{bw-runs-proper-time-vs-t}).

The second reason is that the lapse is a pure coordinate quantity, and
by comparing the lapse one cannot directly make a statement about the
location of gravitational waves.  That is, the Brill wave might have
radiated away before the lapse has fully recovered.  A better (but
still not perfect) quantity to look at is $g_{yy}$ along the $x$ axis.
This is a metric component which is orthogonal to the direction in
which the Brill wave radiates, and hence is connected to the
transverse gravitational wave degrees of freedom in the system.

\begin{figure}
\includegraphics[height=0.45\textheight]{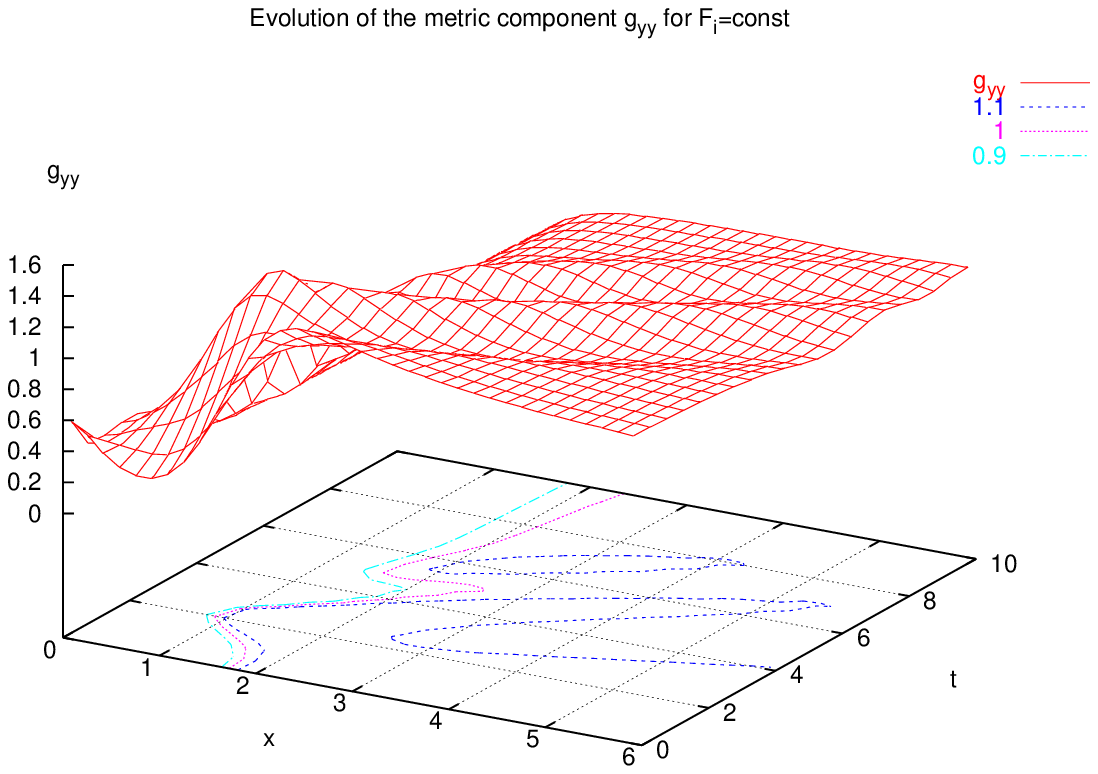}\\
\includegraphics[height=0.45\textheight]{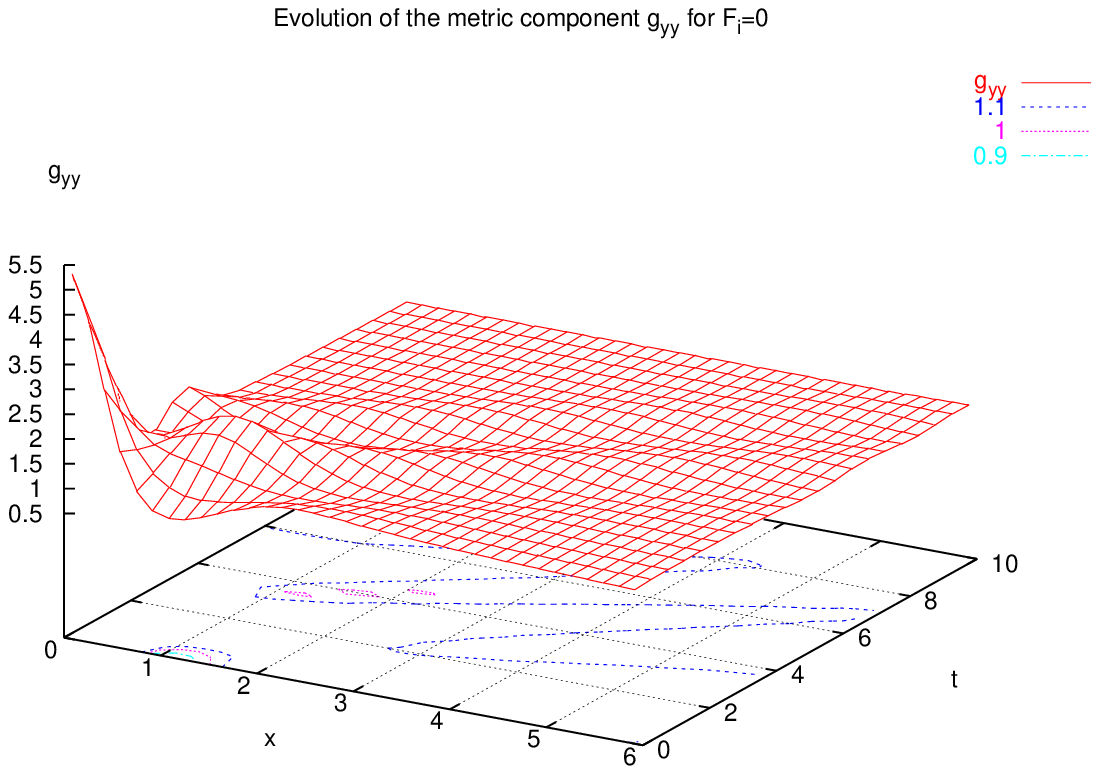}
\caption{\label{bw-runs-gyy-vs-x-t}Brill wave runs: Metric component
$g_{yy}$ vs.\ radius and time for different gauge conditions.  Note
that also $K=0$ everywhere.}
\end{figure}

\begin{figure}
\includegraphics[height=0.45\textheight]{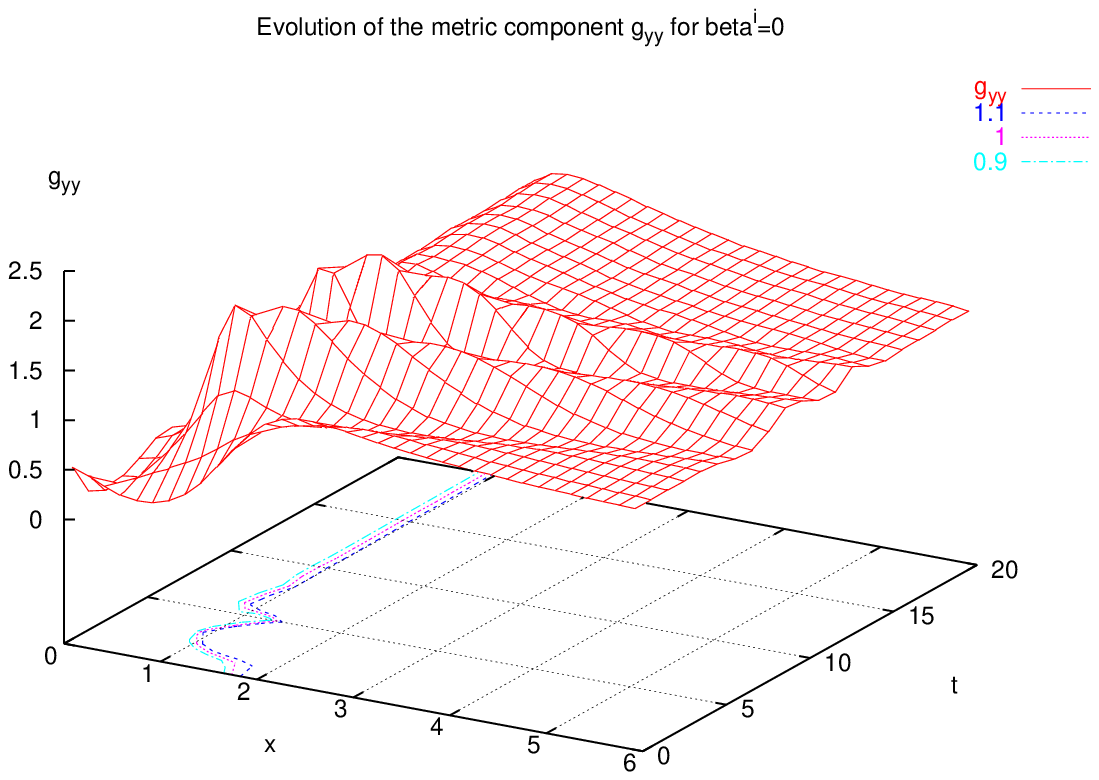}
\caption{\label{bw-runs-gyy-vs-x-t-2}Brill wave runs: Metric component
$g_{yy}$ vs.\ radius and time for zero shift.  Note that also $K=0$
everywhere.}
\end{figure}

Figures \ref{bw-runs-gyy-vs-x-t} and \ref{bw-runs-gyy-vs-x-t-2} show
the evolution of $g_{yy}$ vs.\ time.  One can see that for the gauges
$F_i=\mathrm{const}$ and $F_i=0$ there are two wave trains that leave
the simulation domain at $x=6$ at about $t=10$.  With the coordinate
condition $\beta^i=0$, there are also two wave trains, but it is clear
from the graph that they arrive at $x=6$ substantially earlier than at
$t=20$.  Hence there was not really a factor of two difference between
the coordinate times at which the Brill wave has radiated away to
begin with.

\begin{figure}
\begin{tabular}{rr}
\includegraphics[width=0.45\textwidth]{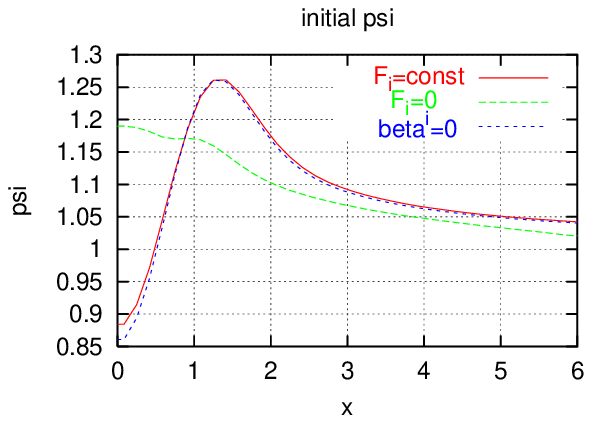}&
\includegraphics[width=0.45\textwidth]{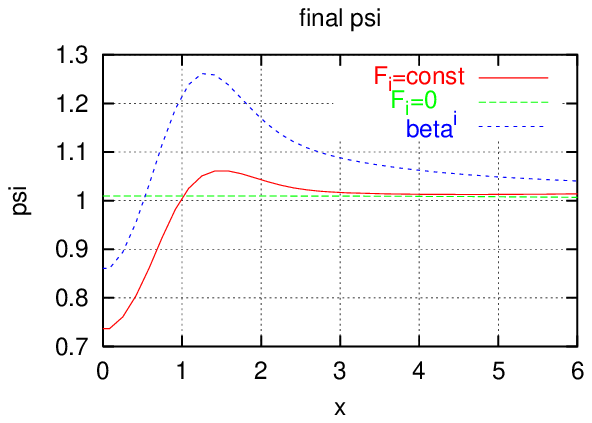}\\
\includegraphics[width=0.45\textwidth]{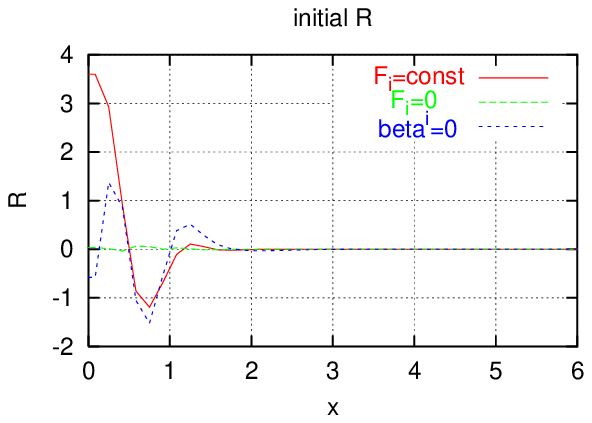}&
\includegraphics[width=0.45\textwidth]{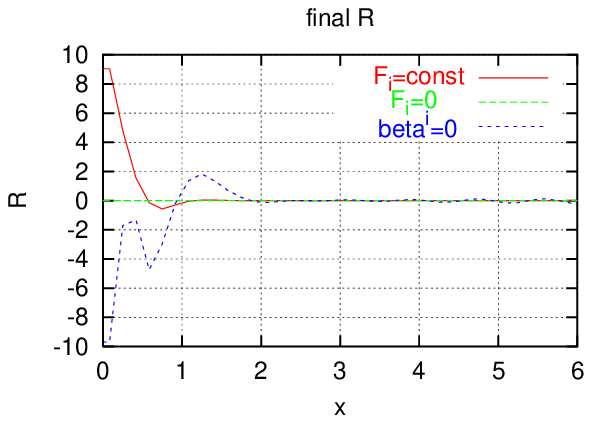}\\
\includegraphics[width=0.45\textwidth]{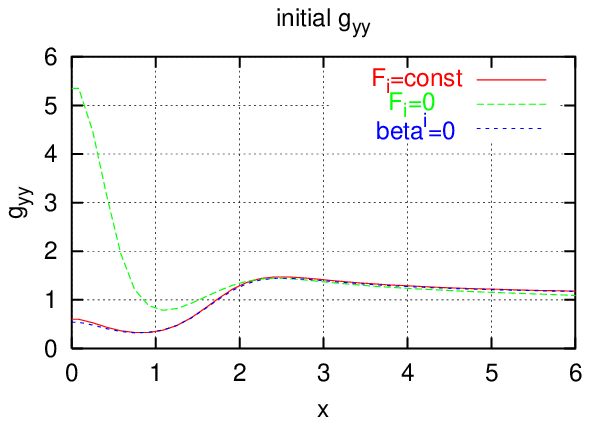}&
\includegraphics[width=0.45\textwidth]{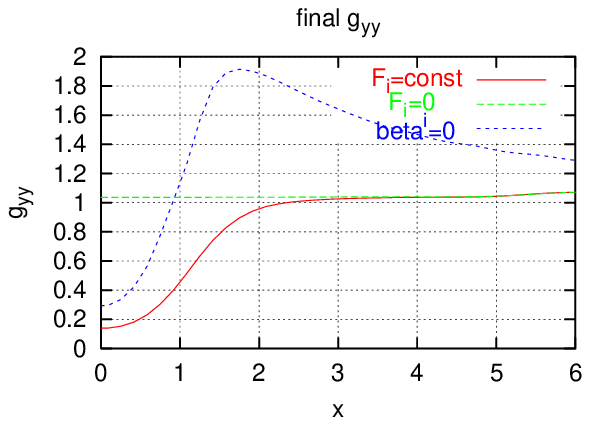}
\end{tabular}
\caption{\label{bw-runs-psi-ricci-gyy}Brill wave runs: Conformal
factor $\psi$, Ricci scalar $R$, and metric component $g_{yy}$ vs.\
radius along the $x$ axis at $t=0$ and at $t=\mathrm{late}$.  Note
that for all shown cases $\alpha=1$ and $\beta^i=0$.  The initial
Ricci scalar $R$ should be identically zero; it shows the magnitude of
the discretisation errors.}
\end{figure}

Figure \ref{bw-runs-psi-ricci-gyy} compares then initial (at $t=0$)
and late time (at $t=10$ and $t=20$, resp.) behaviour of several
quantities, namely the conformal factor $\psi$, the Ricci scalar $R$,
and the transverse metric component $g_{yy}$.  All quantities are
plotted vs.\ the $x$ coordinate.

By construction, the conformal factor $\psi$ is initially the same for
$F_i=\mathrm{const}$ and $\beta^i=0$.  With $K=0$ and $\beta^i=0$, it
is $\partial_t \psi=0$ (see eqn.\ (\ref{evol-conffact})), so that the
final $\psi$ is the same as the initial one in this case.  For $K=0$
and $F_i=0$, flat space has a Minkowski metric, so that $\psi=1$ in
this case.  (This state has not yet completely been reached at
$t=10$.)  Note that the cases $F_i=\mathrm{const}$ and $\beta^i=0$ go
to different static coordinate systems of flat space.

The Ricci scalar should be zero initially, due to the Hamiltonian
constraint and time symmetry.  The Ricci scalar is nonzero because of
discretisation errors.  The final Ricci scalar should be zero for the
$F_i=0$ case.

Again, by construction the initial values of $g_{yy}$ are the same for
$F_i=\mathrm{const}$ and $\beta^i=0$.  The case $F_i=0$ differs,
because it has a different gauge condition enforced onto it.  At late
times, $g_{yy}$ tends to its Minkowski value of one for $F_i=0$.

A ``real'' comparison between the two $F_i=\mathrm{const}$ and
$\beta^i=0$ runs would involve finding a four-coordinate
transformation between the two four-metrics.  The two spacetimes are
identical if and only if such a coordinate transformation exists (up
to numerical errors).  This would be a difficult undertaking, not only
because it involves a huge amount of data (namely two complete
spacetimes).  I know of no proven numerical method to find such a
transformation.  The Lazarus project \cite{lazarus} tries to solve a
similar problem; they compare e.g.\ curvature invariants.

%% \todo{collapsing Brill wave}

%% flush figure floats
\clearpage

\section{Kerr--Schild black hole}

One of the most basic tests of a nonlinear evolution code is a static
or stationary black hole.  I already tested the right hand side of the
evolution equations in various coordinate systems in section
\ref{convtest-static} above.  Here I want to continue these tests by
performing time evolutions with different numerical configurations,
i.e.\ with different resolutions, outer boundary locations, and
different enforced grid symmetries.

I consider the following configurations:

\begin{center}
\begin{tabular}{l|cccc}
title & resolution & spin & location of & symmetries \\
      &            &      & outer bnd.  &            \\\hline
std   & 1/4        & 0    & 4           & octant     \\
dx8   & 1/8        & 0    & 4           & octant     \\
ob8   & 1/4        & 0    & 8           & octant     \\
full  & 1/4        & 0    & 4           & none       \\
spin  & 1/4        & 1/2  & 4           & octant
\end{tabular}
\end{center}

In all cases, the mass of the black hole is $M=1$.  I excise a region
with radius $r=1$ around the singularity.  I use Dirichlet inner and
outer boundary conditions, and I add artificial diffusion with a
coefficient of $C_{\mathrm{SM}} = 0.1$ (see appendix
\ref{artvisc}).  I use an iterative Crank--Nicholson time integrator
with 2 iterations after the initial Euler step (see appendix
\ref{icn}).  I run these configuration up to $t=100$.

With octant symmetry, only one octant of the spacetime is simulated.
The boundary conditions at the zero coordinate planes are given by the
symmetry conditions.  Using these symmetries reduces the memory and
run time requirements by a factor of eight, which is considerable.
However, using or not using these symmetries can influence the
stability of the code, and therefore has to be tested.

I also use other, more physical boundary conditions.  For these runs,
I use radiative outer boundary conditions, or (third-order, or cubic)
extrapolated inner boundary conditions, or both.\footnote{However, the
inner boundary condition for the conformal factor $\psi$ was always
Dirichlet.}  Otherwise, these runs are identical to the std run above:

\begin{center}
\begin{tabular}{l|ll}
title      & outer bc  & excision bc  \\\hline
std        & Dirichlet & Dirichlet    \\
rad        & radiative & Dirichlet    \\
extrap     & Dirichlet & extrapolated \\
extrap-rad & radiative & extrapolated
\end{tabular}
\end{center}

\begin{figure}
\begin{tabular}{rr}
\includegraphics[width=0.45\textwidth]{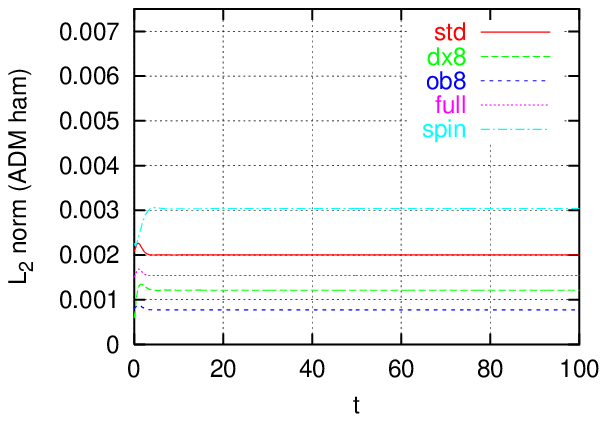}&
\includegraphics[width=0.45\textwidth]{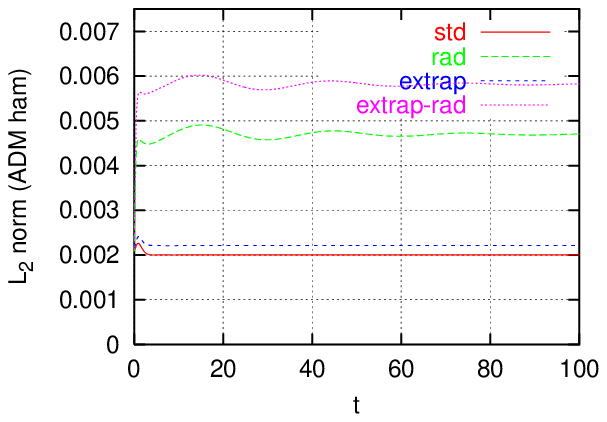}\\
\includegraphics[width=0.45\textwidth]{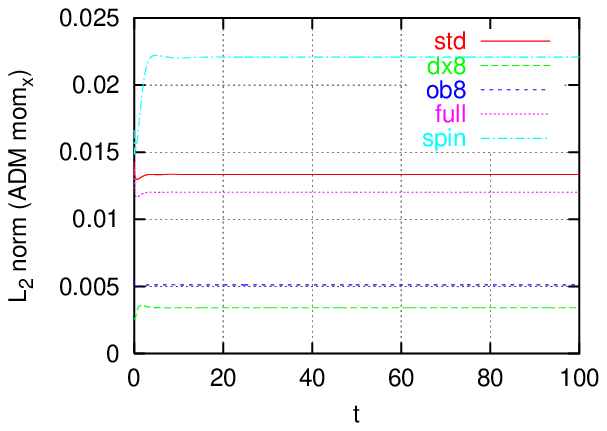}&
\includegraphics[width=0.45\textwidth]{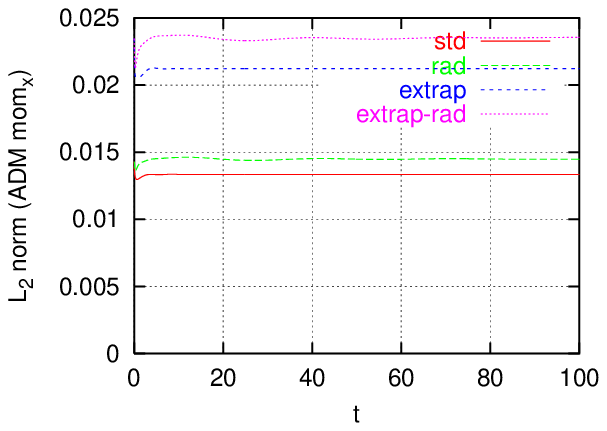}\\
\includegraphics[width=0.45\textwidth]{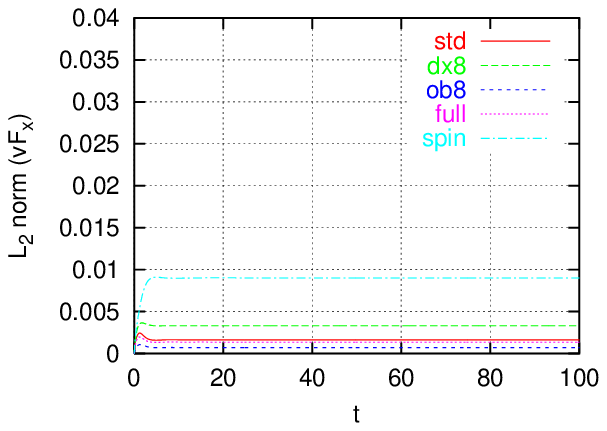}&
\includegraphics[width=0.45\textwidth]{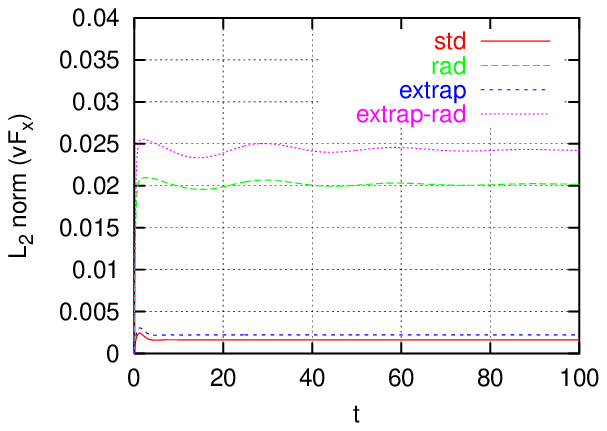}
\end{tabular}
\caption{\label{ks-runs}Kerr--Schild black hole time evolution:
$L_2$-norms of the constraint and gauge violations for various
configurations and boundary conditions.
\emph{std}=standard, \emph{dx8}=higher resolution, \emph{ob8}=larger
domain, \emph{full}=without symmetries, \emph{spin}=with spin,
\emph{rad}=radiative outer boundary, \emph{extrap}=extrapolated inner
boundary}
\end{figure}

I show in figure \ref{ks-runs} the $L_2$ norms of the Hamiltonian and
the momentum constraints and of the gauge violation.  The time
evolutions of the above configurations std, dx8, ob8, full, and spin
show an initial transient, and then settle down close to the initial
data, i.e.\ close to the analytic solution.  They are all stable in
the sense that no growth is visible during the examined time scale.
Because the code is robustly stable, I do not expect instabilities at
later times, although I have no proof for that.
%% An elapsed time of $T=100$ is considered sufficient to collide black
%% holes.

\begin{figure}
\begin{tabular}{rr}
\includegraphics[width=0.45\textwidth]{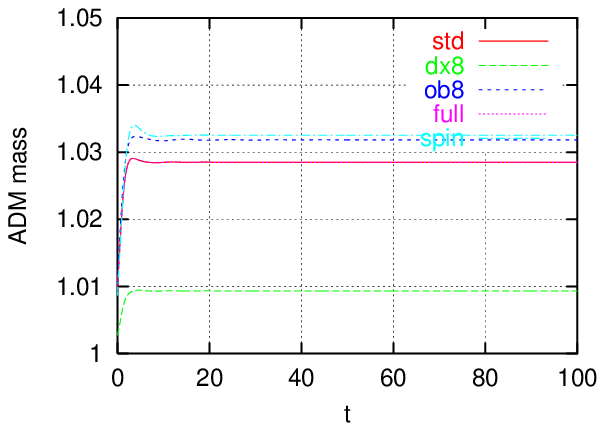}&
\includegraphics[width=0.45\textwidth]{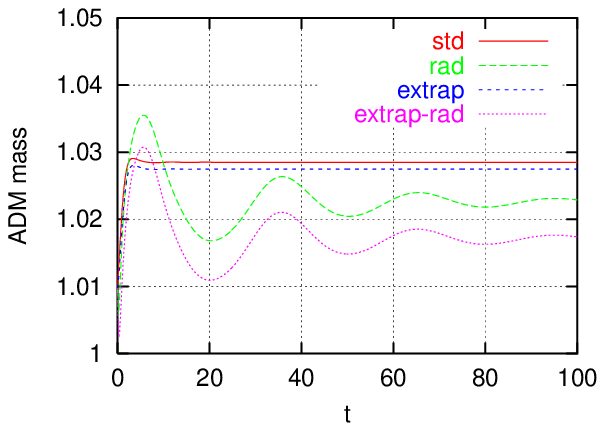}\\
\includegraphics[width=0.45\textwidth]{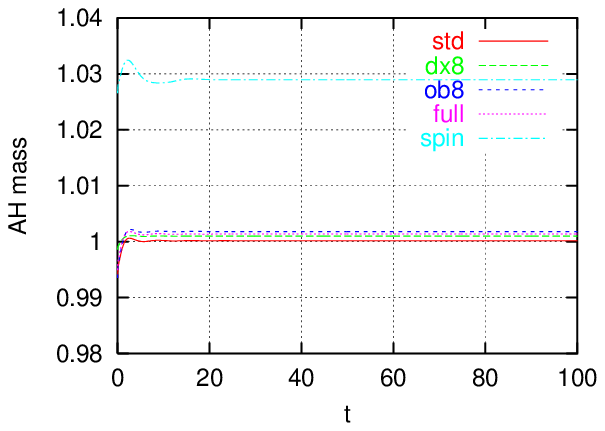}&
\includegraphics[width=0.45\textwidth]{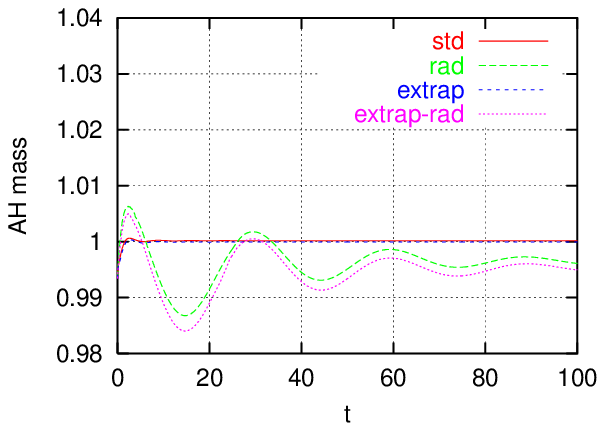}
\end{tabular}
\caption{\label{ks-runs-masses}Kerr--Schild black hole time evolution:
ADM masses and apparent horizon masses.
\emph{std}=standard, \emph{dx8}=higher resolution, \emph{ob8}=larger
domain, \emph{full}=without symmetries, \emph{spin}=with spin,
\emph{rad}=radiative outer boundary, \emph{extrap}=extrapolated inner
boundary}
\end{figure}

The configurations rad and extrap-rad, which have a radiative outer
boundary condition, pick up a slowly decaying global breathing mode.
This error has an amplitude of about $1\%$ in the ADM and apparent
horizon masses, as shown in figure \ref{ks-runs-masses}.  This mode
has not yet decayed at $t=100$.

When comparing simple (std) Kerr--Schild runs with the resolutions
$dx=1/4$, $dx=1/6$, and $dx=1/8$ at $t=100$ along the $x$ axis, I get
following convergence factors (cf):

\vspace{1ex}
\begin{tabular}{l|lllll}
   $x$ & $1.5$ & $2.0$ & $2.5$ & $3.0$ & $3.5$
\\\hline
   cf & $2.26687$ & $2.71461$ & $2.74507$ & $2.73848$ & $2.73511$
\end{tabular}
\vspace{1ex}

A factor of $(1/4^2-1/6^2) / (1/6^2-1/8^2) \approx 2.85714$ in the
above would indicate second order convergence.  The numbers suggest
that the TGR system is able to evolve static black holes in
Kerr--Schild coordinates with excision in a stable and convergent
manner.  The stability of the evolution is independent of the
resolution, the location of the outer boundaries, the symmetries that
are imposed, and the kind of boundary conditions that are used.

%% \todo{explain why ADM masses might be bad}

%% flush figure floats
\clearpage

%% \section{Superposed Kerr--Schild black holes}
%% 
%% ip4, ip8
%% 
%% \begin{figure}
%% \begin{tabular}{rr}
%% \includegraphics[width=0.45\textwidth]{RUNS/kerrschild2/kerrschild2-adm_ham}&
%% \includegraphics[width=0.45\textwidth]{RUNS/kerrschild2/kerrschild2-adm_momx}\\
%% \includegraphics[width=0.45\textwidth]{RUNS/kerrschild2/kerrschild2-vfx}
%% \end{tabular}
%% \caption{Superposed Kerr--Schild black hole evolution:
%% \emph{std}=standard, \emph{dx8}=higher resolution}
%% \end{figure}
%% 
%% \todo{extrapolation inner bc}
%% 
%% \todo{radiative outer bc}

% LocalWords:  eschnett Exp dx ks hyy ayy adm momy pg hc SM zz constr gzz Szil
% LocalWords:  agyi Institut Golm PETSc betax gyy momx std ob extrap bnd TGR ip
% LocalWords:  bc

% -*-LaTeX-*-
% $Header: /home/eschnett/cvs/diss/conclusion.tex,v 1.8 2003/01/10 16:39:49 eschnett Exp $

\chapter{Conclusion}
\label{conclusion}

In this thesis, I set out to find a way to arrive at some more
understanding of the instabilities that one encounters in today's
black hole evolution simulations.  I looked at gauge conditions and
classified them into gauge evolution and gauge fixing conditions,
assuming that gauge modes do play an important role in the
instabilities.  I proposed a gauge fixing condition and an
implementation for it, and examined it in various test problems and
some applications.

Gauge fixing seems to lead to very stable evolutions.  The Tiger code
is robustly stable, and has also sufficient accuracy to treat more
complex configurations.  It remains to be shown that this system also
works for two coalescing black holes.  The computational resources
required by the current implementation do unfortunately not allow such
a test; further algorithmic improvements (the implementation of a
parallel multigrid solver) will probably be necessary for that.

\appendix
% -*-LaTeX-*-
% $Header: /home/eschnett/cvs/diss/equations.tex,v 1.13 2003/01/10 16:39:49 eschnett Exp $

\chapter{Equations}
\label{equations}

Many of the equations in this appendix are only a repetition of
well-known literature.  They serve mostly as a quick reference, and to
define the sign conventions.  For other equations, I give a short
derivation.  I restrict myself to the vacuum case everywhere.

\section{The ADM formalism}

The ADM formalism is one way to transform the Einstein equations into
a system of time evolution equations.

\subsection{Variables}
\label{adm-variables}

The ADM variables \cite{adm} are the three-metric $\gamma_{ij}$, the
extrinsic curvature $K_{ij}$, the lapse $\alpha$ and the shift
$\beta^i$.  The four-metric can be written as
\begin{eqnarray}
   g_{\mu\nu} & = &
   \left( \begin{array}{c|ccc}
      - \alpha^2 + \beta^2 & & \beta_j & \\\hline
      & & & \\
      \beta_i & & \gamma_{ij} & \\
      & & &
   \end{array} \right)
\end{eqnarray}
where $\beta_i = \gamma_{ij} \beta^j$, and $\beta^2 = \beta_j
\beta^j$.  This metric is equivalent to the line element
\begin{eqnarray}
   ds^2 & = & - \alpha^2 dt^2 + \gamma_{ij} \left( dx^i + \beta^i dt
   \right) \left( dx^j + \beta^j dt \right) \\
   & = & \left( - \alpha^2 + \gamma_{ij} \beta^i \beta^j \right) dt^2
   + 2 \gamma_{ij} \beta^j dx^i dt + \gamma_{ij} dx^i dx^j \quad.
\end{eqnarray}
The inverse metric is then
\begin{eqnarray}
   g^{\mu\nu} & = & \frac{1}{\alpha^2}
   \left( \begin{array}{c|ccc}
      -1 & & \beta^i & \\\hline
      & & & \\
      \beta^i & & \alpha^2 \gamma^{ij} - \beta^2 & \\
      & & &
   \end{array} \right)
\end{eqnarray}

The three-metric and extrinsic curvature on a time slice are enough to
completely specify the physics in the whole spacetime.  Lapse $\alpha$
and shift $\beta^i$ can be chosen freely during time evolution and
determine only the coordinate system.  The extrinsic curvature is
defined via the time derivative of the three-metric by eqn.\
(\ref{evol-metric}) below.

\subsection{Time evolution}
\label{adm-time-evolution}

The time evolution of the three-metric (which is also definition of
the extrinsic curvature $K_{ij}$) is given by
\begin{eqnarray}
%%    \label{evol-metric} \partial_t \gamma_{ij} & = & -2 \alpha K_{ij} +
%%    \gamma_{lj} \partial_i \beta^l + \gamma_{il} \partial_j \beta^l +
%%    \beta^l \partial_l \gamma_{ij}
   \label{evol-metric} \partial_t \gamma_{ij} & = & -2 \alpha K_{ij} +
   \gamma_{lj} \beta^l_{,i} + \gamma_{il} \beta^l_{,j} + \beta^l
   \gamma_{ij,l}
\end{eqnarray}

The time evolution of the extrinsic curvature is given by
\begin{eqnarray}
%%    \label{evol-extcurv} \partial_t K_{ij} & = & - \nabla_i \nabla_j
%%    \alpha + \alpha \left[ R_{ij} - 2 K_{il} K^l_j + K K_{ij} \right]
%%    \\
%% %
%%    \nonumber & & {} + K_{lj} \partial_i \beta^l + K_{il} \partial_j
%%    \beta^l + \beta^l \partial_l K_{ij}
   \label{evol-extcurv} \partial_t K_{ij} & = & - \nabla_i \nabla_j
   \alpha + \alpha \left[ R_{ij} - 2 K_{il} K^l_j + K K_{ij} \right]
   \\
   \nonumber & & {} + K_{lj} \beta^l_{,i} + K_{il} \beta^l_{,j} +
   \beta^l K_{ij,l}
\end{eqnarray}

\subsection{Ricci tensor}
\label{adm-ricci}

The connection coefficients to the three-metric are given by
\begin{eqnarray}
   \Gamma^i_{jk} & = & \frac{1}{2} \gamma^{il} \left[ \gamma_{lj,k} +
   \gamma_{lk,j} - \gamma_{jk,l} \right] %% \\
%% %
%%    \Gamma^a_{bc,d} & = & \frac{1}{2} \gamma^{ae}_{,d} \left[
%%    \gamma_{eb,c} + \gamma_{ec,b} - \gamma_{bc,e} \right] \\
%% %
%%    \nonumber & & {} + \frac{1}{2} \gamma^{ae} \left[ \gamma_{eb,cd} +
%%    \gamma_{ec,bd} - \gamma_{bc,ed} \right] \\
%% %
%%    \Gamma^a_{bc} & = & \frac{1}{2} \gamma^{ae} \left[ \gamma_{eb,c} +
%%    \gamma_{ec,b} - \gamma_{bc,e} \right] \\
%% %
%%    \Gamma^a_{bc,d} & = & \frac{1}{2} \gamma^{ae}_{,d} \left[
%%    \gamma_{eb,c} + \gamma_{ec,b} - \gamma_{bc,e} \right] \\
%% %
%%    \nonumber & & {} + \frac{1}{2} \gamma^{ae} \left[ \gamma_{eb,cd} +
%%    \gamma_{ec,bd} - \gamma_{bc,ed} \right] \\
%% %
%%    & = & - \frac{1}{2} \gamma^{ax} \gamma^{ey} \gamma_{xy,d} \left[
%%    \gamma_{eb,c} + \gamma_{ec,b} - \gamma_{bc,e} \right] \\
%% %
%%    \nonumber & & {} + \frac{1}{2} \gamma^{ae} \left[ \gamma_{eb,cd} +
%%    \gamma_{ec,bd} - \gamma_{bc,ed} \right]
\end{eqnarray}

The three-Ricci tensor is e.g.\ given by
\begin{eqnarray}
   R_{ij} & = & \Gamma^k_{ij,k} - \Gamma^k_{ik,j} + \Gamma^k_{jl}
   \Gamma^l_{ik} - \Gamma^k_{lk} \Gamma^l_{ij} %% \\
\end{eqnarray}
It is a bad idea to calculate it in this way numerically, as this
would involve taking derivatives of derivatives.  Instead, one has to
explicitely take second derivatives of the three-metric.

\subsection{Constraints}
\label{adm-constraints}

The Hamiltonian constraint is usually written as
\begin{eqnarray}
\label{ham}
   H & = & R + K^2 - K_{ij} K^{ij}
\end{eqnarray}

The momentum constraint is usually written as
\begin{eqnarray}
\label{mom}
   M_i & = & \nabla^j \left( K_{ij} - \gamma_{ij} K \right)
\end{eqnarray}

As these constraints have to be zero for a physical spacetime, one can
multiply them by an arbitrary nonzero function to arrive at other
formulations for the constraints.  This especially leaves scaling
freedoms, so that a statement saying the constraints are smaller than
a certain value is only meaningful if this is with respect to a
certain given length scale.

One can introduce normalised constraints e.g.\ as
\begin{eqnarray}
   H^{(N)} & = & \frac{H}{|R| + |K^2| + |K_{ij} K^{ij}|} \\
   M_i^{(N)} & = & \frac{M_i}{|\nabla^j K_{ij}| + |\nabla_i K|}
\end{eqnarray}
where the absolute value of a vector $x^i$ is calculated as $|x^i|^2 =
\gamma_{ij} x^i x^j$.  These normalised constraints have a range of
$[0;1]$ and are thus scaling invariant.  They have the disadvantage of
being ill-defined e.g.\ in the Minkowski spacetime, and are
meaningless in perturbations thereof.

\section{The TGR system}

The TGR system is a conformal traceless ADM (ctADM) formalism that is
derived from the ADM formalism by introducing two additional scalar
quantities $\psi$ and $K$, which represent the determinant of the
three-metric and the trace of the extrinsic curvature, respectively.
This also leads to two new constraints, namely just these conditions.

\subsection{Variables}
\label{ctadm-variables}

The variant of conformal traceless ADM formalism (compare
\cite{shibata-nakamura, baumgarte-shapiro}) that I use has the
following five variables:

The conformal factor
\begin{eqnarray}
   \label{def-psi} \psi^4 & = & ( \det \gamma_{ij} )^{1/3}
\end{eqnarray}

The conformal metric
\begin{eqnarray}
   \tilde \gamma_{ij} & = & \psi^{-4} \gamma_{ij} \\
   \label{constr-psi} \textrm{where thus} \quad \det \tilde
   \gamma_{ij} & = & 1
\end{eqnarray}

The trace of the extrinsic curvature:
\begin{eqnarray}
   \label{def-K} K & = & K_i^i
\end{eqnarray}

The traceless extrinsic curvature (this is only an intermediate
quantity):
\begin{eqnarray}
   A_{ij} & = & K_{ij} - \frac{1}{3} \gamma_{ij} K \\
   \textrm{where thus} \quad A_i^i & = & 0
\end{eqnarray}

The conformal traceless extrinsic curvature:
\begin{eqnarray}
   \tilde A_{ij} & = & \psi^2 A_{ij} \\
   \label{constr-A} \textrm{where thus} \quad \tilde A_i^i & = & 0
\end{eqnarray}

%% The traceless conformal three-metric (this is only an intermediate
%% quantity):
%% %
%% \begin{eqnarray}
%%    \bar h_{ij} & = & \tilde \gamma_{ij} - \frac{1}{3} \delta_{ij}
%%    \delta^{kl} \tilde \gamma_{kl} \\
%% %
%%    \textrm{where thus} \quad \delta^{ij} \bar h_{ij} & = & 0
%% \end{eqnarray}
%% 
%% The divergence of the traceless conformal three-metric:
%% %
%% \begin{eqnarray}
%%    \label{def-F} \tilde F_i & = & \bar h_{ij,j} \\
%% \end{eqnarray}

Quantities with a tilde $\tilde \cdot$ have their indices raised and
lowered using the conformal metric $\tilde \gamma_{ij}$.  Equations
(\ref{constr-psi}) and (\ref{constr-A}) form two additional, algebraic
constraints.
%% Quantities with a bar $\bar \cdot$ have their indices
%% raised and lowered using the coordinate metric $\delta_{ij}$.
%% Equation (\ref{def-F}) forms an additional, non-algebraic constraint.

\subsection{Time evolution}
\label{ctadm-time-evol}

The time evolution equations for the ctADM quantities follow in a
straightforward way from their definitions, and from the time
evolution equations of the ADM quantities.

The time evolution of the conformal factor is given by
\begin{eqnarray}
%%    \partial_t \psi & = & - \frac{1}{6} \alpha \psi K + \beta^l
%%    \partial_l \psi + \frac{1}{6} \psi \partial_l \beta^l
   \label{evol-conffact} \partial_t \psi & = & - \frac{1}{6} \alpha \psi K
   + \beta^l \psi_{,l} + \frac{1}{6} \psi \beta^l_{,l}
\end{eqnarray}

The time evolution of the conformal metric is given by
\begin{eqnarray}
%%    \partial_t \tilde \gamma_{ij} & = & -2 \alpha \psi^{-6} \tilde
%%    A_{ij} \\
%% %
%%    \nonumber & & {} + \tilde \gamma_{lj} \partial_i \beta^l + \tilde
%%    \gamma_{il} \partial_j \beta^l + \beta^l \partial_l \tilde
%%    \gamma_{ij} - \frac{2}{3} \tilde \gamma_{ij} \partial_l \beta^l
   \label{evol-confmetric} \partial_t \tilde \gamma_{ij} & = & -2
   \alpha \psi^{-6} \tilde A_{ij} \\
   \nonumber & & {} + \tilde \gamma_{lj} \beta^l_{,i} + \tilde
   \gamma_{il} \beta^l_{,j} + \beta^l \tilde \gamma_{ij,l} -
   \frac{2}{3} \tilde \gamma_{ij} \beta^l_{,l}
\end{eqnarray}

The time evolution of the trace of the extrinsic curvature is given by
\begin{eqnarray}
%%    \label{evol-tracek} \partial_t K & = & - \triangle \alpha + \alpha
%%    \left[ R + K^2 \right] + \beta^l \partial_l K
   \label{evol-tracek} \partial_t K & = & - \triangle \alpha + \alpha
   \left[ R + K^2 \right] + \beta^l K_{,l}
\end{eqnarray}
Here $\triangle$ denotes the covariant Laplace operator $\nabla^l
\nabla_l$.

The time evolution of the traceless extrinsic curvature (an
intermediate quantity) is given by
\begin{eqnarray}
   \partial_t A_{ij} & = & \partial_t K_{ij} - \frac{1}{3} \partial_t
   \left( \gamma_{ij} K \right) \\
   & = & - \nabla_i \nabla_j \alpha + \alpha \left[ R_{ij} - 2 K_{il}
   K^l_j + K K_{ij} \right] \\
   \nonumber & & {} + K_{lj} \beta^l_{,i} + K_{il} \beta^l_{,j} +
   \beta^l K_{ij,l} \\
   \nonumber & & {} - \frac{1}{3} \left[ -2 \alpha K_{ij} +
   \gamma_{lj} \beta^l_{,i} + \gamma_{il} \beta^l_{,j} + \beta^l
   \gamma_{ij,l} \right] K \\
   \nonumber & & {} - \frac{1}{3} \gamma_{ij} \left[ - \triangle
   \alpha + \alpha \left[ R + K^2 \right] + \beta^l K_{,l} \right] \\
   & = & - \left( \nabla_i \nabla_j \alpha \right)^{\mathrm{TF}} +
   \alpha \left( R_{ij} \right)^{\mathrm{TF}} \\
   \nonumber & & {} + \alpha \left[ - 2 A_{il} A^l_j \right] \\
   \nonumber & & {} + A_{lj} \beta^l_{,i} + A_{il} \beta^l_{,j} +
   \beta^l A_{ij,l}
\end{eqnarray}
Here $( X_{ij} )^{\mathrm{TF}}$ denotes the tracefree part of $X_{ij}$
with respect to the physical metric, i.e.\ $X_{ij} - \frac{1}{3}
\gamma_{ij} \gamma^{lm} X_{lm}$.

The time evolution of the conformal traceless extrinsic curvature is
given by
\begin{eqnarray}
   \label{evol-confextcurv} \partial_t \tilde A_{ij} & = & \partial_t
   \left( \psi^2 A_{ij} \right) \\
   & = & 2 \psi \left( - \frac{1}{6} \alpha \psi K + \beta^l \psi_{,l}
   + \frac{1}{6} \psi \beta^l_{,l} \right) A_{ij} \\
   \nonumber & & {} + \psi^2 \left( - \left( \nabla_i \nabla_j \alpha
   \right)^{\mathrm{TF}} + \alpha \left( R_{ij} \right)^{\mathrm{TF}}
   \right) \\
   \nonumber & & {} + \psi^2 \left ( \alpha \left[ - 2 A_{il} A^l_j +
   \frac{1}{3} A_{ij} K \right] \right) \\
   \nonumber & & {} + \psi^2 \left( A_{lj} \beta^l_{,i} + A_{il}
   \beta^l_{,j} + \beta^l A_{ij,l} \right) \\
   & = & - \psi^2 \left( \nabla_i \nabla_j \alpha
   \right)^{\mathrm{TF}} + \alpha \psi^2 \left( R_{ij}
   \right)^{\mathrm{TF}} \\
   \nonumber & & {} - 2 \alpha \psi^{-6} \tilde A_{il} \tilde A^l_j \\
   \nonumber & & {} + \tilde A_{lj} \beta^l_{,i} + \tilde A_{il}
   \beta^l_{,j} + \beta^l \tilde A_{ij,l} + \frac{1}{3} \tilde A_{ij}
   \beta^l_{,l}
\end{eqnarray}

\subsection{Constraints}
\label{ctadm-constraints}

The conformal Hamiltonian constraint can be written as
\begin{eqnarray}
   \tilde H & = & \tilde \triangle \psi - \frac{1}{8} \psi \tilde R -
   \frac{1}{12} \psi^5 K^2 + \frac{1}{8} \psi^{-7} \tilde A_{ij}
   \tilde A^{ij}
\end{eqnarray}

The conformal momentum constraint can be written as
\begin{eqnarray}
   \tilde M_i & = & \tilde \nabla^j \tilde A_{ij} - \frac{2}{3} \psi^6
   \tilde \nabla_i K
\end{eqnarray}

Note that that these two constraints contain different scaling factors
than the ADM constraints (\ref{ham}) and (\ref{mom}).  These two
formulations of the constraints cannot readily be compared to each
other.

\subsection{Enforcing the constraints}
\label{ctadm-enforce-constraints}

In order to enforce the constraints, one can extend the system of
ctADM variables with a vector potential $V_i$ for the conformal
traceless extrinsic curvature \cite[section 2.2.1]{cook}.  This leads
to the modified conformal traceless extrinsic curvature
\begin{eqnarray}
   \label{elim-V}
   \tilde A_{ij} & = & \tilde A^*_{ij} + (\mathrm{\tilde L} V)_{ij}
\end{eqnarray}
where the operator $\mathrm{\tilde L}$ is the longitudinal derivative
with respect to the conformal metric, defined by
\begin{eqnarray}
   (\mathrm{\tilde L} V)_{ij} & = & \tilde \nabla_i V_j + \tilde
   \nabla_j V_i - \frac{2}{3} \tilde \gamma_{ij} \tilde \nabla^k V_k
\end{eqnarray}
which makes the gradient $(\mathrm{\tilde L} V)_{ij}$ symmetric and
tracefree.

Given a conformal metric $\tilde \gamma_{ij}$, a mean curvature $K$
and a background conformal traceless extrinsic curvature $\tilde
A^*_{ij}$, one can enforce the constraints by solving the constraint
equations for the conformal factor $\psi$ and the vector potential
$V_i$.  This vector potential and the background conformal traceless
extrinsic curvature then define the real conformal traceless extrinsic
curvature $\tilde A_{ij}$ through (\ref{elim-V}).

\subsection{Gauge condition}
\label{ctadm-gauge}

I introduce the traceless conformal metric
\begin{eqnarray}
   \bar h_{ij} & = & \tilde \gamma_{ij} - \frac{1}{3} \delta_{ij}
   \delta^{kl} \tilde \gamma_{kl} \quad .
\end{eqnarray}
The bar $\bar \cdot$ is used to indicate that the indices are to be
raised and lowered using the coordinate metric $\delta_{ij}$.

The gauge variable is
\begin{eqnarray}
   F_i & = & \bar h_{ij,j} \quad .
\end{eqnarray}
which is close to what Shibata and Nakamura use in
\cite{shibata-nakamura}.

The gauge condition should constrain the conformal metric, and the
conformal metric already has to satisfy the constraint $\det \tilde
\gamma_{ij} = 1$.  Enforcing the gauge condition will change the
conformal metric, but one has to make sure that it does not change its
determinant.  By making the gauge condition independent of the trace
of the conformal metric, one can later change the trace to adjust its
determinant to one.

%% \subsubsection{An idea}
%% 
%% Instead of having the gauge condition act on the conformal metric, one
%% could envision a condition that acts on the logarithm of the conformal
%% metric.  The logarithm has a trace of zero, and one can make sure that
%% enforcing the gauge consists of adding tracefree quantities to the
%% logarithm.  Thus the trace of the logarithm, and hence also the
%% determinant of the conformal metric will not be affected by enforcing
%% the gauge.  As the gauge condition uses only derivatives of the
%% metric, using a logarithm here is not expensive either, as the
%% logarithm will never have to be calculated.
%% 
%% This would lead to the definition
%% %
%% \begin{eqnarray}
%%    F_i & = & \partial_j \log \tilde \gamma_{ij} \\
%% %
%%    \nonumber & \stackrel{?}{=} & \delta_{ik} \delta_{jl} \tilde
%%    \gamma^{kl}_{,j}
%% \end{eqnarray}
%% %
%% which looks very close to what has been used by Baumgarte and Shapiro
%% \cite{baumgarte-shapiro}.

\subsection{Enforcing the gauge condition}
\label{ctadm-enforce-gauge}

Enforcing the gauge condition on the trace of the extrinsic curvature
$K$ is trivial.

The method that I use to enforce the gauge on the conformal metric is
similar to the method used to enforce the constraints.  I extend the
system of ctADM variables with a vector potential $W_i$ for the
traceless conformal metric.  This leads to the modified traceless
conformal metric
\begin{eqnarray}
   \label{elim-W}
   \bar h_{ij} & = & \bar h^*_{ij} + (\bar {\mathrm{L}} W)_{ij}
\end{eqnarray}
where the operator $\bar {\mathrm{L}}$ is the longitudinal derivative
with respect to the coordinate metric, i.e.\ using partial
derivatives.  It is defined by
\begin{eqnarray}
   (\bar {\mathrm{L}} W)_{ij} & = & W_{j,i} + W_{i,j} - \frac{2}{3}
   \delta_{ij} W_{k,k}
\end{eqnarray}
which makes the gradient $(\bar {\mathrm{L}} W)_{ij}$ symmetric and
tracefree.

The gauge condition is then
\begin{eqnarray}
   \label{def-F}
   F_i %% & = & \bar h_{ij,j} \\
%% %
%%    & = & \bar h^*_{ij,j} + W_{j,ij} + W_{i,jj} - \frac{2}{3}
%%    \delta_{ij} W_{k,kj} \\
%% %
   & = & \bar h^*_{ij,j} + W_{i,jj} + \frac{1}{3} W_{j,ij}
\end{eqnarray}

Given a background traceless conformal metric $\bar h^*_{ij}$, one can
enforce the gauge condition on the metric by solving the gauge
equation (\ref{def-F}) for the vector potential $W_i$.  This vector
potential and the background traceless conformal metric, together with
the constraint $\det \tilde \gamma_{ij} = 1$ from which the trace of
the conformal metric is calculated, then define the real conformal
metric $\tilde \gamma_{ij}$ through (\ref{elim-W}).

\subsection{Determining lapse and shift}
\label{coordinate-conditions}

The gauge condition implicitly determines the lapse and shift.

One differentiates the gauge conditions with respect to time, and then
uses the time evolution equations for the trace of the extrinsic
curvature $\partial_t K$ and for the conformal metric $\partial_t
\tilde \gamma_{ij}$ to arrive at a set of coupled elliptic equations
for the lapse and shift.

The equation for the lapse $\alpha$ is immediately given by
(\ref{evol-tracek}).  The equation for the shift $\beta^i$ is largish,
but not very enlightening, and I omit most of it here.  The fact that
it is so complicated seems to point to the fact that there should be a
more elegant gauge condition.  The terms containing second derivatives
of the shift are
\begin{eqnarray}
   \label{shift-condition} \partial_t F_i & \simeq & \tilde
   \gamma_{ik} \beta^k_{,jj} + \frac{1}{3} \tilde \gamma_{jk}
   \beta^k_{,ij} - \frac{2}{3} \tilde \gamma_{ij} \beta^k_{,jk} +
   \frac{2}{9} \tilde \gamma_{jj} \beta^k_{,ik} \\
%
%%    & = & ( \bar h_{ik} + \frac{1}{3} \delta_{ik} \bar h )
%%    \beta^k_{,jj} + \frac{1}{3} ( \bar h_{jk} + \frac{1}{3} \delta_{jk}
%%    \bar h ) \beta^k_{,ij} \\
%% %
%%    \nonumber && {} - \frac{2}{3} ( \bar h_{ij} + \frac{1}{3}
%%    \delta_{ij} \bar h ) \beta^k_{,jk} + \frac{2}{9} ( \bar h_{jj} +
%%    \frac{1}{3} \delta_{jj} \bar h ) \beta^k_{,ik} \\
%% %
   & = & \tilde \gamma_{ik} \beta^k_{,jj} + \frac{1}{3} \tilde
   \gamma_{jk} \beta^k_{,ij} - \frac{2}{3} \bar h_{ij} \beta^k_{,jk}
\end{eqnarray}
These terms look sufficiently close to being elliptic that one can
hope that the equation is well-posed.  The numerical experiment shows
that this seems to be the case.

\chapter{Numerics}
\label{numerics}

\section{Spatial discretisation}

I discretise the quantities on a Cartesian grid.  I approximate the
partial derivatives by second order centred derivatives.  This is
among the most simple possible discretisation methods.

\section{Time integration}

\label{icn}
\label{midpoint}
I integrate in time with an explicit second order integrator, using
the iterative Crank--Nicholson scheme \cite{icn}.  For that I usually
use two iterations after the initial Euler step.  For some experiments
I use only one iteration after the Euler step, which turns this scheme
into the midpoint rule, or a second order Runge--Kutta.  This is one
of the most simple possible second-order time integrators.

Given the ordinary differential equation
\begin{eqnarray}
   \frac{d}{dt} f(t) & = & u \left[ t, f(t) \right]
\end{eqnarray}
the midpoint rule uses the time stepping scheme
\begin{eqnarray}
   f_n^{(1)} & := & f_{n-1} + h\; u \left[ t_{n-1}, f_{n-1} \right]
\\
   f_n^{(2)} & := & f_{n-1} + h\; u \left[ \frac{1}{2} (t_{n-1} +
   t_n), \frac{1}{2} (f_{n-1} + f_n^{(1)}) \right]
\\
   f_n & := & f_n^{(2)}
\end{eqnarray}
to advance from $f_{n-1}$ to $f_n$ with the step size $h$.
$f_n^{(1)}$ is an intermediate quantity.

Similarly, the iterative Crank--Nicholson method uses the time
stepping scheme
\begin{eqnarray}
   f_n^{(1)} & := & f_{n-1} + h\; u \left[ t_{n-1}, f_{n-1} \right]
\\
   f_n^{(i)} & := & f_{n-1} + h\; u \left[ \frac{1}{2} (t_{n-1} +
   t_n), \frac{1}{2} (f_{n-1} + f_n^{(i-1)}) \right]
\\
   f_n & := & \lim_{i \to \infty} f_n^{(i)}
\end{eqnarray}
where the limes is usually only taken up to $i=3$.  The first step is
a simple Euler step.  All other right hand sides are evaluated at the
half-time $t_{n-1} + \frac{h}{2}$.

\subsection{Artificial viscosity}
\label{artvisc}

The advection terms in the time derivatives, i.e.\ those terms
involving the shift $\beta^i$, are unstable when discretised as above
and integrated with the midpoint rule.  I therefore add artificial
diffusion to the right hand sides of the time evolution equations.  I
implement two kinds of artificial diffusion, of which one depends on
the shift:
\begin{eqnarray}
   (\partial_t f)_{\mathrm{AD}} & = & \sum_i \left( C_{\mathrm{ADV}}
   \frac{| \beta^i |}{2} + C_{\mathrm{SM}} \right) dx^i f_{,ii}
\end{eqnarray}
with $0 \le C_{\mathrm{ADV}} \lesssim 1$ and $0 \le C_{\mathrm{SM}}
\lesssim 1$.  The quantity $dx^i$ is the grid spacing.  These two
viscosities are explained below.

The first kind, ADV, which depends on the shift, transforms the
centred differencing advection terms into first order upstream
differencing terms when $C_{\mathrm{ADV}} = 1$.  In one dimension this
follows as follows:
\begin{eqnarray}
   \beta \partial^{\mathrm{us}}_x f & \equiv & \Theta(\beta) \beta
   \frac{f{\scriptstyle (x+h)} - f{\scriptstyle (x)}}{h} +
   \Theta(-\beta) \beta \frac{f{\scriptstyle (x)} - f{\scriptstyle
   (x-h)}}{h} \\
   & = & \left( \beta + |\beta| \right) \frac{f{\scriptstyle (x+h)} -
   f{\scriptstyle (x)}}{2 h} \\
   & & {} - \left( -\beta + |-\beta| \right) \frac{f{\scriptstyle (x)}
   - f{\scriptstyle (x-h)}}{2 h} \\
   & = & \beta\, \frac{f{\scriptstyle (x+h)} - f{\scriptstyle (x-h)}}{2
   h} + |\beta| \frac{f{\scriptstyle (x+h)} - 2 f{\scriptstyle (x)} +
   f{\scriptstyle (x-h)}}{2 h} \\
   & = & \beta\, \partial_x f + \frac{|\beta|}{2}\, h\, \partial_{xx}
   f
\end{eqnarray}
where $h$ is the grid spacing.  The first line is the first order
upstream difference.  I use the Heaviside--function $\Theta(x)$ which
has the value $0$ for $x<0$ and $1$ for $x>0$.  It is also $2 |x|
\Theta(x) = x + |x|$.

The second kind of artificial diffusion, SM, is a generic smoothing
operator, and is independent of the shift.  It is necessary is some
simulations.  I found that, without this diffusion, the system of
equations is unstable against high frequency noise, and the
amplification rate depends on the grid resolution.  This points to
either a bad discretisation scheme, or an ill-posed problem already at
the physical level.  I did not want to examine this kind of
instability any further, so I decided to remove it by using some
suitable artificial diffusion.

Both kinds of artificial diffusion terms remove high frequency modes,
making the simulation stable against such noise.  They also dissipate
energy, dampening the amplitude of gravitational waves.  They
introduce a second order error into the system, turning it into a
first order correct system only.

\section{Elliptic integration}

In order to transform the constraints and the gauge conditions into
elliptic equations, I add vector potentials to the set of equations.
That means that enforcing the constraints requires going back and
forth between different representations, and that introduces
discretisation errors.

\subsection{Variable transformations}

The step of introducing the vector potentials does not introduce
errors, as the vector potentials are set to zero initially.  However,
going back to the original system then does introduce discretisation
errors.  This can be demonstrated by using the prototype of a
constraint equation
\begin{eqnarray}
   \label{firsteqn} F_i & = & \partial_j h_{ij} \quad.
\end{eqnarray}
I first introduce a vector potential $W_i$, leading to
\begin{eqnarray}
   \label{newvar} h_{ij} & = & h^*_{ij} + \partial_j W_i \quad.
\end{eqnarray}
This turns the constraint equation into
\begin{eqnarray}
   \label{secondeqn} F'_i & = & \partial_j h^*_{ij} + \partial_{jj}
   W_i \quad,
\end{eqnarray}
which is now elliptic in $W_i$.  It can be solved to arbitrary
accuracy.  Note that $F_i$ and $F'_i$ have fundamentally different
properties.

After solving $F'_i$ for $W_i$, one calculates the new $h_{ij}$ from
its definition (\ref{newvar}).  This requires taking a first
derivative of $W_i$.  In order to evaluate the original constraint
$F_i$, one has to take a first derivative of $h_{ij}$, which contains
then two successive first derivatives of $W_i$.  This is, on the
discrete level, quite different from the constraint $F'_i$, which
involves taking one second derivative of $W_i$.  Both these steps
introduce different discretisation errors.  That means, even after
solving $F'_i$ to very high accuracy, $F_i$ will still contain an
error of the order of the discretisation error.  Furthermore, taking
two successive first derivatives tends to introduce a high frequency
noise.

This problem might be eliminated by using staggered grids.  On a
staggered grid, two consecutive first derivatives are the same as one
second derivative.

\subsubsection{Double first derivatives}

Using two consecutive first derivatives for solving the equations
would be unstable.  A second order centred first derivative has the
stencil weights
\begin{equation}
\label{stencil-first}
\begin{array}{ccc}
-\frac{1}{2} & 0 & +\frac{1}{2}
\end{array}
\end{equation}
while a second derivative has the weights
\begin{equation}
\label{stencil-second}
\begin{array}{ccc}
+1 & -2 & +1
\end{array}
\end{equation}
Taking two consecutive first derivatives leads to an effective stencil
with the weights
\begin{equation}
\label{stencil-firstfirst}
\begin{array}{ccccc}
+\frac{1}{4} & 0 & -\frac{1}{2} & 0 & +\frac{1}{4}
\end{array}
\end{equation}
Stencil (\ref{stencil-firstfirst}) contains two zeros.  If one
calculates the derivative of e.g.\ an even-numbered grid point, then
only other even-numbered grid points contribute.  This decouples odd-
and even-numbered grid points.  If one uses this stencil while solving
an elliptic equation, then one obtains effectively two independent
solutions, namely one for the odd, and one for the even grid points.
These solutions are coupled through their initial data and boundary
conditions only and will differ slightly.  Going from one grid point
to the next, one jumps between these two solutions, and that is
equivalent to a high frequency mode.

\subsection{Elliptic solvers}

Solving elliptic equations is expensive.  The elliptic solver needs
the majority (more than 90\%) of the total computing time in the Tiger
code.  This is partly my fault, as I use methods that are known to be
subideal and very slow.

Currently I use PETSc \cite{petsc-home-page, petsc-manual,
petsc-efficient} to solve the elliptic equations.  PETSc uses a
Newton-like method to reduce the nonlinear equations to linear ones,
and then uses Krylov-subspace methods to solve these.  I calculate the
Jacobian that is necessary for the Newton-like method numerically.
This is known to be a very expensive operation; it is about ten to
twenty times slower than a hand-coded Jacobian.  On the other hand,
hand-coded Jacobians would need to be written and debugged.

I also have a simple iterative Jacobi solver for testing purposes.
This solver is only a toy, and is only usable for very small problems.

I estimate that a nonlinear multigrid solver would be several orders
of magnitude faster than the way I (ab)use PETSc, as it does not need
to calculate the Jacobian.  An experimental implementation of a
nonlinear full approximation storage multigrid solver for the
two-dimensional Poisson equation with excision shows promising
results.

\section{Coding equations}
\label{coding-equations}

As mentioned in section \ref{tiger}, I chose to code the equations by
hand instead of using a symbolic algebra package.  This only makes
sense because it is possible to make the computer code look
sufficiently close to an explicit mathematical notation.  The standard
mathematical notation used in general relativity is full of
context-sensitive abbreviations, such as the Einstein sum convention,
that cannot be used in an unambiguous notation.

One of the key points to obtain a readable source code is to make use
of the local character of the equations, leading to a \emph{pointwise}
programming style.  A tensor such as $K_{ij}$ is really a tensor field
$K_{ij}(x^k)$, which becomes a grid function $K_{ij}(x,y,z)$ after
discretisation.  Dealing with a set of five indices $(i,j,x,y,z)$ at
the same time leads to an awkward notation.  Hence I chose to deal
with one grid point at a time only.

As an example, I present the code for one of the more complicated
equations, namely the time evolution equation of the extrinsic
curvature.  It is given in appendix \ref{adm-time-evolution} as:
\begin{eqnarray}
   \partial_t K_{ij} & = & - \nabla_i \nabla_j \alpha + \alpha \left[
   R_{ij} - 2 K_{il} K^l_j + K K_{ij} \right] \\
   \nonumber & & {} + K_{lj} \beta^l_{,i} + K_{il} \beta^l_{,j} +
   \beta^l K_{ij,l}
\end{eqnarray}

When translated into a Fortran subroutine for Cactus, this equation
becomes
{\small
\begin{verbatim}
subroutine calc_extcurv_dot &
     (gu, ri, kk, dkk, alfa, ggalfa, beta, dbeta, kk_dot)
  CCTK_REAL, intent(in)  :: gu(3,3)
  CCTK_REAL, intent(in)  :: ri(3,3)
  CCTK_REAL, intent(in)  :: kk(3,3), dkk(3,3,3)
  CCTK_REAL, intent(in)  :: alfa, ggalfa(3,3)
  CCTK_REAL, intent(in)  :: beta(3), dbeta(3,3)
  CCTK_REAL, intent(out) :: kk_dot(3,3)
  integer :: i,j,k,l
  ! K_ij,t = - D_i D_j alpha
  !          + alpha [ R_ij - 2 K_ik K^kj + K K_ij ]
  !          + K_kj d_i beta^k + K_ik d_j beta^k + beta^k d_k K_ij
  do i=1,3
     do j=1,3
        kk_dot(i,j) = - ggalfa(i,j) + alfa * ri(i,j)
        do k=1,3
           do l=1,3
              kk_dot(i,j) = kk_dot(i,j) &
                   - 2*alfa * gu(k,l) * kk(i,k) * kk(j,l) &
                   +   alfa * gu(k,l) * kk(k,l) * kk(i,j)
           end do
           kk_dot(i,j) = kk_dot(i,j) &
                + kk(k,j) * dbeta(k,i) + kk(i,k) * dbeta(k,j) &
                + beta(k) * dkk(i,j,k)
        end do
     end do
  end do
end subroutine calc_extcurv_dot
\end{verbatim}
}

In the above routine, the variables \texttt{gu}, \texttt{ri}, etc.\
represent the corresponding tensors at a single point in the time
slice only.  This routine is called once for every grid point for
every right hand side evaluation.  Such a coding convention needs
powerful compilers and optimisers to yield good performance, and
likely cannot be used on vector machines.  This also shows that
programming styles are heavily influenced by the current software and
hardware technology.  However, I strove for clarity and simplicity
rather than for performance.

% LocalWords:  eschnett Exp PETSc petsc handcoded Jacobians gu

% -*-LaTeX-*-
% $Header: /home/eschnett/cvs/diss/defs.tex,v 1.15 2003/01/10 16:39:49 eschnett Exp $

\chapter{Definitions}
\label{defs}

\section{Glossary of terms}

\begin{description}

   \item[Agave:] A numerical code for the vacuum Einstein equations,
   descending from the Grand Challenge

   \item[Cactus:] A framework for solving three-dimensional
   time-dependent systems of partial differential equations, see
   \cite{cactus-grid, cactus-tools, cactus-webpages}

   \item[coordinate condition:] A condition that determines lapse
   $\alpha$ and shift $\beta^i$ (equivalent to the term \emph{gauge
   condition})

   \item[excision:] A method to treat singularities.  One cuts a
   region of the spacetime that contains the singularity out of the
   simulation domain, introducing an inner boundary.  This is possible
   as long as the boundary is located within the event horizon.  As no
   physical information can propagate out of the event horizon, the
   details of how the excision boundary is handled cannot influence
   the simulation outside the event horizon

   \item[gauge condition:] A condition that determines lapse $\alpha$
   and shift $\beta^i$ (equivalent to the term \emph{coordinate
   condition})

   \item[gauge evolution condition:] A choice of lapse $\alpha$ and
   shift $\beta^i$ that imposes no restrictions on the three-metric
   $\gamma_{ij}$ and extrinsic curvature $K_{ij}$.  For example,
   geodesic slicing ($\alpha=1$) is a gauge evolution condition

   \item[gauge fixing condition:] A condition on the three-metric
   $\gamma_{ij}$ and extrinsic curvature $K_{ij}$ that also leads to a
   prescription for lapse $\alpha$ and shift $\beta^i$ via the ADM
   time evolution equations.  For example, maximal slicing ($K=0$) is
   a gauge fixing condition

   \item[Maya:] A numerical code, descending from Agave, see
   \cite{maya}

   \item[performance:] (of a code) the quality of the result of a
   simulation, i.e.\ a measure of the magnitude of the numerical
   errors.  This has nothing to do with the speed of the code on a
   particular hardware

   \item[punctures:] A method to treat singularities.  The singularity
   is contained in a conformal factor $\psi$, and all other quantities
   can be chosen so that they remain finite.  This is the case e.g.\
   for Brill--Lindquist black holes (see section
   \ref{brill-lindquist}).  The singularity is placed in between grid
   points.  The conformal factor and its derivatives are known
   analytically.  With maximal slicing ($K=0$) and normal coordinates
   ($\beta^i=0$), the conformal factor is then constant in time

   \item[robust stability:] A practical definition for the notion of
   stability of a code that can be tested empirically.  See
   \cite{robust-stability, robust-stability-2}

   \item[Tiger:] A numerical code, descending from Maya

   \item[time slice:] Another term for a spacelike hypersurface

\end{description}

\section{Abbreviations}

\begin{description}

   \item[ADM:] Arnowitt--Deser--Misner.  An initial--boundary--value
   formulation for the Einstein equations, see \cite{adm}

   \item[AEI:] the Albert--Einstein--Institut in Golm near Potsdam,
   Germany

   \item[BH:] black hole

   \item[BSSN:] Baumgarte--Shapiro--Shibata--Nakamura.  An ADM-like
   initial--bound\-ary--value formulation for the Einstein equations,
   see \cite{nakamura1, nakamura2, shibata-nakamura,
   baumgarte-shapiro}

   \item[ctADM:] conformal traceless ADM system

   \item[KST:] the Kidder--Scheel--Teukolsky system; see \cite{kst}

   \item[MPI:] Max--Planck--Institut, e.g.\ the AEI, which is also
   called the Max--Planck--Institut f\"ur Gravitationsphysik

   \item[MTW:] the book \emph{Gravitation} by Misner, Thorne, and
   Wheeler; see \cite{mtw}

   \item[NS:] neutron star

   \item[PSU:] Penn State University

   \item[TAT:] Theoretische Astrophysik T\"ubingen, the Abteilung
   f\"ur Theoretische Astrophysik in the Institut f\"ur Astronomie und
   Astrophysik of the Fakult\"at f\"ur Mathematik und Physik at the
   Eber\-hard--Karls--Universit\"at T\"ubingen (yes, this is in
   Germany)

   \item[TGR:] the name of the T\"ubingen General Relativity (Tiger)
   code

\end{description}

\section{Symbols}
\label{symbols}

It goes without saying that Greek letters denote four-vector indices
$0 \ldots 3$, while Latin letters denote three-vector indices $1
\ldots 3$.  I follow the sign conventions of MTW \cite{mtw}.

\begin{list}{}{
% \bigskip, \medskip
%\setlength{\parskip}{0ex}
\setlength{\itemsep}{0ex}
\setlength{\leftmargin}{4em}
\setlength{\rightmargin}{1em}
\setlength{\itemindent}{0em}
\setlength{\labelsep}{1em}
\setlength{\labelwidth}{3em}
}

   \item[\textsf{\textbf{Four-quantities:}}]

   \item[$\eta_{\mu\nu}$] Minkowski metric.  $\eta_{\mu\nu} =
   \mathrm{diag}(-1,+1,+1,+1)$

   \item[$g_{\mu\nu}$] four-metric

   \item[$G_{\mu\nu}$] Einstein tensor

   \item[$T_{\mu\nu}$] energy density tensor

   \bigskip

   \item[\textsf{\textbf{Three-quantities:}}]

   \item[$\delta_{ij}$] identity tensor.  $\delta_{ij} =
   \mathrm{diag}(+1,+1,+1)$

   \item[$\gamma_{ij}$] three-metric, see appendix \ref{adm-variables}

   \item[$K_{ij}$] extrinsic curvature, see appendix
   \ref{adm-time-evolution}

   \item[$\alpha$] lapse, see appendix \ref{adm-variables}

   \item[$\beta^i$] shift, see appendix \ref{adm-variables}

   \item[$\Gamma^i_{jk}$] connection coefficients, see appendix
   \ref{adm-ricci}

   \item[$\Gamma^i$] contracted connection coefficients.  $\Gamma^i =
   \gamma^{jk} \Gamma^i_{jk}$

   \item[$R_{ij}$] Ricci tensor, see appendix \ref{adm-ricci}

   \item[$H$] Hamiltonian constraint, see appendix
   \ref{adm-constraints}

   \item[$M_i$] momentum constraint, see appendix
   \ref{adm-constraints}

   \bigskip

   \item[\textsf{\textbf{Conformal quantities:}}]

   \item[$\tilde \cdot$] a tilde denotes a conformal quantity, which
   has its indices raised and lowered with the conformal metric

   \item[$\psi$] $\psi^4$ is the conformal factor, see appendix
   \ref{ctadm-variables}

   \item[$\tilde \gamma_{ij}$] conformal three-metric, see appendix
   \ref{ctadm-variables}

   \item[$K$] trace of the extrinsic curvature, see appendix
   \ref{ctadm-variables}

   \item[$A_{ij}$] traceless extrinsic curvature, see appendix
   \ref{ctadm-variables}

   \item[$\tilde A_{ij}$] conformal traceless extrinsic curvature, see
   appendix \ref{ctadm-variables}

   \item[$V_i$] vector potential for the extrinsic curvature, see
   appendix \ref{ctadm-enforce-constraints}

   \bigskip

   \item[\textsf{\textbf{Coordinate quantities:}}]

   \item[$\bar \cdot$] a bar denotes a coordinate quantity, which has
   its indices raised and lowered with the coordinate metric (i.e.\
   $\delta_{ij}$ for Cartesian coordinates)

   \item[$\bar h_{ij}$] traceless conformal three-metric, see appendix
   \ref{ctadm-gauge}

   \item[$F_i$] gauge condition for the metric, akin to a constraint
   variable, see appendix \ref{ctadm-gauge}

   \item[$W_i$] vector potential for the traceless three metric, see
   appendix \ref{ctadm-enforce-gauge}

\end{list}

% LocalWords:  eschnett Exp ij diag jk

%% \include{publications}
% -*-LaTeX-*-
% $Header: /home/eschnett/cvs/diss/postfix.tex,v 1.8 2002/10/28 13:25:07 eschnett Exp $

%% \cleardoublepage
%% \listoffigures

%% \cleardoublepage
%% %% \addcontentsline{toc}{chapter}{Index}
%% \printindex

\cleardoublepage
%% \addcontentsline{toc}{chapter}{Bibliography}
% base styles are: abbrv, alpha, apalike, ieeetr, plain, siam, unsrt
% natbib styles are: abbrvnat, plainnat, unsrtnat
% ams styles are: amsalpha, amsplain
% good styles are: plain, amsplain
%% \bibliographystyle{amsplain}     % numbered alphabetically
%% \bibliographystyle{abbrv}        % numbered alphabetically, only initials
%% \bibliographystyle{alpha}        % initials + year
%% \bibliographystyle{apalike}      % authors + full year
%% \bibliographystyle{ieeetr}      % numbered alphabetically
%% \bibliographystyle{plain}      % numbered alphabetically
%% \bibliographystyle{siam}      % numbered alphabetically, small caps
%% \bibliographystyle{unsrt}      % numbered alphabetically, unsorted
%% \bibliographystyle{plainnat}      % authors + full year, bogus
\bibliographystyle{amsalpha}      % initials + year
\bibliography{nr}

\clearpage
\pagestyle{plain}

\begin{quote}
And crawling on the planet's face\\
Some insects called the human race\\
Lost in time, and lost in space\\
And meaning.
\end{quote}
\hspace{5em}RHPS \cite{rhps}

\end{document}